\documentclass[3p,english,12pt,times,number,nopreprintline]{elsarticle}
\usepackage{formatting-style}

\begin{document}

\begin{frontmatter}

\title{Spectral Dependence}

\author[kaust]{Hernando Ombao \orcidID{0000-0001-7020-8091}}
\cortext[cor]{Corresponding author}
\corref{cor}
\author[oslomet]{Marco Pinto \orcidID{0000-0001-6495-1305}}
\address[kaust]{Statistics Program,
King Abdullah University of Science and Technology (KAUST), 
Saudi Arabia}
\address[oslomet]{Institutt for maskin, elektronikk og kjemi, Oslo Metropolitan University, Norway.}

\begin{abstract}
This paper presents a general framework for modeling dependence in multivariate time series.
Its fundamental approach relies on decomposing each signal in a system into various 
frequency components and then studying the dependence properties through these oscillatory activities. 
The unifying theme across the paper is to explore the strength of dependence and possible 
lead-lag dynamics through filtering. The proposed framework is capable of representing both 
linear and non-linear dependencies that could occur instantaneously or after some delay 
(lagged dependence). Examples for studying dependence between oscillations are illustrated 
through multichannel electroencephalograms. These examples emphasized that some of 
the most prominent frequency domain measures such as coherence, partial coherence, 
and dual-frequency coherence can be derived as special cases under this general framework. 
This paper also introduces related approaches for modeling dependence through 
phase-amplitude coupling and causality of (one-sided) filtered signals.
\end{abstract}

\begin{keyword}
Causality\sep
Cross-coherence\sep
Dual frequency coherence\sep
Fourier transform\sep
Harmonizable processes\sep
Multivariate time series\sep
Spectral Analysis.
\end{keyword}

\end{frontmatter}

\tableofcontents

\section{Introduction}

Multivariate time series data, sets of a discretely sampled sequence of
observations, are the natural approach for analyzing phenomena that
display simultaneous, interacting, and time-dependent stochastic
processes. As a consequence, they are actively studied in a wide variety
of fields: environmental and climate science
\citep{AirQuality-Sethi-2020, AssessmentRoleRenewable-Usman-2020, EarthSystemData-Mahecha-2020, EnvironmentalConsequencesPopulation-Pham-2020, GPARXBasedStructuralDamage-Tatsis-2020, ClusteringNonlinearNonstationary-Harvill-2017},
finance
\citep{EffectHiddenSource-Koutlis-2021, ChangepointMethodsMultivariate-Hlavka-2020, PriceSpilloversRare-Reboredo-2020, OilPricesPolicy-Qin-2021, MultipleChangePoints-Bai-2020, PortfolioOptMixture-Ravagli-2020},
computer science and engineering
\citep{RegularizedEstHigh-Lin-2018, MultivariateGPVARModels-Avendano-Valencia-2020, RvwOutlierAnomaly-Blazquez-Garcia-2020, FaultTolerantEarlyClassif-Gupta-2020, SAXARMDeviantEvent-Park-2020, ApplyingDL-Huang-2021},
public health
\citep{PeerPressureOveruse-Giorno-2020, DynamicsLifeExpectancy-Aburto-2020, EstimatingLongrun-Baum-2020, LifeExpectancyBirth-MartinCervantes-2020, AgeStructuralTransitions-Lin-2020},
and neuroscience
\citep{PersonTimeVaryingVector-Chen-2020, EEGConnectivityPattern-Steinmann-2020, EffectiveEEGConnectivity-Goyal-2020, GrangerCausalInference-Soleimani-2020, GrangerCausalityInference-Manomaisaowapak-2020, HightructureLearning-Suotsalo-2020, LinearNonlinearEEGBased-Schoenberg-2020, LearningCommonGranger-Manomaisaowapak-2020, OpticalBrain-Pinto-Orellana-2020, BrainWaveNetsAre-Nascimento-2020, DetectingDynamicCommunity-Ting-2021, ChangepointDetection-Jiao-2021, SemiParamTimeSeries-Maia-2020, SpectralProp-Sundararajan-2020, ConexConnectLearningPatterns-Guerrero-2021, SpectralApproach-Ombao-2019}.
Considering the inherent complexity of those studied phenomena, one of
the most common challenges and tasks is identifying and explaining the
interrelationship between the various components of the multivariate
data. Thus, the purpose of this paper is to provide a summary of the
various characterizations of dependence between the elements of a
multivariate time series. The emphasis will be on the spectral measures 
of dependence which essentially examines the cross-relationships between 
the various oscillatory activities in these signals. These measures will be 
demonstrated, for the most part, through the oscillations derived from 
linear filtering. 

Suppose ${\bf X}(t) = [X_1(t), \ldots, X_P(t)]'$ is a multivariate
time series with $P$ components. This abstraction is capable of
providing a general framework that describes a broad range of scenarios
and phenomena. For instance, in environmental studies, components
$X_1, \ldots, X_P$ could represent recordings from various air
pollution sensors at a fixed location. Similarly, the same framework
could describe wind velocity recordings at $P$ different geographical
locations. In a neuroscience experiment, a component $X_i$ could be
the measurement of brain electrical, or hemodynamic, activity from a
specific sensor (electrode) which is placed either on the scalp or on the surface
of the brain cortex. The key question that we will address through
various statistical models and data analysis tools is to understand the
a) nature of marginal dependence between $X_p(t)$ and $X_q(t)$ or b)
between $X_p(t)$ and $X_q(t)$ conditional on the other components in
the data.

This work is largely motivated from a neuroscience perspective. The
brain circuitry can be conceived as the integration of a sensory, motor,
and cognitive system that receives, processes, and reacts to external
impulses or internal auto-regulation activities
\citep[pp.~80-91]{BrainArchitectureUnderstanding-Swanson-2003}.
Communication and feedback between those structures enable brain
functions. For instance, memory is believed to rely on the hippocampus
because it is at the center in the signal flow from and to the cortical
areas, such as the orbitofrontal cortex, olfactory bulb, and superior
temporal gyrus
\citep[p.~160-165]{ArchitectureEntorhinalCortex-Witter-2017, BrainOrganizationMemory-McGaugh-1990}.
From a macro perspective, those memory flows also imply a high activity
in the temporal brain region
\citep[p.~164]{BrainOrganizationMemory-McGaugh-1990}. From an analytic
perspective, these signal interactions can be studied as undirected
dependencies (functional connectivity) $X_p(t)\Leftrightarrow X_q(t)$;
or directed, or causal, networks (effective connectivity)
$X_p(t)\Rightarrow X_q(t)\,\wedge\,X_q(t)\Rightarrow X_p(t)$.
Additionally, current imaging techniques allow us to understand those
interactions at different biological levels: (a.) through neuron
hemoglobin changes (energy consumption) using functional magnetic
resonance imaging (fMRI) or functional near-infrared spectroscopy
(fNIRS); (b.) through measurements of the electrical activity at the
scalp, (electroencephalogram, EEG), at the cortical surface
(electrocorticogram, ECoG), or in the extracellular environment (local
field potentials, LFPs).

\begin{figure}[t]
\centering%
\includegraphics[width=0.9\textwidth]{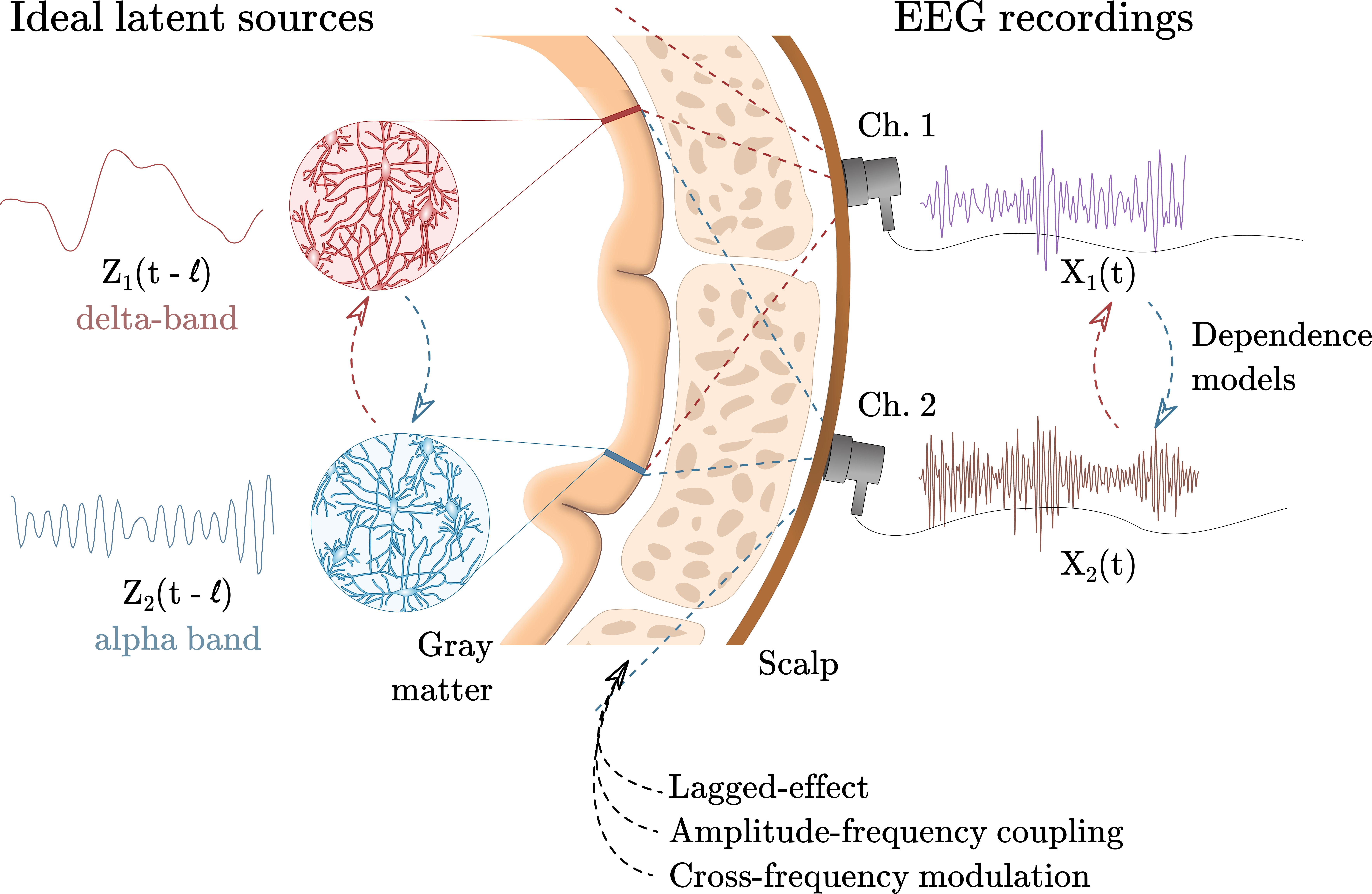}
\caption{%
Abstraction of dependence in biomedical signals. It is assumed that hypothetical latent sources
could generate neural oscillations (with their own frequency characteristics)
that can interact among them before being recorded as the scalp.
Different abstractions that model the sources' interactions and
the mixing models are explained in this paper.
\label{fig:graphical-abstract}
}
\end{figure}

We should emphasize that there are several models that aim to represent
the different types of dependence (Figure~\ref{fig:graphical-abstract}).
The most common measure of dependence is cross-covariance (or
cross-correlation). In the simple case where $\EX{X_p(t)} = 0$ and
$\VAR{X_p(t)} = 1$ for all $p=1, \ldots, P$ and all time $t$, then
the cross-correlation between $X_p$ and $X_q$ is
\begin{equation}
\rho_{pq}(0) = \EX{X_p(t) X_q(t)} = \int u_p u_q f_{pq}(u_p, u_q)\; du_p du_q
\end{equation}
where $f_{pq}(u_p, u_q)$ is the joint probability density function of
$(X_p(t), X_q(t))$ and the integral is evaluated globally over the
entire support of $(X_p(t), X_q(t))$. Cross-correlation provides a
simple metric that measures the linear synchrony between a pair of
components across the entire support of their joint distribution. When
$\vert \rho_{pq}(0) \vert$ has a value close to $1$, we
conclude that there is a strong linear dependence between $X_p$ and
$X_q$. It is obvious that cross-correlation does not completely
describe the nature of dependence between $X_p$ and $X_q$. First,
dependence goes beyond linear associations, and hence this paper
addresses some types of non-linear dependence between components.
Second, dependence may vary across the entire support. That is, the
association between $X_p$ and $X_q$ at the "center" of the
distribution may be different from its "tails" (e.g., extrema). Third,
since time series data can be viewed as superpositions of sine and
cosine waveforms with random amplitudes, the natural aim is to identify
the specific oscillations that drive the linear relationship. This will
be the main focus of the models and methods that will be covered in this
paper.

To consider spectral measures of dependence (i.e., dependence between
oscillatory components), our starting point will be the Cram\'er
representation of stationary time series. Under stationarity, we can
decompose both $X_p$ and $X_q$ into oscillations at various
frequency bands. The key elements of the Cram\'er representation are the
Fourier basis waveforms
$\bigl\{ \exp(i 2 \pi \omega t), \omega \in (-\frac 1 2, \frac 1 2) \bigr\}$ and the
associated random coefficients $\{ dZ(\omega) \}$ which is an
orthogonal increment random process that satisfies
$\EX{dZ(\omega)} = 0$ and $\cov(dZ(\omega), dZ(\omega')) \ne 0$ for
$\omega \ne \omega'$. The Cram'er representation for
$(X_p(t), X_q(t))$ is given by
\begin{align}
\left ( 
\begin{matrix}
X_p(t) \\
X_q(t) \\
\end{matrix}
\right) =& 
\left( 
\begin{matrix} 
\exp(i 2 \pi \omega t)\, dZ_p(\omega) \\
\exp(i 2 \pi \omega t)\, dZ_q(\omega) 
\end{matrix}
\right)
\end{align}
and $X_q$ by examining the nature of synchrony between the
$\omega$-oscillation in $X_p$ and the $\omega$-oscillation in
$X_q$ which are, respectively, $\exp(i 2 \pi \omega t) dZ_p(\omega)$
and $\exp(i 2 \pi \omega t) dZ_q(\omega)$. This idea will be further
developed in the next sections of this paper. We emphasize that
stationarity assumption can be held in resting-state conditions
\citep[p.~188]{HandbookNeuroimagingData-Ombao-2017} or within reasonable
short time intervals \citep[p.~20]{EEGSigProcessing-Sanei-2007}.

Most brain signals exhibit non-stationary behavior, which may be
reflected changes in either (a.) the mean level, or (b.) the
variance at some channels, or (c.) the cross-covariance structure
between some pairs of channels. Note here that (a.) is a condition on
the first moment while (b.) and (c.) are conditions on the second
moment. Moreover, (a.) and (b.) are properties within a channel, while
(c.) is a property that describes dependence between a pair of channels.
There is no measure that completely describes the nature of
dependence between channels. The most common pair of measures consists
of the cross-covariance and cross-correlation, whose equivalent measures
in the frequency domain are the cross-spectrum and cross-coherence,
respectively. This paper will focus on the frequency domain measures and
thus dig deeper into being able to identify the oscillations that drive
the dependence between a pair (or group) of channels. Under
non-stationarity, the dependence structure can change over time. Our
approach here is to slide a localized window across time and estimate
the spectral properties within each window. This approach is proposed in
\citep{Priestley:1965} and then reformulated in
\citep{Dahlhaus:2012locally} to establish an asymptotic framework for
demonstrating the theoretical properties of the estimators.

\section{EEG spectral characteristics}\label{sec:eeg-dataset}
To illustrate the various spectral dependence measures, we shall focus
on the analysis of EEG signals. EEG is a noninvasive imaging technique
that collects electrical potential at the scalp gathered from
synchronized responses of groups of neurons (``signal sources'')
\citep[p.~4]{HandbookEEGInterpretation-Tatum-2008} \citep[p.~555]{HandbookNeuroimagingData-Ombao-2017}
that are perpendicular to the scalp and dynamically organized in
neighborhoods with scales of a few centimeters
\citep[p.~5]{AtlasEEGPatterns-Stern-2013}. Naturally, EEG is affected by
the volume conduction of the signals over tissue and skull
\citep[p.~144]{ElecFieldsBrain-Nunez-2006}. Despite its limitations, the
portability and inexpensiveness allow the integration of EEG in clinical
settings and cognitive experiments that require naturalistic
environments (in contrast, fMRI experiments require the participants to
be in a supine position in a restricted space). In addition, the EEG
high temporal resolution enables to capture the temporal dynamics of the
neuronal activity. Therefore, through spatial, temporal, and
morphological EEG patterns, some conditions can be diagnosed \citep[p.3,
19-22]{AtlasEEGPatterns-Stern-2013}. For instance, a specific metric
obtained from EEG frequency properties measured on the channel Cz is
considered an FDA-approved clinical method to assess attention deficit hyperactivity disorder (ADHD)
\citep{IntegrationEEGBiomarker-Snyder-2015}. Thus, EEGs have been
extensively used in cognitive neuroscience, neurology, and psychiatry to
study the neurophysiological basis of cognition and neuropsychiatric disorders: motor abilities
\citep{RestingstateCorticalConnectivity-Wu-2014}, anesthetic
similarities with comma \citep{GeneralAnesthesiaSleep-Brown-2010},
encephalopathy \citep{RvwFractalDimension-Jacob-2019}, schizophrenia
\citep{EEGConnectivityPattern-Steinmann-2020}, addictions
\citep{EEGFreqBands-Newson-2019}, spectrum autism disorder
\citep{EEGFreqBands-Newson-2019, EEGEEGSig-Ibrahim-2018}, depression
\citep{ElectroencephalogramEEGSig-Mahato-2019, MajorDepressionDetection-Liao-2017, NonInvasiveEEG-Mantri-2015},
and ADHD
\citep{IntegrationEEGBiomarker-Snyder-2015, EEGClassifADHD-Mohammadi-2016}.
Here, we will illustrate some methods that characterize the dynamics of
the inter-relationships between the activity measured at different
channels.

\begin{figure}
\centering%
\includegraphics[width=\textwidth,height=6cm]{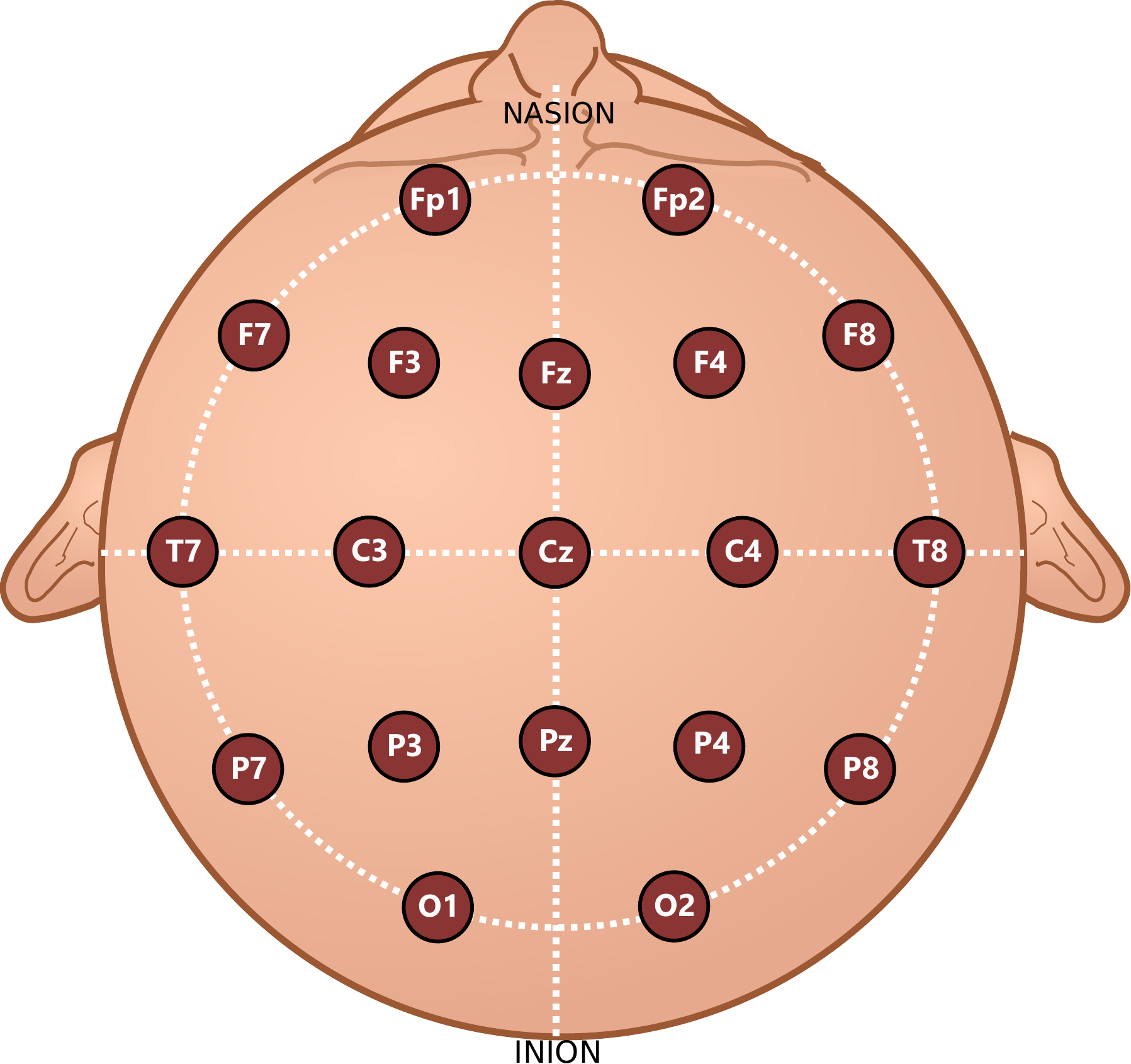}
\caption{Electrode positioning in the 10-20 standard layout.}
\label{fig:layout-1020}
\end{figure}

In this paper, we use the EEG dataset related to a mental health
disorder collected by Nasrabadi et al.
\citep{EEGDataADHD-MotiNasrabadiAli-2020}. This data comprises EEGs,
sampled at 128Hz, from 60 children with attention deficit hyperactivity
disorder (ADHD) and 60 children with no registered psychiatric disorder
as a control group. These electrical recordings were collected from 19
channels evenly distributed on the head in the 10-20 standard layout (Figure~\ref{fig:layout-1020}).
Average recording from both ear lobes (A1 and A2) was used as electrode
references. The experiment was intended to show potential differences in
the brain response under visual attention task
\citep{EEGDataADHD-MotiNasrabadiAli-2020, DetectingADHDChildren-Allahverdy-2011, EEGClassifADHD-Mohammadi-2016}.
Therefore, the 120 participants were exposed to a series of images that
they should count. The number of images in each set ranged from 5 to 16
with a reasonable size in order to allow them to be recognizable and
countable by the children. Each collection of pictures was displayed
without interruptions in order to prevent distraction from the subjects.

\begin{figure}
\centering%
\includegraphics[width=1\textwidth,height=\textheight]{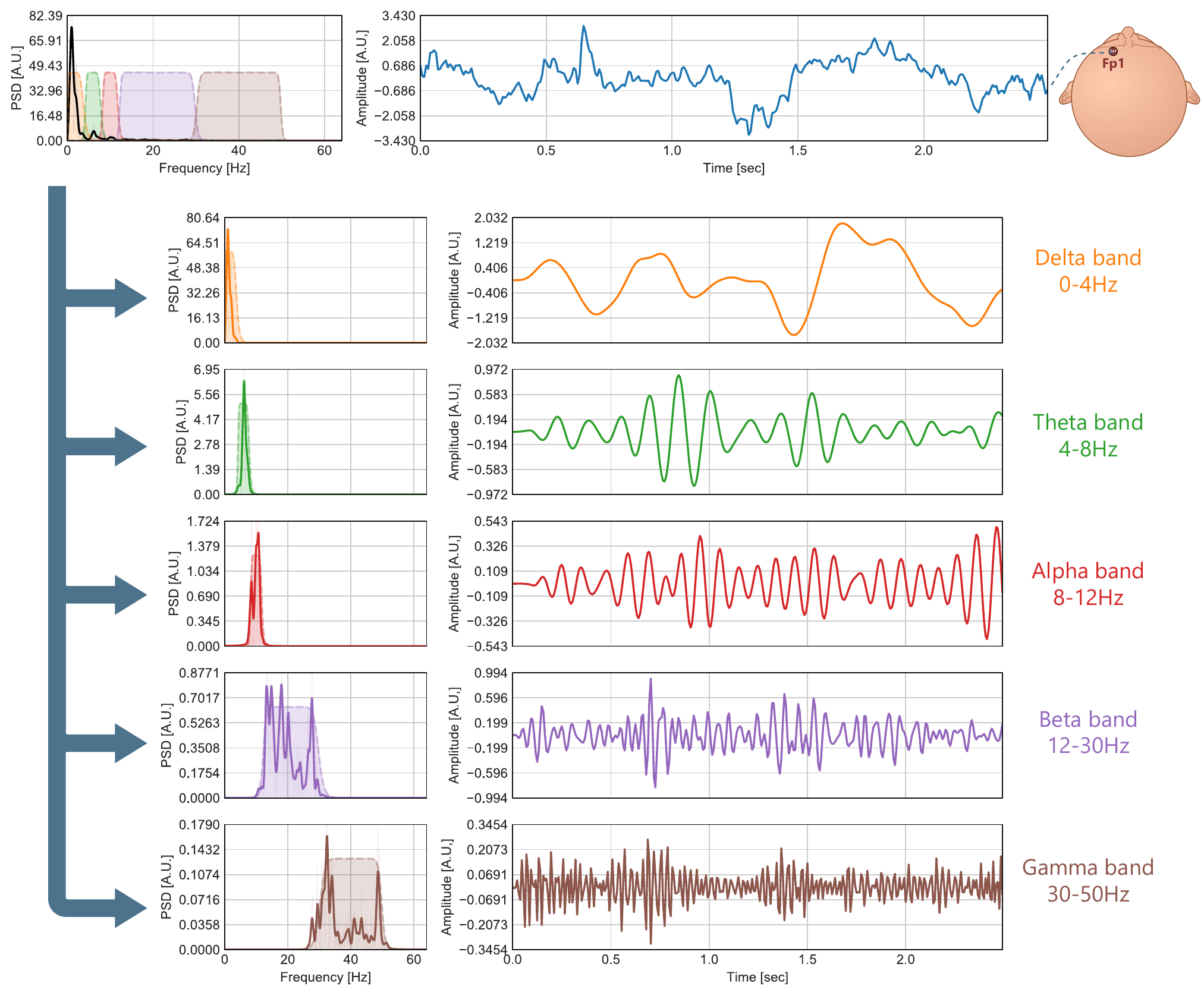}
\caption{Frequency decomposition of an EEG, collected at channel Fp1,
into the five brain rhythms: delta, theta, alpha, beta and gamma.}
\label{fig:eeg-decomposition}
\end{figure}

Prior to analysis, EEGs undergo significant pre-processing steps. These
procedures are not unique to EEGs, or brain signals, as most data often
need some cleaning before statistical modeling. Several attempts have
been performed to standardize those pre-processing steps
\citep{PREPPipelineStandardized-Bigdely-Shamlo-2015, AutorejectAutomatedArtifact-Jas-2017, AutomagicStandardizedPreprocessing-Pedroni-2019, MarylandDevelopmental-Debnath-2020, EEGIntegratedPlatform-Desjardins-2021}.
In general, pre-processing aims to increase the quality of the recorded
signal by \citep{HowSensitiveAre-Robbins-2020}: (a.) removing the effect
of the electrical line (electrical interference at 50Hz or 60Hz due to
the electrical source); (b.) removing artifacts due to eye movements,
eye blinks or muscular movements; (c.) detecting, removing or repairing
bad quality channels; (d.) filter non-relevant signal components; and
(e.) re-referencing the signal to improve topographical localization. In
this scenario, filtering is a crucial step that removes portions of the
signal that could not be related to cognitive or physiological
processes, such as extremely high-frequency components. In this dataset,
we applied a band-pass filter on the frequency interval (0.0-70.0) Hz
and segmented the signal into the main ``brain rhythms''
\citetext{\citealp[p.610]{HandbookNeuroimagingData-Ombao-2017}; \citealp[p.~33-34,
169-170,
413-414]{AtlasEEGPatterns-Stern-2013}; \citealp[p.~12]{ElecFieldsBrain-Nunez-2006}}:
frequency range into the delta band: (0.5, 4.0) Hertz, theta band: (4.0, 8.0)
Hz, alpha band: (8.0, 12.0) Hertz, beta band: (12.0, 30.0) Hertz, and gamma band:
(30.0, 50.0) Hertz. A non-causal second-order Butterworth filter was used
to perform this frequency filtering: Figure~\ref{fig:eeg-decomposition}
shows an example of this decomposition where the shadow regions show the
magnitude response of the applied filter. Finally, the EEGs are also
often segmented into epochs of sixty seconds.

\section{Coherence and Partial Coherence}

In this section, we formally describe two of the most common measures of 
spectral dependence, namely, coherence and partial coherence. A 
formal description for these measures is derived under the context of the 
Cram\'er representation of weakly stationary processes. We also presented
an alternative interpretation of these dependence metrics as the squared correlation 
between filtered signals. Finally, these measures are derived for the 
general case where the signals exhibit non-stationarity, e.g., under conditions
where the dependence between signals may evolve over time. 

\subsection{Coherence and Correlation via the Fourier transform}

Suppose that ${\bf X(t)}$ is a $P-$dimensional weakly stationary process with 
mean $\ex {\bf X}(t) = \underline{0}$ for all time $t$ and sequence of covariance 
matrices $\left\{ \Sigma(h), \ h=0, \pm 1, \pm 2, \ldots \right\}$ where $\Sigma(h)$ is 
defined by 
\begin{equation*}
\Sigma(h) = \cov\left({\bf X}(t+h), {\bf X}(t)\right) = \ex\left( {\bf X}(t+h) {\bf X}'(t)\right). 
\end{equation*}
Here, we assume absolute summability (over lag $h$) of every element of 
$\Sigma(h)$. Let $\sigma_{pq}(h)$ be the $(p,q)$-th entry of the matrix 
$\Sigma(h)$. Then, $\sum_{h=-\infty}^{\infty} \vert \sigma_{pq}(u) \vert < \infty$ 
for all $(p,q)$. This condition will ensure that auto-correlation and cross-correlation 
decay sufficiently fast to $0$.

We need to highlight that there is a one-to-one relationship between the covariance matrix $\Sigma(h)$ 
and the spectral matrix via the Fourier transform. The spectral matrix, denoted 
${\bf f}(\omega)$, is defined as
\begin{equation*}
{\bf f}(\omega)
  = \sum_{h=-\infty}^{\infty} \Sigma(h) \exp(-i 2 \pi\,\omega\,h)
    \quad \text{ where } 
    \omega \in \Omega_0 = \left(-0.5, 0.5\right).
\end{equation*}
${\bf f}(\omega)$ is a $P \times P$ Hermitian semi-positive definite matrix.

From a univariate perspective, the matrix ${\bf f}(\omega)$ can be sufficient to 
describe the spectral properties of each one of their components. For instance,
the auto-spectrum (univariate spectrum) of the $p$-th channel, which is
denoted as $f_{pp}(\omega)$, can be obtained from the $(p,p)$-th entry of ${\bf f}(\omega)$.

In addition, the autocovariance sequence $\sigma_{pp}(h)$ can be derived from 
auto-spectrum $f_{pp}(\omega)$ via the inverse-Fourier transform
\begin{equation*}
\sigma_{pp}(h)
  = \int_{\omega\in\Omega_0}
    f_{pp}(\omega)
    \, \exp \left( i 2 \pi\,\omega,h \right)
    \, d\omega.
\end{equation*}

Note that as a special case, when $h=0$, then $\sigma_{pp}(0) = \var\left(X_p(t)\right)$.
This corollary provides the intuition that the auto-spectrum is decomposing the signal variance
across all frequencies $\omega\in\Omega_0=(-0.5, 0.5)$:
\begin{equation}\label{Eq:VarSpecDecomp}
\sigma_{pp}(0)
  = \int_{\omega\in\Omega_0}
    f_{pp}(\omega)\,d\omega
  = \var\left(X_p(t)\right).
\end{equation}

Now, we can introduce correlation as a dependence metric.
Correlation between two components $X_p$ and $X_q$ at time lag $h$ is defined by 
\begin{equation*}
r_{p,q}(h) = \cor\left(X_p(t+h), X_q(t)\right)
  = \frac {\cov\left(X_p(t+h), X_q(t)\right)}
          {\sqrt{\var X_p(t)\,\var X_q(t)}}
  = \frac {\sigma_{pq}(h)}
          {\sqrt{\sigma_{pp}(0)\,\sigma_{qq}(0)}}.
\end{equation*}

It is clear that $r_{p,q}(h)\in[-1,1]$ will reach its extreme value when one of the signals
is proportional to the other. Thus, correlation $r_{p,q}(h)$ is known for being the simplest
measure the quantifies the linear dependence, or synchrony, at a lag $h$ between the pair of time series. 

Furthermore, the spectral matrix ${\bf f}(\omega)$ provides more information about the interactions
of their components. For instance, it allows us to identify the cross-spectrum
between any pair of components, $X_p$ and $X_q$, through its $(p,q)$-th entry.
In a similar manner to the correlation, in the time domain, we can define another measure that quantifies
the similarity between the simultaneous spectral response of $X_p$ and $X_q$. This metric is known as 
coherency, and it is formally defined as
\begin{equation} \label{Eq:coherency}
\tau_{pq}(\omega) 
  = \frac {f_{pq}(\omega)}
          {\sqrt{f_{pp}(\omega) f_{qq}(\omega)}}
\end{equation}

However, it is more common to use a derived metric: the cross-coherence that is the square magnitude of the coherency:
\begin{equation} \label{Eq:coherence}
\rho_{pq}(\omega)
  = \left| \tau_{pq}(\omega)  \right|^2
  = \frac {\left| f_{pq}(\omega) \right|^2}
          {f_{pp}(\omega) f_{qq}(\omega)}.
\end{equation}
Cross-coherence $\rho_{pq}(\omega)\in[0,1]$ will achieve its maximum value only when both compared time series have a proportional
cross-frequency response.

\subsection{Coherence and Correlation using the Cram\'er representation}

We now define coherency and coherence in the robust framework of the Cram\'er 
representation (CR) as an alternative of the definitions based on the Fourier transform of the covariance matrix.
The CR of a zero-mean $P-$dimensional weakly stationary process ${\bf X(t)}$ is given by
\begin{equation*}
  {\bf X}(t)
    = \int_{-0.5}^{0.5}
      \exp\left(i 2\pi\,\omega\,t\right)
      \,d{\bf Z}(\omega)
\end{equation*}
where $d{\bf Z}(\omega) = \left[dZ_1(\omega), \ldots, dZ_P(\omega)\right]'$ is a 
vector of random coefficients associated with the Fourier waveform 
$\exp(i 2 \pi \omega t)$ for each of the $P$ components. Here, 
$\{d{\bf Z}(\omega)\}$ is a random process defined on $(-0.5, 0.5)$
that satisfies 
$\ex \left[ d{\bf Z}(\omega) \right] = 0$ for all frequencies $\omega$ and 
\begin{eqnarray*}
\cov\left(d{\bf Z}(\omega), d{\bf Z}(\omega')\right)
  & = & \left\{
        \begin{array}{cc}
            0,                          & \omega \ne \omega' \\
            {\bf f}(\omega) d\omega,    & \omega = \omega'
        \end{array}
  \right.
\end{eqnarray*}

From the above formulation, we note that the correlation and modulus-squared 
correlation with the coefficients are, in fact, coherency and 
coherence as they were defined in Equations~\ref{Eq:coherency} 
and ~\ref{Eq:coherence}:
\begin{eqnarray*}
  \cor\left[ d{\bf Z}_p(\omega), d{\bf Z}_q(\omega) \right]
    & = & \frac {f_{pq}(\omega)}
                {\sqrt{f_{pp}(\omega) f_{qq}(\omega}} \\
  \left\|\cor\left[d{\bf Z}_p(\omega), d{\bf Z}_q(\omega)\right] \right\|^2
    & = & \frac {\left|f_{pq}(\omega)\right|^2}
                {f_{pp}(\omega) f_{qq}(\omega)}.
\end{eqnarray*}

Recall that the CR represents a time series as a linear mixture of infinitely
many sinusoidal waveforms (Fourier waveforms with random amplitudes).
This perspective allows us to provide a different interpretation of the above-mentioned
dependence metrics. Consider the components $X_p$ and $X_q$,
both of them contain Fourier oscillations in a continuum of frequencies, and
let us now focus only on the specific frequency $\omega$ in the two components:
\begin{eqnarray*}
X_{p, \omega}(t)
  & = & \exp\left(i 2 \pi \omega t\right)\,dZ_p(\omega) \\
X_{q, \omega}(t)
  & = & \exp\left(i 2 \pi \omega t\right)\,dZ_q(\omega).
\end{eqnarray*}
Assuming that these random coefficients have zero mean, the 
variance at the $\omega$-oscillatory activity of $X_p$ is 
\begin{equation*}
\var\left[ X_{p, \omega}(t) \right]
  = \left|
      \exp(i 2 \pi\,\omega\,t)
    \right|^2
    \var\left[ dZ_p(\omega) \right].
\end{equation*}
Since the random coefficients $\left\{ dZ_p(\omega) \right\}$ are uncorrelated across 
$\omega \in (-0.5, 0.5)$, it follows that 
\begin{equation*}
\VAR{X_p(t)}
  = \int
    \VAR{X_{p, \omega}(t)}
  = \int
    \VAR{dZ_p(\omega)}.
\end{equation*}

Based on this relationship, we can introduce an alternative interpretation of
the spectral decomposition denoted along with Equation~\ref{Eq:VarSpecDecomp}:
the total variance of a weakly stationary signal $X_p$ at any time point can be
viewed as an infinite sum of the variance of each of the random coefficients.
Furthermore, we can think of the relationship between the variance of the random
coefficients and the spectrum as follows
\begin{equation*}
\VAR{dZ_p\left( \omega \right)}
  = f_{pp} \left( \omega \right)
    \, d\omega.
\end{equation*}

Now, let us study the relationship between the signals $X_p$ and $X_q$ 
as a function of the oscillations. 
The covariance between the $\omega-$oscillation in $X_p$ and the $\omega-$oscillation in $X_q$ is 
\begin{eqnarray*}
  \COV{X_{p, \omega}(t), X_{p, \omega}(s)}
    & = &
    \exp\left(i 2 \pi \omega (t-s)\right)
    \, \COV{dZ_p(\omega), dZ_q(\omega)}.
\end{eqnarray*}
Moreover, since $\vert \exp(i 2 \pi \omega t) \vert = 1$ for all $t$,
the variances of these respective $\omega$-oscillations are
\begin{equation*}
\VAR{X_{p, \omega}(t)}
  = \VAR{dZ_p(\omega)}
\quad
{\mbox{and}}
\quad
\VAR{X_{q, \omega}(t)}
  = \VAR{dZ_q(\omega)}.
\end{equation*}
In addition, the correlation between these $\omega$-oscillations is 
\begin{equation*}
\CORR{X_{p, \omega}(t), X_{p, \omega}(s)}
  = \exp\left(i 2 \pi \omega(t-s)\right)
    \CORR{dZ_p(\omega), dZ_q(\omega)}
\end{equation*}
and the respective modulus-squared correlation is given by
\begin{equation*}
\left|
  \CORR{X_{p, \omega}(t), X_{p, \omega}(s)}
\right| ^ 2
  = \left|
      \CORR{dZ_p(\omega), dZ_q(\omega)}
    \right| ^ 2
\end{equation*}
which is identical to the definition of coherence given in Equation~\ref{Eq:coherence}.

In summary, it is possible to affirm that coherence between a pair of 
weakly stationary signals at frequency $\omega$ is the square-magnitude of the 
correlation between the $\omega$-oscillations of these signals.

\subsection{Coherence and filtering} 

In practical EEG analysis, coherence between a pair of channels is defined
and estimated at some frequency bands - rather than in a singleton frequency. 
The standard frequency bands are delta $(0.5, 4.0)$ Hertz, theta $(4.0, 8.0)$ Hertz, 
alpha $(8.0, 12.0)$ Hertz, beta $(12.0, 30.0)$ Hertz and gamma $(30.0, 50.0)$ Hertz. 
This segmentation of the frequency axis has been widely accepted for many decades. 
However, there is a growing direction towards a more specific (narrower) frequency
band analysis and a more data-adaptive approach to determining (a.) the number of frequency peaks, 
(b.) the location of these peaks, and (c.) the bandwidth associated with each one of them.
This will be necessary for a finer differentiation between experimental conditions and patient diagnosis groups.
The immediate task at hand is to point out the connection between linear filters and spectral and 
coherence estimation. 

To estimate coherence, the first step is to apply a linear filter on time series 
components $X_p$ and $X_q$ so that the resulting filtered time series will have
spectra whose power is concentrated around a pre-specified band $\Omega$.
In essence, a $k$-th order linear frequency-filter is comprised of a set of coefficients $\left\{a_0,\ldots,a_K\right\}$
and $\left\{b_1,\ldots,b_K\right\}$ under the constraint
$\sum_{k} \vert a_k \vert < \infty$ and $\sum_{k} \vert b_k \vert < \infty$.
The filtered signal $y^{*}(t)$ is obtained as a linear combination of the previous
values of the unfiltered time series $y(t),\ldots,y(t-K)$ and its previous values $y^{*}(t-1),\ldots,y^{*}(t-K)$:
\begin{equation}
y^{*}(t)=\sum_{k=0}^{M}b_{k}y(t-k)-\sum_{k=1}^{M}a_{k}y^{*}(t-k)
\end{equation}

Filters with $a_k=0\,\forall k$ are known as finite impulse response (FIR) filters
or infinite impulse response (IIR) filters when $\exists a_k\ne0$ for some $k$.
For an extensive analysis of linear time-invariant filtering, we refer to \citep{StatisticalDigitalSig-Hayes-1996}.
Here, we will focus on the FIR family of filters.
Consider the filter so that $\left\{c_k, k=0, \pm 1, \pm 2, \ldots\right\}$.
A one-sided linear filter will be used to examine causality between the different
oscillations, a further discussion about the reasons behind this condition is given in Section \ref{sec:spectral-causality}.

The Fourier transform of the sequence of FIR filter coefficients $\{ c_j \}$ is 
\begin{equation*}
C\left(\omega\right)
  = \sum_k \ c_k \exp(-i 2 \pi \omega\,k)
\end{equation*}
which is called the frequency response function so that $\vert C(\omega) \vert$.
The set of filter coefficients are selected so that $C(\omega)$ has a peak
that is concentrated in the neighborhood of the frequency band $\Omega$ (band-pass region). 
In the frequency intervals outside of $\Omega$ (stop-band region),$\vert C(\omega) \vert$ is expected to has relatively small values. 
Thus, to extract the component of  $X_p$ and $X_q$ that is associated with the 
$\Omega$-oscillation, we apply 
the linear filter $\left\{c_j, j=0, \pm 1, \pm 2, \ldots \right\}$ to obtain the convolution
\begin{eqnarray*}
X_{p, \Omega}(t)
  & = &
  \sum_{j=-\infty}^{\infty} c_j X_p(t-j) \\
X_{q, \Omega}(t)
  & = &
  \sum_{j=-\infty}^{\infty} c_j X_q(t-j).
\end{eqnarray*}

The auto-spectra of each filtered series, $\{ X_{p, \Omega} \}$ and  $ \{ X_{q, \Omega} \}$, is
\begin{eqnarray*}
f_{pp, \Omega}(\omega)
  & = &
  \left| C(\omega) \right|^2
  \, f_{pp}(\omega) \\ 
f_{qq, \Omega}(\omega)
  & = &
  \left| C(\omega) \right|^2
  \, f_{qq}(\omega),
\end{eqnarray*}
respectively, which has spectral power attenuated outside of the band $\Omega$. 

\begin{figure}
\centering%
\includegraphics[width=1\textwidth,height=\textheight]{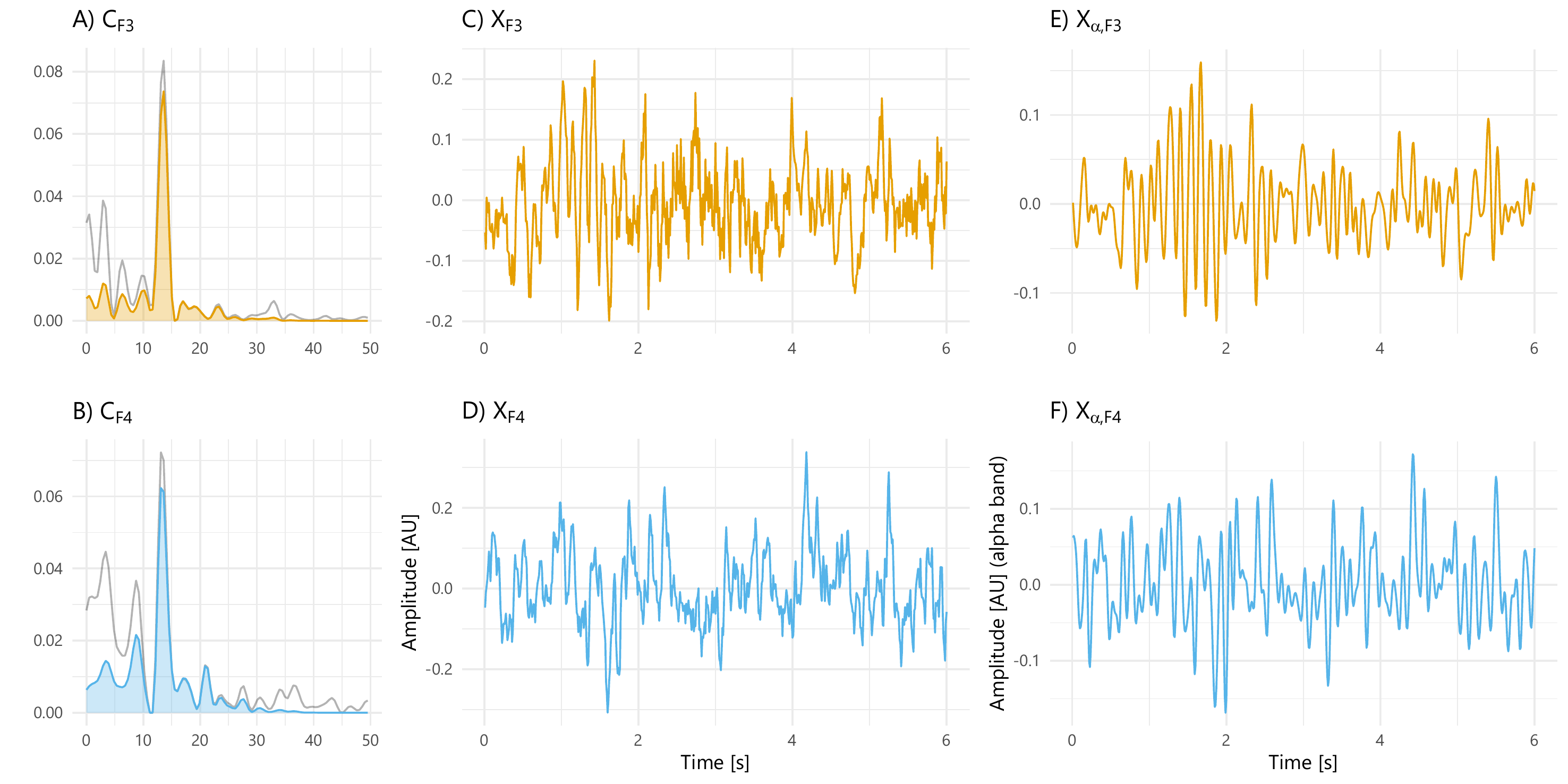}
\caption{%
Alpha-filtering of EEG signals in channels F3 and F4: %
A,B) Power spectrum density $C_{F4}$ and $C_{F3}$ after applying a band-pass filter. %
Note that the power is concentrated only in the interval 8-12Hz.
The gray line in the background corresponds to the unfiltered spectra.%
C,D) Signals $X_{F3}$ and $X_{F4}$ without before filtering;
and E.F) Alpha waves in both channels: $X_{F3,\alpha}$ and $X_{F4,\alpha}$.
}
\label{fig:eeg-decomposition-F3F4}
\end{figure}

\begin{example}[EEG example]
Recall the EEG-ADHD dataset described in Section \ref{sec:eeg-dataset}.
Let us focus in the signals recorded at channels F3 and F4, collected from the control subject S041.
We denoted them by $X_p$ and $X_q$ (Figure \ref{fig:eeg-decomposition-F3F4}.C-D).
These biomedical signals are sampled at $128$ Hertz and with a 60-second epoch.
In consequence, the total number of time points available for the analysis is $T = 128 \times 60 = 7680$. 

We extract the alpha-band component ($\Omega=8-12$Hertz) using a 10-th order FIR band-pass filter defined by the coefficients:
\begin{eqnarray*}
\left\{
c_i
\right\}
  & = & \left\{
        -0.0272, -0.0468, -0.0423, 0.0771, 0.2677,
      \right. \\
  &\, & \left.
        0.3629, 0.2677, 0.0771, -0.0423, -0.0468, -0.0272
      \right\}
\end{eqnarray*}
The filtering results are depicted in Figure \ref{fig:eeg-decomposition-F3F4}.E-F with the spectrum after the filtering process
in Figure \ref{fig:eeg-decomposition-F3F4}.A-B.
\end{example}



\begin{figure}
\centering%
\includegraphics[width=1\textwidth,height=\textheight]{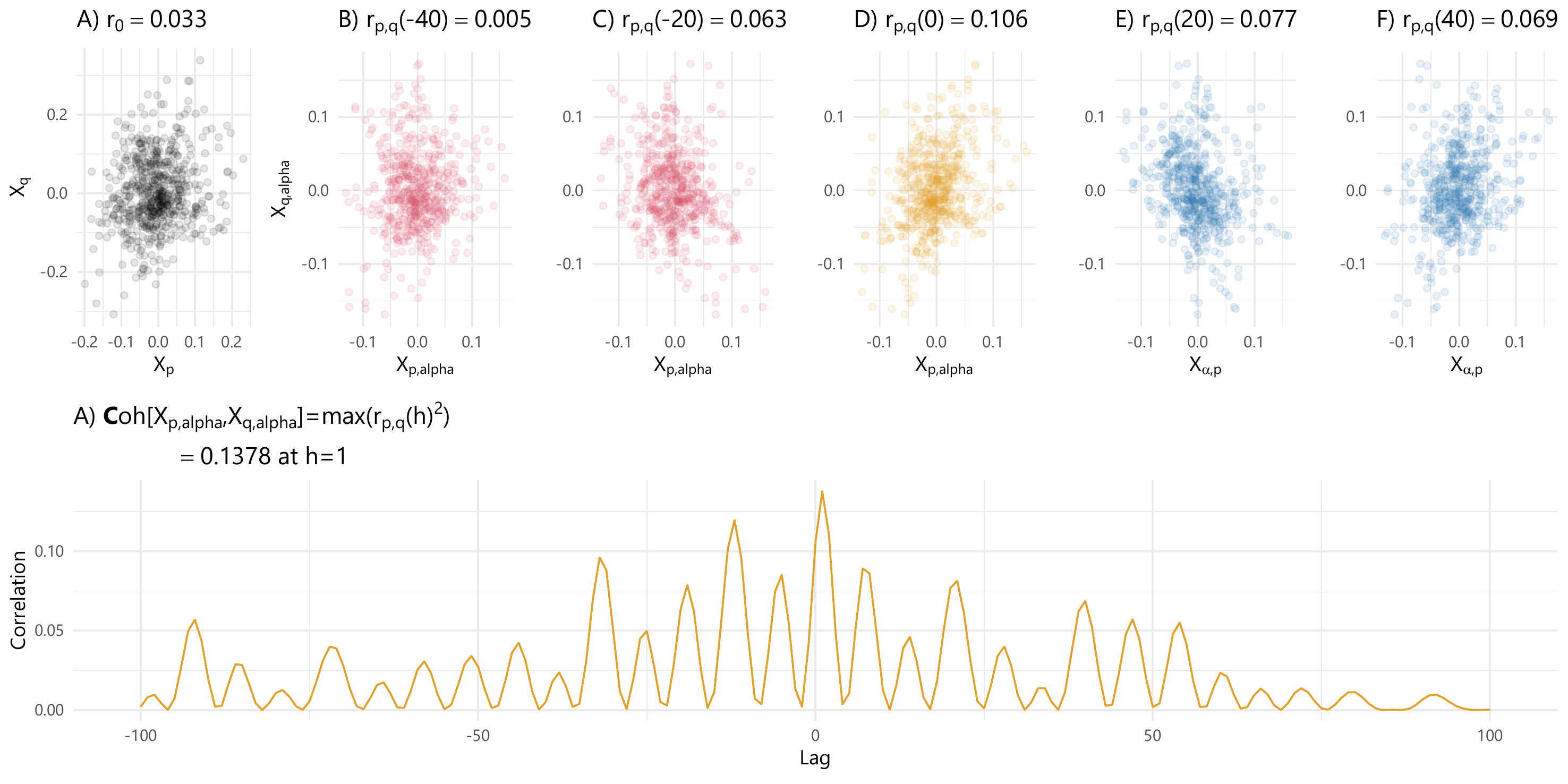}
\caption{%
Linear relationships between the channels F3 and F4: %
A) Scatterplot of the unfiltered signals, with a correlation of 0.033. %
B-F) Scatterplots of alpha-filtered channels at different lags: -40, -20, 0, 20, and 40. %
Note that lags are denoted from reference to F3 ($X_p$).%
Thus, negative lags imply that F4 is leading F3 during the correlation estimation.
G) Cross-correlation as a function of the lags. %
The maximum cross-correlation (0.138) is located at lag 1.
}
\label{fig:eeg-decomposition-F3F4}
\end{figure}

The next step is to study the dependence between the time series $X_p$ and 
$X_q$ through their filtered components. In practice, dependence can be examined 
at various frequency bands but in this example, we focus only on the 
alpha band for illustration purposes.


From \citet{Ombao:2008hc}, we can estimate the band-coherence, i.e., the coherence
between the two EEG channels, $X_p$
the estimated coherence between the EEG channels $X_p$ 
and $X_q$ at the alpha-band is 
\begin{equation}
  \rho_{p,q}(\alpha) \propto \max_{\ell} \left| \sum_{t} X_{p, \alpha}(t) \, X_{q, \alpha}(t-\ell)  \right|^2.
\end{equation}

\begin{figure}
\centering%
\includegraphics[width=1\textwidth,height=\textheight]{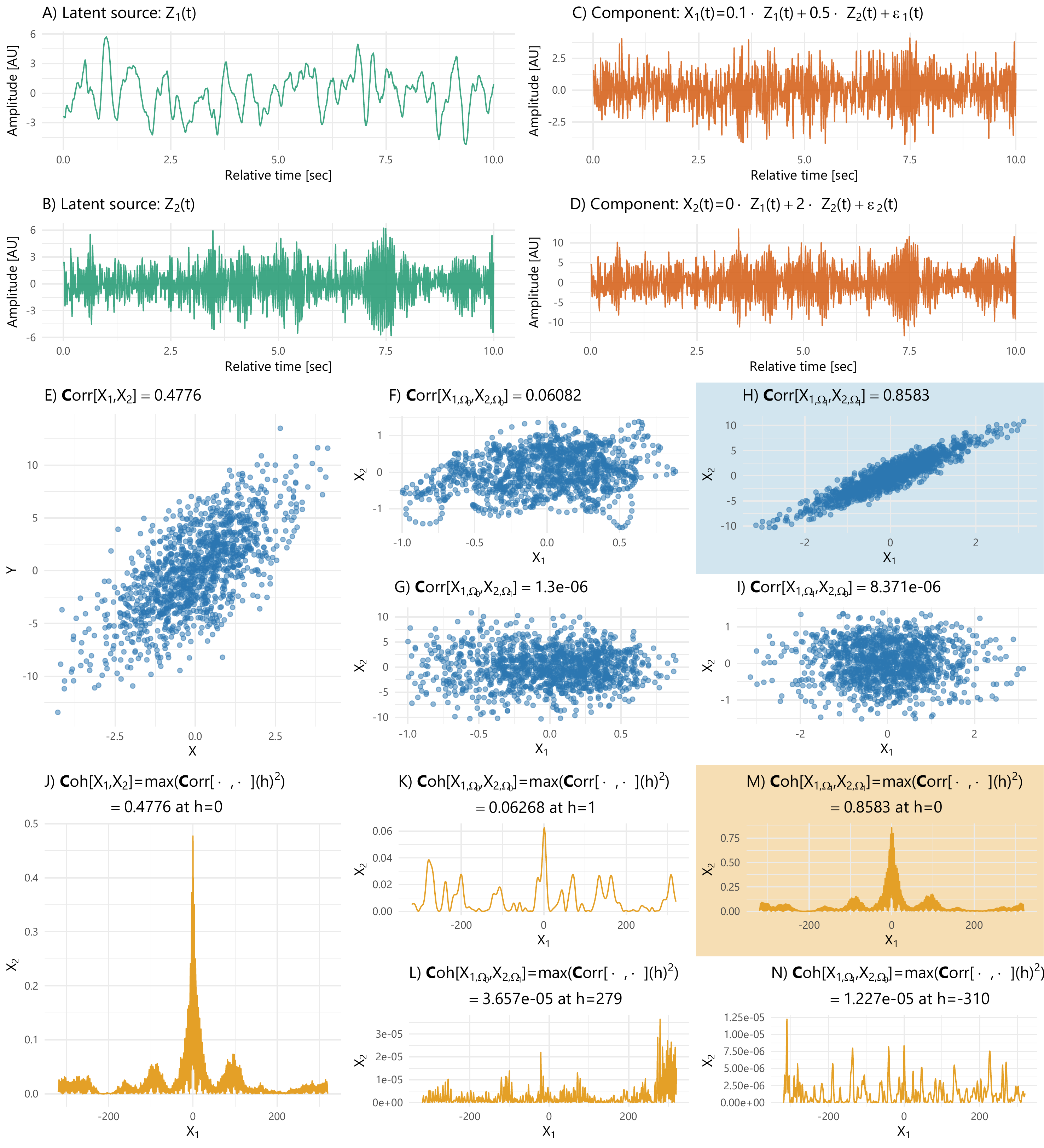}
\caption{%
Simulated example \ref{ex:instantaneous-mixture-coh}: system with an instantaneous mixture with latent sources ($Z_1(t)$, $Z_3(t)$) and mixed components ($X_1(t)$, $X_2(t)$). Note the differences in the correlation between the unfiltered, and the $\Omega_0$- and $\Omega_1$-filtered components. $\CORR{X_1(t),X_2(t)}=0.4776$ while $\CORR{X_{\Omega 0}(t),X_{\Omega 0}(t)}=0.0608$ and $\CORR{X_{\Omega 1}(t),X_{\Omega 1}(t)}=0.8583$.
}
\label{fig:eeg-linear-mixture}
\end{figure}

\begin{example}[Contemporaneous mixture]\label{ex:instantaneous-mixture-coh}
We now illustrate that 
coherence gives information beyond cross-correlation (or the square cross-correlation). 
Consider the setting where 
there are two latent sources $Z_1$ and $Z_2$ where $Z_1$ is a high-frequency source 
and $Z_2$ is a low-frequency source. Define the observed time series $X_1$ and 
$X_2$ to be mixtures of these two sources, e.g., 
\begin{eqnarray}
\left ( 
\begin{matrix}
X_1(t) \\
X_2(t)
\end{matrix}
\right )
& = & 
\left ( 
\begin{matrix}
c_{11} & c_{12} \\
c_{21} & c_{22}
\end{matrix}
\right ) \ \
\left ( 
\begin{matrix}
Z_1(t) \\
Z_2(t)
\end{matrix}
\right ) \ + \  
\left ( 
\begin{matrix}
\epsilon_1(t) \\
\epsilon_2(t)
\end{matrix}
\right ).
\end{eqnarray} \label{Eq:ModelMix1}
Suppose that $c_{21} = 0$ and the other entries of the mixing matrix are all non-zero. 
This implies that $X_1$ contains both low-frequency component $Z_1$ and high-frequency 
component $Z_2$. However, $X_2$ contains only the high-frequency component $Z_2$. 
Thus, it is the high-frequency component that drives the dependence between 
$X_1$ and $X_2$. The scatterplot of $X_2(t)$ vs $X_1(t)$ in Figure \ref{fig:eeg-linear-mixture}.E shows 
that these two time series are correlated and the sample cross-correlation is computed 
to be 0.4776. However, correlation is limited in the information it can convey about 
the relationship between a pair of signals. For instance, it does not indicate what frequency 
band(s) drive that relationship. 

To now investigate deeper the relationship between $X_1$ and $X_2$, we apply a
low-pass filter (on band $\Omega_0$) and a high-pass filter (on band $\Omega_1$) and 
denote these filtered signals to be 
\begin{equation*}
X_{1, \Omega_0}(t), \  X_{1, \Omega_1}(t), \ X_{2, \Omega_0}(t), \  X_{2, \Omega_1}(t). 
\end{equation*}

Under stationarity, the random coefficients in the Cram\'er representation are uncorrelated 
across frequencies (i.e., $\cov [dZ_1(\omega), dZ_2(\omega')]=0$ when $\omega \ne \omega'$). 
Thus, 
\begin{eqnarray*}
\cov[X_{1, \Omega_0}(s), X_{2, \Omega_1}(t)] & = & 0 \\
\cov[X_{1, \Omega_1}(s), X_{2, \Omega_0}(t)] & = & 0.
\end{eqnarray*}
For some non-stationary processes, there could be possible linear and non-linear 
dependence between different frequency bands.
For the moment, we focus only on the 
correlation between the low-frequency components $X_{1, \Omega_0}(s)$ and 
$X_{2, \Omega_0}(t)$ and the high-frequency components 
$X_{1, \Omega_1}(s)$ and $X_{2, \Omega_1}(t)$. The lag-$0$ scatterplots are 
shown in \ref{fig:eeg-linear-mixture}.F-I. It is clear here that the linear relationship between 
the low-frequency components is weaker than the dependence on high-frequency components. 
This was to be expected from the data generating model in Equation~\ref{Eq:ModelMix1}, 
which specifies that $X_1$ and $X_2$ both share the common high-frequency latent 
source. Assuming that the sample mean of these oscillations are all $0$, then the 
coherence estimate at the low-frequency band is the squared cross-correlation between 
the $X_{1, \Omega_0}$ and $X_{2, \Omega_0}$, i.e., 
\begin{eqnarray*}
\widehat{\rho}_{12}(\Omega_0) = \frac{ \left \{ \sum_{t=1}^{T} (X_{1, \Omega_0}(t) X_{2, \Omega_0}(t)) \right
\}^2}{ \sum_{t=1}^{T} (X_{1, \Omega_0}(t))^2  \sum_{t=1}^{T} (X_{2, \Omega_0}(t))^2\ }.
\end{eqnarray*}
The coherence estimate at the high-frequency band is computed similarly. The 
estimated values for coherence at the low and high-frequency bands are, respectively, 
0.0608 and 0.8583.
\end{example}


\begin{figure}
\centering%
\includegraphics[width=1\textwidth,height=\textheight]{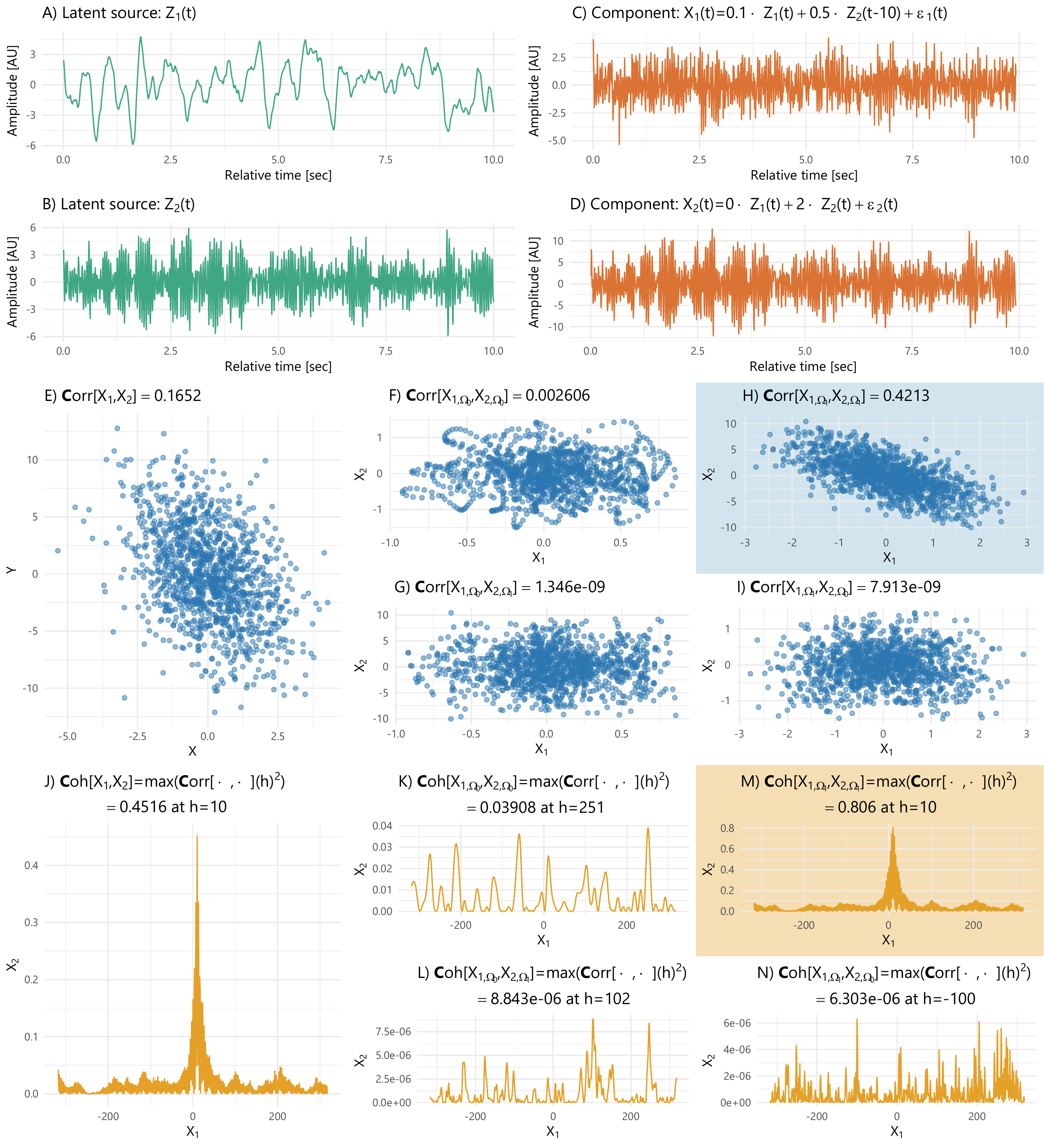}
\caption{%
Simulated example \ref{ex:lagged-mixture-coh}: system with latent sources ($Z_1(t)$, $Z_3(t)$) and lagged mixed components ($X_1(t)$, $X_2(t)$). Note the high contrast between correlation and coherence in low- and high-frequency components. $\CORR{X_{\Omega 0}(t),X_{\Omega 0}(t)}=0.0026$ and $\CORR{X_{\Omega 1}(t),X_{\Omega 1}(t)}=0.4213$ while $\COH{X_{\Omega 0}(t),X_{\Omega 0}(t)}=0.039$ and $\COH{X_{\Omega 1}(t),X_{\Omega 1}(t)}=0.8060$. Furthermore, the maximum correlation in the high-frequency components is located at lag $10$.
}
\label{fig:eeg-lagged-mixture}
\end{figure}

\begin{example}[Lagged mixture] \label{ex:lagged-mixture-coh}
The data generating process in the 
previous example (Equation~\ref{Eq:ModelMix1}) assumes an instantaneous mixture, 
i.e., the observed signals at a specific time $t$, $\{X_1(t), X_2(t)\}$ depend explicitly on 
the latent processes $Z_1(t)$ and $Z_2(t)\}$ also at the {\it same} time $t$. 
However, we can extend this model towards
cases when there is some lag in the mixtures, e.g., the latent source has a delayed 
effect on some components of the observed signals. 
As Nunez et al. pointed out \citep{EEGFConnectivity-Nunez-2015}, axons can have
propagation speeds in the range 600-900$\frac{cm}{s}$, and considering the average distance
from the cortex to the scalp is 14.70 mm for middle-aged humans \citep{ScalpCortexDistance-Lu-2019},
we presume that delays of a few milliseconds (from any neural source to the scalp) can be feasible in the EEGs.
Statistically, in order to handle these lagged 
mixtures, we introduce the backshift operator $B^{h}$ (where $h=0, 1, \ldots$) to be 
$B^h Z(t) = Z(t-h)$. Consider now the lagged mixture 
\begin{eqnarray}\label{eq:ex-lagged-mixture}
\left ( 
\begin{matrix}
X_1(t) \\
X_2(t)
\end{matrix}
\right )
& = & 
\left ( 
\begin{matrix}
c_{11}B^{h_{11}} & c_{12}B^{h_{12}} \\
c_{21}B^{h_{21}} & c_{22}B^{h_{22}}
\end{matrix}
\right ) \ \
\left ( 
\begin{matrix}
Z_1(t) \\
Z_2(t)
\end{matrix}
\right ) \ + \  
\left ( 
\begin{matrix}
\epsilon_1(t) \\
\epsilon_2(t)
\end{matrix}
\right ).
\end{eqnarray}\label{Eq:ModelMix2}

Here, suppose that the mixture weight $c_{21}=0$; and the lags for the 
latent sources are $h_{12}=0$ and $h_{22} = 10$. Thus, the two observed 
time series are 
\begin{eqnarray*}
X_1(t) & = & c_{11} Z_1(t) + c_{12} Z_2(t-10) + \epsilon_2(t)\\
X_2(t) & = & c_{22} Z_2(t) + \epsilon_1(t).
\end{eqnarray*}

As in the previous example, the observed signals are driven 
by the high-frequency latent source $Z_2$, but the effect of 
$Z_2$ on $X_2$ is delayed by $10$ time units. Consider now 
the scatterplots of the high-frequency filtered time series at 
(a.) lag $0$: $X_{1, \Omega_1}(t)$ vs. $X_{2, \Omega_1}(t)$ and 
(b.) lag $10$: $X_{1, \Omega_1}(t)$ and $X_{2, \Omega_1}(t-10)$. 
In Figure \ref{fig:eeg-lagged-mixture}, it is clear that the linear relationship at 
lag $10$ appears stronger compared to the contemporaneous 
correlation. Denote the cross-correlation estimate at lag $\ell$ to be 
\[
\widehat{r}_{12}(\Omega_1, \ell)
  = \frac{\left\{
             \sum_{t=1}^{T} (X_{1, \Omega_1}(t) X_{2, \Omega_0}(t-\ell))
          \right\}^2}
        { \sum_{t=1}^{T} (X_{1, \Omega_1}(t))^2  \sum_{t=1}^{T} (X_{2, \Omega_1}(t))^2\ }.
\]
Then the estimated coherence at the frequency band $\Omega_1$ is 
\begin{eqnarray*}
\widehat{\rho}_{12}(\Omega_0) = \max_{\ell = 0, \pm 1, \ldots} \vert \widehat{r}_{12}(\Omega_1, \ell) \vert^2.
\end{eqnarray*}
\end{example}


\subsection{AR(2) processes - discretized Cram\'er representation}

One can approximate the Cram\'er representation of weakly stationary 
processes by a representation based on latent sources with identifiable 
spectra where each spectrum has its own unique peak frequency and 
bandwidth (or spread). Here, we will consider the class of weakly stationary 
second-order autoregressive, or simply AR(2), models to serve as "basis" 
latent sources. A process $Z(t)$ is AR(2) if it admits a representation 
\begin{equation*}
Z(t) - \phi_1 Z(t-1) - \phi_2 Z(t-2) = W(t) 
\end{equation*}
where $\left\{W(t)\right\}$ is white noise with $\EX{W(t)} = 0$ and $\VAR{W(t)} = 
\sigma_W^{2}$ and the AR(2) coefficients $\phi_1$ and $\phi_2$ must 
satisfy that the roots of the AR(2) polynomial function (denoted $u_1, u_2$)
\begin{equation*}
\Phi(u) = 1 -\phi_1 u - \phi_2 u^2
\end{equation*}
must satisfy $\vert u_{c} \vert >1$ for both $c=1,2$. In particular, we will consider 
the subclass of $AR(2)$ models whose roots are non-real complex-valued so that 
they are complex-conjugates of each other $u_2 = u_1^*$ and thus they can be 
reparametrized as 
\begin{equation*}
u_1 = M \exp(i 2 \pi \psi) \ \ {\mbox{and}} \ \ u_2 = M \exp(-i 2 \pi \psi)
\end{equation*}
where $M > 1$ and $\psi \in (-0.5, 0.5)$. Note that the AR(2) model can be 
parametrized by the coefficients $(\phi_1, \phi_2)$ or by the roots 
$(u_1, u_2)$ or by the magnitude and phase of the roots $(M, \psi)$. In fact, 
the one-to-one relationship between the coefficients and the roots is given by
\begin{equation} \label{eq:AR2-oscillator}
\phi_1
  = \frac{2}{M}
    \cos\left( 2 \pi \psi \right)
    \quad
    {\mbox{and}}
    \quad
\phi_2
  = -\frac{1}{M^2}
\end{equation}

\begin{figure}
\centering%
\includegraphics[width=1\textwidth,height=\textheight]{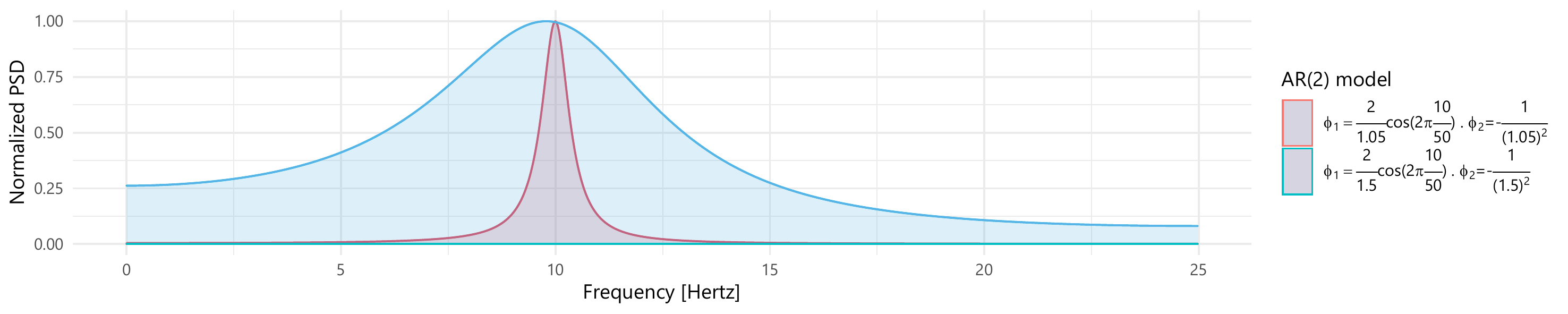}
\caption{%
Autoregressive process spectrum $AR(2)$ as a function of the magnitude of its roots. The closer the magnitude is to one, the narrower the spectrum.
}
\label{fig:simulation-psd-ar2}
\end{figure}

One very important and interesting property of an AR(2) process with 
complex-valued roots is that its spectrum has a peak at $\psi$ and the 
spread of this peak is governed by the magnitude $M$. When the root 
magnitude $M \longrightarrow 1+$ then the bandwidth of the peak around 
$\psi$ becomes narrower. Conversely, when $M$ becomes much larger
than $1$ then the bandwidth around $\psi$ becomes wider (Figure \ref{fig:simulation-psd-ar2}). 

\begin{example}[AR(2) with the peak at the alpha band]
We now describe how to specify an AR(2) process whose spectrum has a peak at 
$10$ Hertz. Assume that the sampling rate is $100$ Hertz with a consequent Nyquist frequency of $50$ Hertz. The roots of the AR(2) processes are then 
complex-valued with phase $\psi = \frac{10}{50}$ and magnitude $L$. 
To model an AR(2) spectra with a narrowband response around $10$ Hertz, 
we set the magnitude of the roots to be $L=1.05$. Moreover, to model
a broadband response, we set $L=1.50$.

Therefore, the AR(2) coefficient parameters for the narrowband process are
\begin{equation}
\phi_1
  = \frac{2}{1.05}
    \cos\left( 2 \pi \frac{10}{50} \right)
    \quad
    {\mbox{and}}
    \quad
\phi_2
  = -\frac{1}{(1.05)^2}
\end{equation}
while, for the broadband signal, the AR(2) coefficients are given by
\begin{equation}
\phi_1
  = \frac{2}{1.50}
    \cos\left( 2 \pi \frac{10}{50} \right)
    \quad
    {\mbox{and}}
    \quad
\phi_2
  = -\frac{1}{(1.50)^2}.
\end{equation}

The corresponding auto-spectra of these two AR(2) processes are displayed in Figure \ref{fig:simulation-psd-ar2}.
\end{example}

We now construct a representation for ${\bf X}(t)$ that is a linear mixture of 
uncorrelated AR(2) latent processes $Z_1, \ldots, Z_K$ whose spectra 
are identifiable  with peaks within the bands $\Omega_1$, $\ldots,$ 
$\Omega_K$.  Define $g_k(\omega)$ to be the spectrum of the latent 
AR(2) process $Z_k$. These spectra will be standardized so that 
$\int_{-0.5}^{0.5} g_k(\omega) d\omega = 1$ for all $k=1, \ldots, K$. 
Moreover, these spectra have peak locations that are at unique 
frequencies and that these are sufficiently separated. The choice of 
the number of components $K$ may be guided by the standard in neuroscience 
where the $K=5$ bands are modeling components in the delta, theta, alpha, beta, and gamma band (as they were defined in a previous section).
Another approach is developed in \cite{BMARD}, which
data-adaptively selects $K$, the peak locations, and bandwidths. 
Therefore, the mixture of AR(2) latent sources is given by 
\begin{eqnarray}
\left ( 
\begin{matrix}
X_1(t) \\
X_2(t) \\
\ldots \\
X_P(t)
\end{matrix}
\right )  =  
\left ( 
\begin{matrix}
A_{11} & A_{12} & \ldots & A_{1K} \\
A_{21} & A_{22} & \ldots & A_{2K} \\
\ldots & {} & {} & {} \\
A_{P1} & A_{P2} & \ldots & A_{PK}
\end{matrix}
\right ) \
\left( 
\begin{matrix}
Z_1(t) \\
Z_2(t) \\
\ldots \\
Z_K(t)
\end{matrix}
\right)
\end{eqnarray}. 

Consider components $X_p$ and $X_q$. Suppose that, for a particular 
latent source $Z_k$, the coefficients $A_{pk} \ne 0$ and 
$A_{qk} \ne 0$. Then, these two signals share $Z_k$ as the same common 
component, and that $X_p$ and $X_q$ are coherent at frequency 
band $\Omega_k$. 

This stochastic representation in terms of the AR(2) processes as building 
blocks is motivated by the result from \cite{GaoSinica} for univariate processes 
where the essential idea is the following: define the true spectrum of 
a weakly stationary process $X_p$ to be $f_{pp}(\omega)$ and set 
$f_{pp,K}^{A}(\omega)$ to be the mixture (or weighted average) of the spectrum 
of the $K$ AR(2) latent processes that gives the minimum discrepancy, i.e., 
\begin{equation}
f_{pp,K}^{A}
  = \mbox{inf}_{S}
    \int
    \left|
      f_{pp}\left( \omega \right)
      - S_{pp,K}^{A} \left( \omega \right)
    \right|
    \,d\omega
\end{equation}
over all candidate spectra from a mixture of $K$ AR(2) processes, $S_K$. This discrepancy decreases as $K$ increases, and therefore, the model provides a better approximation.

\subsection{Partial coherence}

\begin{figure}
\centering%
\includegraphics[width=0.6\textwidth,height=\textheight]{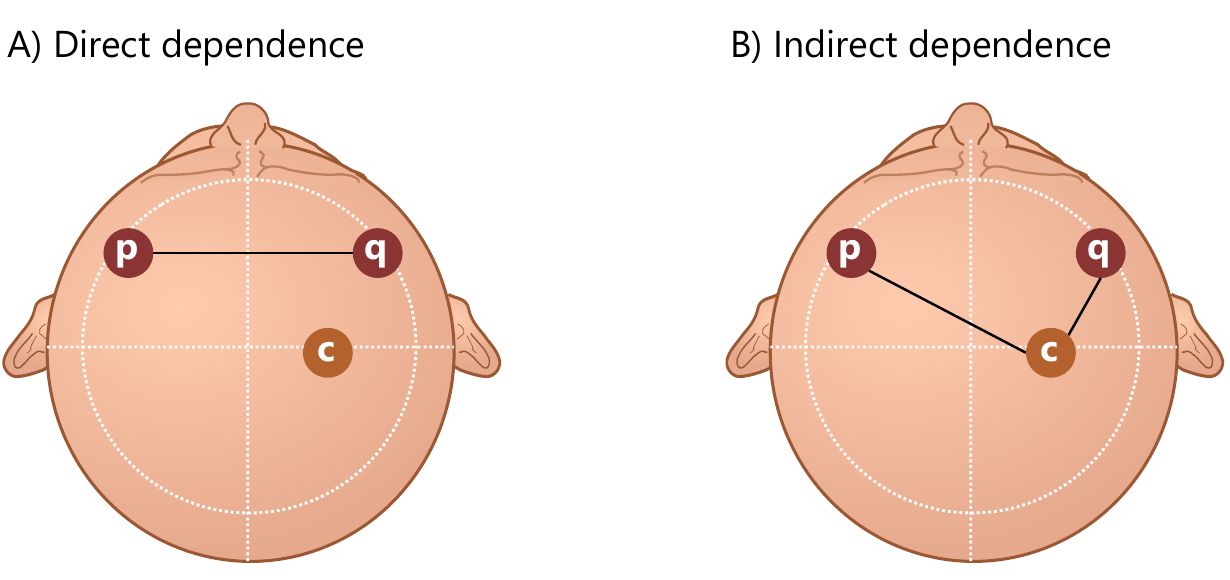}
\caption{%
Direct and indirect dependence between two channels $p$ and $q$
Indirect links could appear if both channels depend on a common channel $c$. 
}
\label{fig:topography-pqc}
\end{figure}


One of the key questions of interest is determining whether the dependence 
between two components $X_p$ and $X_q$ is pure or it is indirect through 
another component $X_c$ (or a set of channels $c \ne p,q$). In Figure \ref{fig:topography-pqc}, 
we show the distinction between the pure vs. indirect dependence between 
$X_p$ and $X_q$. In Figure \ref{fig:topography-pqc}.B, if we remove the link between 
$X_c$ and $X_p$ and the link between $X_c$ and $X_q$, then there is no 
longer any dependence between $X_p$ and $X_q$.

A standard approach in the time domain is to calculate the partial correlation 
between $X_p$ and $X_q$ (\citep{LatentVariable-Fried-2005, PartialOscGraphs-Fried-2005}), which is essentially the cross-correlation between 
them after removing the linear effect of $X_c$ on both $X_p$ 
and $X_q$. This is also achieved by taking the inverse of the covariance 
matrix, denoted $\Sigma^{-1}(0)$ and then standardizing this matrix by a 
pre- and post-multiplication of a diagonal matrix whose elements are the 
inverse of the square root of the diagonal elements of $\Sigma^{-1}(0)$. 
The procedure for the frequency domain follows in a similar manner, as 
outlined in \citet{fiecas2011generalized}: define the inverse of the spectral 
matrix to be ${\bf g}(\omega) = {\bf f}^{-1}(\omega)$ and let ${\bf h}(\omega)$ 
be a diagonal matrix whose elements are $\frac{1}{\sqrt{g_{rr}(\omega)}}$. 
Next, define the matrix 
\begin{equation}\label{Eq:PartialCoh}
\Lambda(\omega) = - {\bf h}(\omega) {\bf g}(\omega) {\bf h}(\omega).
\end{equation}

Then, partial coherence (PC) between components $X_p$ and $X_q$ at frequency 
$\omega$ is $\vert \Lambda_{pq}(\omega) \vert^2$. 

This particular 
characterization of PC requires inverting the spectral matrix. 
This can be time computationally demanding when the dimension $P$ is large, and it also can be prone 
to numerical errors when the condition number of the spectral matrix 
${\bf f}(\omega)$ is very high (i.e., the ratio of the largest to the smallest 
eigenvalue is large). This happens when there is a high degree of 
multicollinearity between the $\omega-$oscillations of the components of 
${\bf X}$. To alleviate this problem, \cite{fiecas2010functional} and 
\cite{fiecas2011generalized} developed a class of spectral shrinkage procedures. 
The procedure consists in constructing a well-conditioned estimator for the spectral matrix 
and hence produces a numerically stable estimate of the inverse. The starting 
point is to construct a "smoothed" periodogram matrix which is a 
nonparametric estimator of ${\bf f}(\omega)$. From a stretch of time series 
${\bf X}(1), \ldots, {\bf X}(T)\}$ \ ($T$ even), compute the Fourier coefficients 
\begin{equation*}
{\bf d}(\omega_k) = \sum_{t=1}^{T} {\bf X}(t) \exp(-i 2 \pi \omega_k t)
\end{equation*}
at the fundamental frequencies $\omega_k = \frac{k}{T}$ where 
$k = - (\frac{T}{2}-1), \ldots, \frac{T}{2}).$ The $P \times 
P$ periodogram matrix is 
\begin{equation*}
{\bf I}(\omega_k) = \frac{1}{T} {\bf d}(\omega_k) {\bf d}^*(\omega_k)
\end{equation*}
where ${\bf d}^*$ is the complex-conjugate transpose of ${\bf d}$. 

Though the periodogram matrix is asymptotically unbiased, it is not 
consistent. We can mitigate this issue by constructing a smoothed periodogram matrix 
estimator 
\begin{equation*}
\widetilde{\bf f}(\omega) = \sum_{\ell} Q_b(\omega_{\ell} - \omega) {\bf I}(\omega_{\ell})
\end{equation*}
where the kernel weights $Q_b(u)$ are non-negative and sum to $1$. The bandwidth 
$b$ can be obtained using automatic bandwidth selection 
methods for periodogram smoothing, such as the least-squares in \cite{LeePeriod} 
or the gamma-deviance-GCV in \cite{GammaGCV}. The shrinkage targets are either 
the scaled identity matrix or a parametric matrix (that can be derived from a VAR model).

Suppose that a VAR$(L)$ model is fit to the signal ${\bf X}(t), t=1, \ldots, T\}$ 
$\Phi(B) {\bf X}(t) = {\bf W}(t)$  where  ${\bf W}(t)$ is a zero-mean $P-$variate 
white noise with $\cov {\bf W}(t) = \Sigma_{\bf W}$ and 
$\Phi(B) = I - \Phi_1 B - \ldots - \Phi_L B^{L}$. Then the spectrum of this VAR($L$) process is 
\begin{equation*}
{\bf h}(\omega) = \Phi(\exp(-i 2 \pi \omega)) \ \Sigma_{\bf W} \  \Phi^*(\exp(-i 2 \pi \omega))
\end{equation*}.

The parametric estimate of the spectral matrix, denoted ${\widetilde{\bf h}}(\omega)$, is 
obtained replacing $\Phi_{\ell}$, $\ell=1, \ldots, L$, by the maximum likelihood or conditional 
maximum likelihood estimators. The shrinkage estimator for the spectrum takes the 
form 
\begin{eqnarray*}
\widehat{\bf f}(\omega) = W_1(\omega) \widetilde{I}(\omega) + W_2(\omega)) \widetilde{\bf h}(\omega)
\end{eqnarray*}
where the weights $W_{s}(\omega)$ fall in $(0,1)$ and $W_1(\omega) + W_2(\omega)$ for 
each $\omega$. Moreover, the weight for the smoothed periodogram is proportional to the 
mean-squared error of the parametric estimator, i.e., $W_1(\omega) \propto \ex \lvert \widetilde{h}(\omega) \ - \ {\bf f}(\omega) \vert^2$.

This method can be interpreted as a spectrum estimator 
that automatically chooses the estimator with better 
"quality". Therefore, when the parametric estimator 
is "poor" (high MSE), the weights shift the estimator 
towards the nonparametric estimator. However, when the 
parametric model gives a good fit, the shrinkage shift 
to the parametric spectral estimate. Thus, the resulting 
spectral estimator has, in general, a good condition 
number and can be used further for the estimation of 
PC where the quality of the spectral information is critical.

An alternative view to the above approach in constructing partial coherence is through analyzing the
oscillations. For simplicity, we consider the $\omega$-oscillations for channels $X_p$, 
$X_q$ and $X_c$, which we denote to be $X_{p, \omega}$, 
 $X_{q, \omega}$ and  $X_{c, \omega}$, respectively. At this stage, we shall consider only the 
 contemporaneous (i.e., zero-phase or zero-lag) partial cross-correlation
\begin{equation*}
 \cor \{ X_{p, \omega}(t), X_{q, \omega}(t) \ \vert \ X_{c, \omega}(t) \}. 
\end{equation*}

To proceed, regress of $X_{p, \omega}(t)$ against $X_{c, \omega}(t)$ and 
extract the residuals, denoted $R_{p \dot c, \omega}(t)$ and also $X_{q, \omega}(t)$ 
against $X_{c, \omega}(t)$, and extract the residuals, named as $R_{q \dot c, \omega}(t)$.

Then, the zero-phase partial coherence between $X_p$ and $X_q$ at frequency 
$\omega$ is the quantity 
\begin{equation*}
\rho_{pq \dot c}(\omega)
  = \left\vert \COR{R_{p \dot c, \omega},  R_{q \dot c, \omega}} \right\vert^2. 
\end{equation*}

\begin{figure}
\centering%
\includegraphics[width=1\textwidth,height=\textheight]{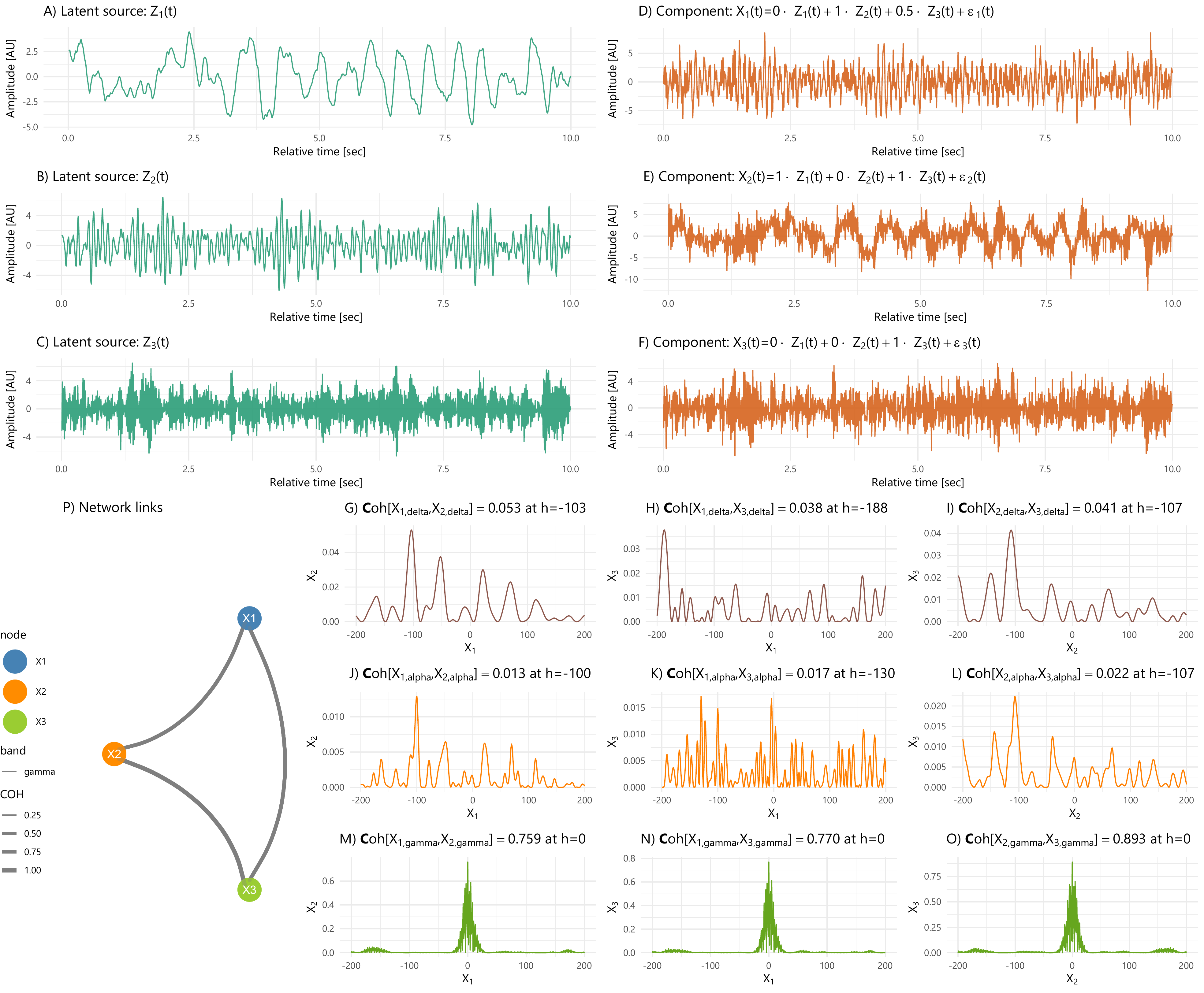}
\caption{%
Coherence network of a system with a dominating gamma component (Example \ref{ex:only-gamma-net}). Three latent sources (delta-, alpha- and gamma-band waves) are linearly into combined three observed components $X_1$, $X_2$ and $X_3$. As expected, the network map (P.) denotes that all components are related within the gamma-band. The correlation across lags is also displayed to emphasize the coherence magnitudes.
}
\label{fig:coh-latent-3-comp}
\end{figure}

\begin{figure}
\centering%
\includegraphics[width=1\textwidth,height=\textheight]{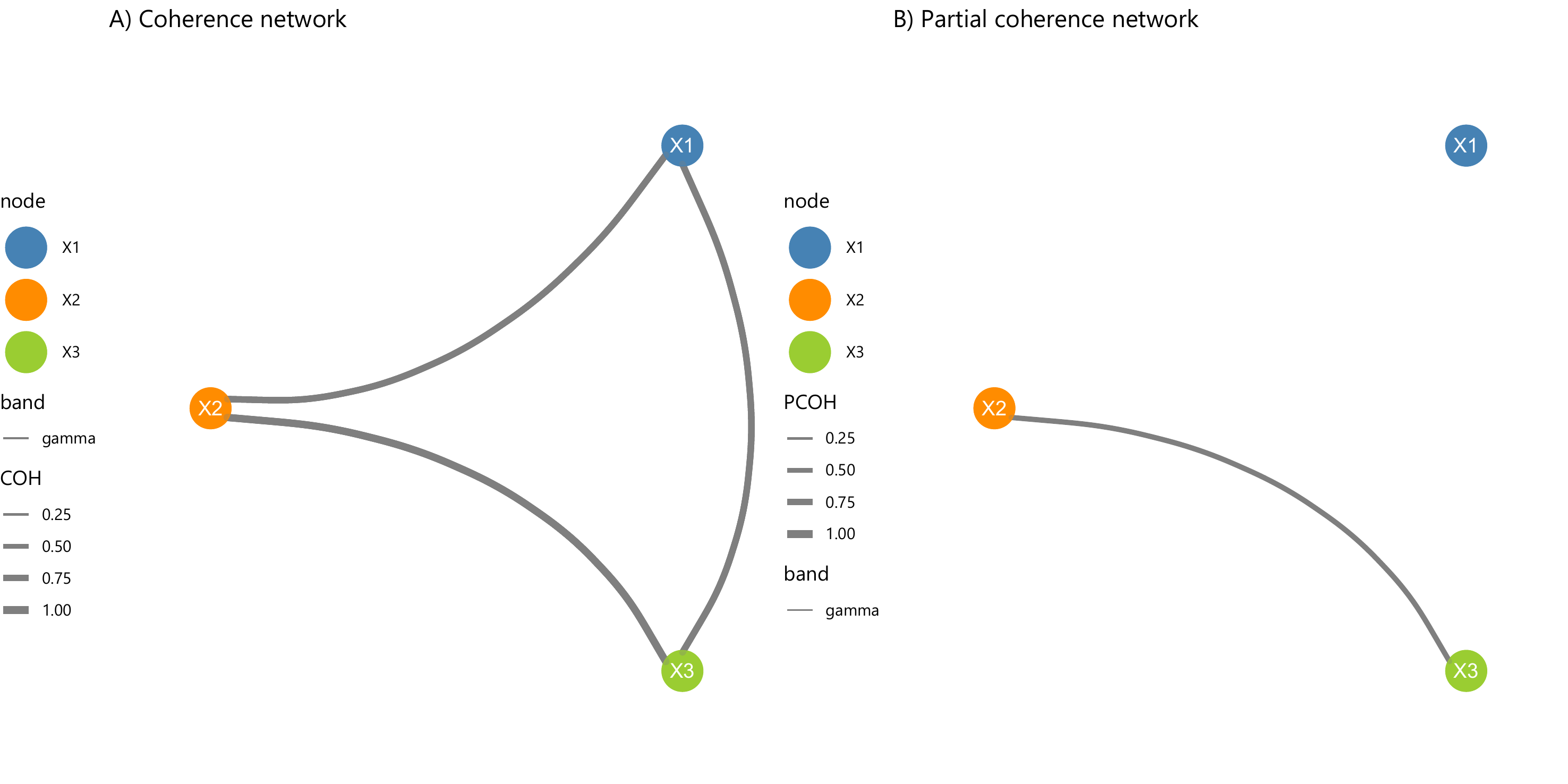}
\caption{%
Comparison between coherence and partial coherence in Example \ref{ex:only-gamma-net}.
}
\label{fig:comparison-coh-pcoh}
\end{figure}

\begin{example}[Partial coherence on a gamma-interacting system]\label{ex:only-gamma-net}
Consider the setting where the 
independent latent processes are specific AR(2)'s that mimic the delta, alpha, and gamma 
activity which we denote by $Z_{\delta}(t)$, $Z_{\alpha}(t)$ and $Z_{\gamma}(t)$. 
Suppose that the observed time series are $X_1(t)$, $X_2(t)$ and $X_3(t)$, which are 
defined by the mixture 
\begin{eqnarray} \label{Eq:ModelMix3}
\left ( 
\begin{matrix}
X_1(t) \\
X_2(t) \\
X_3(t) 
\end{matrix}
\right ) & = &  
\left ( 
\begin{matrix}
a_{1\delta} & a_{1\alpha} & a_{1\gamma} \\
a_{2\delta} & a_{2\alpha} & a_{2\gamma} \\
a_{3\delta} & a_{3\alpha} & a_{3\gamma}
\end{matrix}
\right )
\left (
\begin{matrix}
Z_{\delta}(t) \\
Z_{\alpha}(t) \\
Z_{\gamma}(t)
\end{matrix}
\right ).
\end{eqnarray}

We now examine the dependence between $X_1$ and $X_2$ 
under the following setting. First, suppose that $X_3$  contains 
the gamma-oscillatory activity $Z_{\gamma}$ purely, that is, 
$a_{3\delta} = a_{3\alpha} = 0$. Next, suppose that $X_1$ 
contains only $Z_{\alpha}$ and $Z_{\gamma}$, that is, 
$a_{1\delta} = 0$; and $X_2$ contains only $Z_{\delta}$ 
and $Z_{\gamma}$, that is, $a_{2\alpha} = 0$. A realization 
of such a system is shown in Figure \ref{fig:coh-latent-3-comp}.

Under this construction,
$X_1$ and $X_2$ have zero coherence at the 
$\delta$ and $\alpha$ frequency band. However, $X_1$ and 
$X_2$ both contain the common gamma-oscillatory activity 
and hence have a non-zero coherence at the gamma-band.
In our particular example, $\COH{X_{1,\gamma}, X_{2,\gamma}}=0.759$,
$\COH{X_{1,\gamma}, X_{3,\gamma}}=0.770$, and $\COH{X_{2,\gamma}, X_{3,\gamma}}=0.893$. 

However, if we compare coherence and partial coherence
(Figure \ref{fig:comparison-coh-pcoh}), we can unveil some 
indirect dependence effects. Partial coherence between $X_1$ and $X_2$ 
(conditional on $X_3$) at the gamma-band is zero. This result indicates that if
we remove $X_3$ (the gamma-band activity) from $X_1$ and 
$X_2$, then their "residuals" will no 
longer contain any common latent source. 
\end{example}

\begin{figure}
\centering%
\includegraphics[width=1\textwidth,height=\textheight]{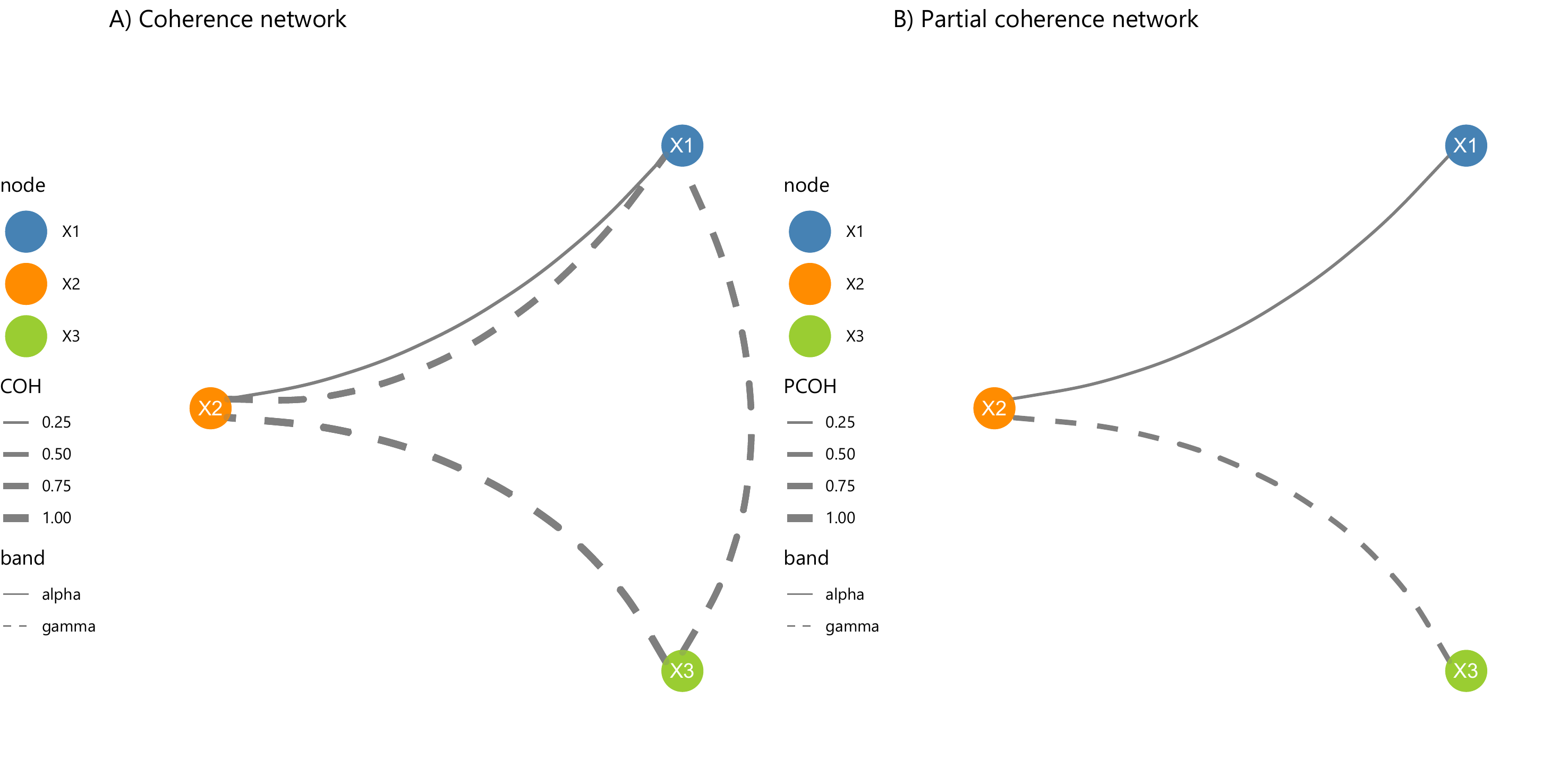}
\caption{%
Comparison between coherence and partial coherence in Example \ref{ex:gamma-alpha-net}.
}
\label{fig:comparison-coh-pcoh-alpha-gamma}
\end{figure}

\begin{example}[Partial coherence]\label{ex:gamma-alpha-net}
As a continuation 
of Example \ref{ex:only-gamma-net}, 
suppose that (a.) $X_3$  contains only gamma-oscillatory activity 
$Z_{\gamma}$, i.e., $a_{3\delta} = a_{3\alpha} = 0$; (b.) $X_1$ 
contains only $Z_{\alpha}$ and $Z_{\gamma}$: 
$a_{1\delta} = 0$; and (c.) $X_2$ contains $Z_{\delta}$, $Z_{\alpha}$ and 
$Z_{\gamma}$.

In such a dependence structure, we can observe the following frequency-dependent effects on $X_1$ and $X_2$:
\begin{itemize}
\item At the delta-band, the coherence between $X_1$ and $X_2$ 
is zero; and hence the partial coherence (conditioned on $X_3$) is also 
zero;
\item At the alpha-band, the coherence and partial coherence between 
$X_1$ and $X_2$ are both non-zero; 
\item At the gamma band, the coherence between $X_1$ and $X_2$ is 
non-zero but the partial coherence is zero.
\end{itemize}

These observations can be summarized in the coherence and partial coherence networks of Figure \ref{fig:comparison-coh-pcoh-alpha-gamma}.

\end{example}




\subsection{Time-varying coherence and partial coherence}

As noted, many brain signals exhibit non-stationarity. In some cases, the 
autospectra varies over time which indicates that the contributions of the 
various oscillations to the total variance change across the entire 
recording. In others, the autospectra might remain constant, but the 
strength and nature of the association between components can 
change. In fact, in \citet{fiecas2016}, the coherence between a pair 
 of tetrodes from a local field potential implanted in a monkey evolves 
 both within a trial and even across trials in an experiment. Here, we will 
 follow the models in \cite{Priestley:1965} and in \cite{Dahlhaus:2012locally} 
 to define and estimate the time-varying coherence and partial coherence. 
 
Recall that for a weakly stationary process, we define and estimate the spectral dependence quantities through the Cram\'er representation 
\begin{equation*}
 {\bf X}(t) = \int_{-0.5}^{0.5} \exp(i 2 \pi \omega t) d{\bf Z}(\omega)  
\end{equation*}
 where the random increments are uncorrelated across $\omega$ and 
 satisfy $\ex d{\bf Z}(\omega) = 0$, $\var d{\bf Z}(\omega) = {\bf f}(\omega) d\omega$.
An equivalent representation is 
\begin{equation*}
 {\bf X}(t) = \int {\bf A}(\omega) \exp(i 2 \pi \omega t) d{\bf U}(\omega)
\end{equation*}
where $d{\bf Z}(\omega) = {\bf A}(\omega) d{\bf U}(\omega)$; 
 ${\bf A}(\omega)$ is the transfer function matrix; $\var d{\bf U}(\omega) = 
 {\bf I} d\omega$; and the spectral matrix is 
\begin{equation*}
 {\bf f}(\omega) = {\bf A}(\omega) \ {\bf A}^*(\omega). 
\end{equation*}
 
The Priestley-Dahlhaus model for a locally stationary process allows the 
transfer function to change slowly over time. This idea was first proposed 
in \cite{Priestley:1965}, but later refined in \cite{Dahlhaus:2012locally} which 
developed an asymptotic framework for constructing a mean-squared 
consistent estimator for the time-varying spectral matrix. A simplified 
expression of the locally stationary time series model ${\bf X}(t), t=1, 
\ldots, T\}$ is
\begin{equation}\label{Eq:Dahlhaus}
{\bf X}(t)
  = \int_{-0.5}^{0.5}
    {\bf A}\left(\frac{t}{T}, \omega\right)
    \exp\left(i 2 \pi \omega t\right) d{\bf U}(\omega) 
\end{equation} 
 and the time-varying spectral matrix defined on rescaled time $\nicefrac{t}{T} \in (0,1)$ and 
 frequency $\omega \in (-0.5, 0.5)$ is 
\begin{equation*}
{\bf f}\left(\frac{t}{T}, \omega\right)
  = {\bf A}\left(\frac{t}{T}, \omega\right)%
    \ %
    {\bf A}^*\left(\frac{t}{T}, \omega\right). 
\end{equation*}

Based on this time-varying spectral matrix, the time-varying coherence 
between components $X_p$ and $X_q$ at rescaled time $\nicefrac{t}{T}$ and frequency 
$\omega$ is 
\begin{equation*}
\rho_{pq}\left(\frac{t}{T}, \omega\right)
  = \frac {\left\vert f_{pq}(\frac{t}{T}, \omega) \right\vert^2}
          { f_{pp}\left(\frac{t}{T}, \omega\right)
            f_{qq}\left(\frac{t}{T}, \omega\right)
          }.
\end{equation*}

In a similar vein, the time-varying partial coherence 
between components $X_p$ and $X_q$ (conditioned on $X_c$) 
at rescaled time $\nicefrac{t}{T}$ and frequency $\omega$ is derived in a similar manner
from Equation~\ref{Eq:PartialCoh}. The standard approach to estimating these 
time-varying spectral dependence measures is computing time-localized 
periodogram matrices and then smoothing these across frequency. 

There are other models and methods for analyzing non-stationary time series. 
One class of methods divides the time series into quasi-stationary blocks: 
dyadic piecewise stationary time series models in \cite{Adak:1998time}; 
dyadic piecewise aggregated AR(2) in \cite{MarcoDASAR2020, MarcoDASAR2021}; 
adaptive information-theoretic based segmentation in \cite{Davis:2006structural}. 
Another class of methods gives stochastic representations in terms of 
time-localized functions as building blocks like wavelets in \cite{Nason:2000}; 
or SLEX (smooth localized complex exponentials) in \cite{Ombao:2001automatic} 
and \cite{ombao2002slex}. For multivariate non-stationary time series, 
\cite{Ombao:2005} develop a procedure for selecting the best SLEX basis for 
signal representation. Using this basis, the estimate for the time-varying 
spectrum and coherence are then derived.



\section{Dual-frequency dependence}\label{Sec:DualFreq}

In this paper, our approach to modeling dependence between the 
components of a multivariate time series ${\bf X}$ is through the 
different oscillatory activity. This was motivated by the Cram\'er 
representation ${\bf X}(t) = \int_{-0.5}^{0.5} \exp(i 2 \pi \omega t) d{\bf Z}(\omega)$. 
For Gaussian weakly stationary processes, the random increments are 
independent across frequencies. Hence, for $\omega \ne \omega'$,
the $\omega-$oscillations in component $X_p$ and the $\omega'$-oscillations in 
$X_q$ are, by default, independent. However, there are situations when 
the interesting dependence structure is between different frequency oscillations. 

Lo\'eve, in \cite{Loeve1955}, introduced the class of harmonizable processes, which 
now allow for dependence between oscillatory activities at different oscillations. 
Here, we shall explore different characterizations of dependence between 
$X_{p, \omega}(t) = \exp(i 2 \pi \omega t) dZ_p(\omega)$ and 
$X_{q, \omega'}(t) = \exp(i 2 \pi \omega' t) dZ_q(\omega')$. The first natural 
measure of dependence is the linear association between these two oscillations, 
which we call the dual-frequency coherence
\begin{equation*}
\rho_{(p, \omega), (q, \omega')}
  = \COR{ X_{p, \omega}(t), X_{q, \omega'}(t) }
\end{equation*}.

This opens up many possibilities of characterizations of dependence that are 
beyond that by classical coherence. In this paper, we will further generalize 
this notion of dual-frequency coherence to the situation where this
measure could evolve over time. For example, during the course of one trial 
recording of an electroencephalogram, it is possible for coherence between 
the theta and gamma oscillations to be reduced immediately upon the 
presentation of a stimulus - but could strengthen over the course of a trial. 
We will summarize the notion of time-dependent dual-frequency coherence 
in Section~\ref{Sec:evol-dual-freq}. In addition, note that coherence, 
partial coherence, and dual-frequency coherence all capture only the 
linear dependence between the various oscillations. Here, we will examine 
other non-linear measures of dependence, such as the phase-amplitude 
coupling. In Section~\ref{Sec:phase-amp}, we will illustrate how the 
amplitude of a gamma-oscillatory activity in channel $p$ might be change 
according to the phase of the alpha-activity in channel $q$. One such 
example could be the increase in the gamma-oscillation amplitude 
when the theta oscillation reaches its peak.

\subsection{Evolutionary dual-frequency coherence}\label{Sec:evol-dual-freq}

The modeling framework for developing the notion of a dual-frequency that 
evolves over time was proposed in \cite{GorrostietaDFCoh} via a 
frequency-discretized harmonizable process which we now describe.
Here, we shall focus on the practical aspect of modeling and analyzing 
dual-frequency coherence in the observed signals. 
The technical details such as the asymptotic theory required for defining the 
population-specific quantity are developed in  \cite{GorrostietaDFCoh}. 

Suppose that we observe the time series ${\bf X}$ and focus on a local 
window centered time point $t$ with $N$ observations. Define the {\it local} 
Fourier coefficient vector at frequency $\omega$ at time point $t$ to be 
\begin{eqnarray}\label{STFT}
{\bf d}(t,\omega) & = & \frac{1}{\sqrt{N}} \sum_{s=t - (\frac{N}{2}-1)}^{t + \frac{N}{2}}
{\bf X}(s) \exp(-i 2 \pi \omega s)\ , \label{Eq:TrialDFT} 
\end{eqnarray} 
and define the local dual-frequency periodogram matrix at frequencies $\omega_j$ 
and $\omega_k$ to be
\begin{eqnarray} \label{Eq:Localperiodogram}
{\bm I}(t; (\omega_j, \omega_k)) = {\bm d}(t, \omega_j) {\bm d}(t, \omega_k) .
\end{eqnarray}
When the data consists of several time-locked trials, one can compute 
trial-specific local dual-frequency periodogram matrices and then average 
them (across trials) in order to obtain some population-specific measure of 
the evolutionary dual-frequency spectral matrix. In the absence of replicated 
trials (i.e., the data is only from a single trial), then one can smooth the local 
dual-frequency periodogram matrices over time $t$ within that single 
trial.  Denote the averaged dual-frequency periodogram matrix to be 
$\widehat{{\bf f}}(t, \omega_j, \omega_k)$. 
One measure of the strength of linear dependence between the
$\omega_j$-oscillations at component $X_p$ and the $\omega_k$-oscillations at component
$X_q$ is the time-localized dual-frequency coherence
\begin{eqnarray} \label{Eq:AveLocalcoherence}
\widehat{\rho}_{(p, \omega_j), (q, \omega_k)}(t)
  & = \frac{\vert \widehat{f}_{(p,\omega_j), (q,\omega_k)}(t) \vert^2}
           {\widehat{f}_{(p,\omega_j), (p,\omega_j)}(t)
            \widehat{f}_{(q,\omega_k, (q,\omega_k)}(t)
           }.
\end{eqnarray}

Consistent with the approach adopted in the paper, we investigate linear 
dependence between the oscillations by linear filtering. Consider two 
frequency bands $\Omega_1$ and $\Omega_2$ and the zero-mean filtered 
signals to be $X_{1, \Omega_1}$ and $X_{2, \Omega_2}$. We will provide 
the time-varying dual-frequency coherence estimate at a local time $t$ by 
first computing the local cross-covariance and local variance estimate over 
the window $\{t-L, \ldots, t+L\}$ (for some $L>0$) as follows
\begin{eqnarray*}
\widehat{f}_{(1, \Omega_1),(2, \Omega_2)}(t) & = \frac{1}{N}\sum_{s=t - (\frac{N}{2}-1)}^{t + \frac{N}{2}}X_{1, \omega_1}(s) X_{2, \omega_2}(s) \\
\widehat{f}_{(1, \Omega_1),(1, \Omega_1)}(t) & = \frac{1}{N}\sum_{s=t - (\frac{N}{2}-1)}^{t + \frac{N}{2}} X_{1, \omega_1}(s) X_{1, \omega_1}(s) \\
\widehat{f}_{(2, \Omega_2),(2, \Omega_2)}(t) & = \frac{1}{N}\sum_{s=t - (\frac{N}{2}-1)}^{t + \frac{N}{2}} X_{2, \omega_2}(s) X_{2, \omega_2}(s). 
\end{eqnarray*}

The estimated local dual-frequency coherence at local time $t$ is 
\begin{eqnarray}
\widehat{\rho}_{(1, \Omega_1),(2, \Omega_2)}(t) 
  = \frac{\vert\widehat{f}_{(1, \Omega_1),(2, \Omega_2)}(t)\vert^2}
         {\widehat{f}_{(1, \Omega_1),(1, \Omega_1)}(t) \widehat{f}_{(2, \Omega_2),(2, \Omega_2}(t) }
\end{eqnarray}

\subsection{Phase-Amplitude coupling}\label{Sec:phase-amp}

In the previous sections, we examined coherence and partial coherence - both 
of which aim to measure the strength of linear dependence between a pair of 
channels or components $X_p$ and $X_q$ through the oscillatory activity at 
the {\it same} frequency. In Section \ref{Sec:DualFreq}, we examine dependence 
at different frequencies and how this dependence may change over time through 
the evolutionary dual-frequency coherence. However, all of these only examine 
{\it linear} dependence. 
In the neuroscience literature, it was acknowledged that inter-frequency modulation could
also appear as a different category of dependence in the brain networks.
Similarly to coherence, this dependence type implies interactions within the frequency domain:
spectral components in $\Omega_H$ are {\it modulated} by
components resonating at frequencies in $\Omega_L$.
When the modulator $\Omega_H$ works at a high frequency, 
and the modulated signal $\Omega_L$, this type of dependence is named phase-amplitude coupling (PAC).
Baseline PAC networks in the human brain changes since birth \citep{DevelopmentalChangesEEG-Mariscal-2019}, 
and specific delta-theta PAC patterns can be altered due to
anesthetic effects \citep{DeltaWavesDifferently-Molaee-Ardekani-2007}.

A common method to quantify PAC in time series was introduced by Tort et al. \citep{PhaseAmplitudeCoupling-Tort-2010}
through the modulation index (MI) for univariate signals.
We should emphasize that this metric is not the same as the homonym "modulation index" defined for amplitude-modulated systems \citep[p.~591]{SigSystems-Oppenheim-1997a}.
To estimate MI, let $x_\Omega$ be a filtered time series with its spectrum concentrated only in the interval $\Omega$. Then, its analytic signal $y_\Omega$ will be defined via the Hilbert transform:
\begin{equation*}
y_{\Omega}\left(t\right)=x_{\Omega}\left(t\right)+j\left(\frac{1}{\pi t}\ast x_{\Omega}\left(t\right)\right),
\end{equation*}

The analytic $y_\Omega(t)$ can be expressed using an exponential form,
\begin{equation*}
y_{\Omega}\left(t\right)=A_{\Omega}\left(t\right)\exp\left(j\phi_{\Omega}\left(t\right)\right).
\end{equation*}
such that the instantaneous phase and amplitude are identified through $\phi_\Omega(t)$ and $A_\Omega(t)$, respectively.

Now obtain $y_{\Omega_L}(t)$ and $y_{\Omega_H}(t)$ for a target low-frequency range $\Omega_L$ and a high-frequency 
interval $\Omega_H$ such as delta and gamma, respectively. Now, consider the joint signals $T=\bigl(A_{\Omega_L}, \phi_{\Omega_H}\bigr)$.
The original algorithm, described by Tort et al., suggests creating a partition of the phase domain $Q$:
$Q=\cup_{j=1}^{N}Q_{j}\quad Q_{j}=\left[2\pi\frac{j-1}{N}2\pi\frac{j}{N}\right)$
and
estimate the marginal mean value in each partition
\begin{equation*}
P(j)
  = \mathbb{E}\left(
      T\mid \phi_{\Omega_H}=\phi
    \right)
    \quad
    \phi\in Q_j
\end{equation*}

Under the non-modulation hypothesis, $P(j)$ should resemble a uniform distribution $P_U(j)=1\,\forall j$.
MI is then introduced as a normalized measure of the divergence between of the difference between $P(j)$ and the uniform alternative $P_U(j)$:
\begin{equation*}
\text{MI}=\frac{1}{\log\left(N\right)}D_{KL}\left(P,U\right)
\end{equation*}
where $D_{KL}(\cdot)$ is the Kullback-Leibler divergence:
\begin{equation*}
D_{KL}\left(P,Q\right)=\sum_{k=1}^{N}P\left(k\right)\log\left(\frac{P\left(i\right)}{Q\left(i\right)}\right)
\end{equation*}

Note that in those cases where no specific frequency band is known,
we suggest starting exploring PAC phenomena in the data using the "smoothed" signals (low-frequency) and their residuals (high-frequency components).
We refer to \citep{Periodic-Fried-2012,RTSigProcessing-Schettlinger-2009,MedianFiltersExtensions-Fried-2011} for a comprehensive description of time-memory efficient estimation methods.

\begin{figure}
\centering%
\includegraphics[width=0.9\textwidth,height=\textheight]{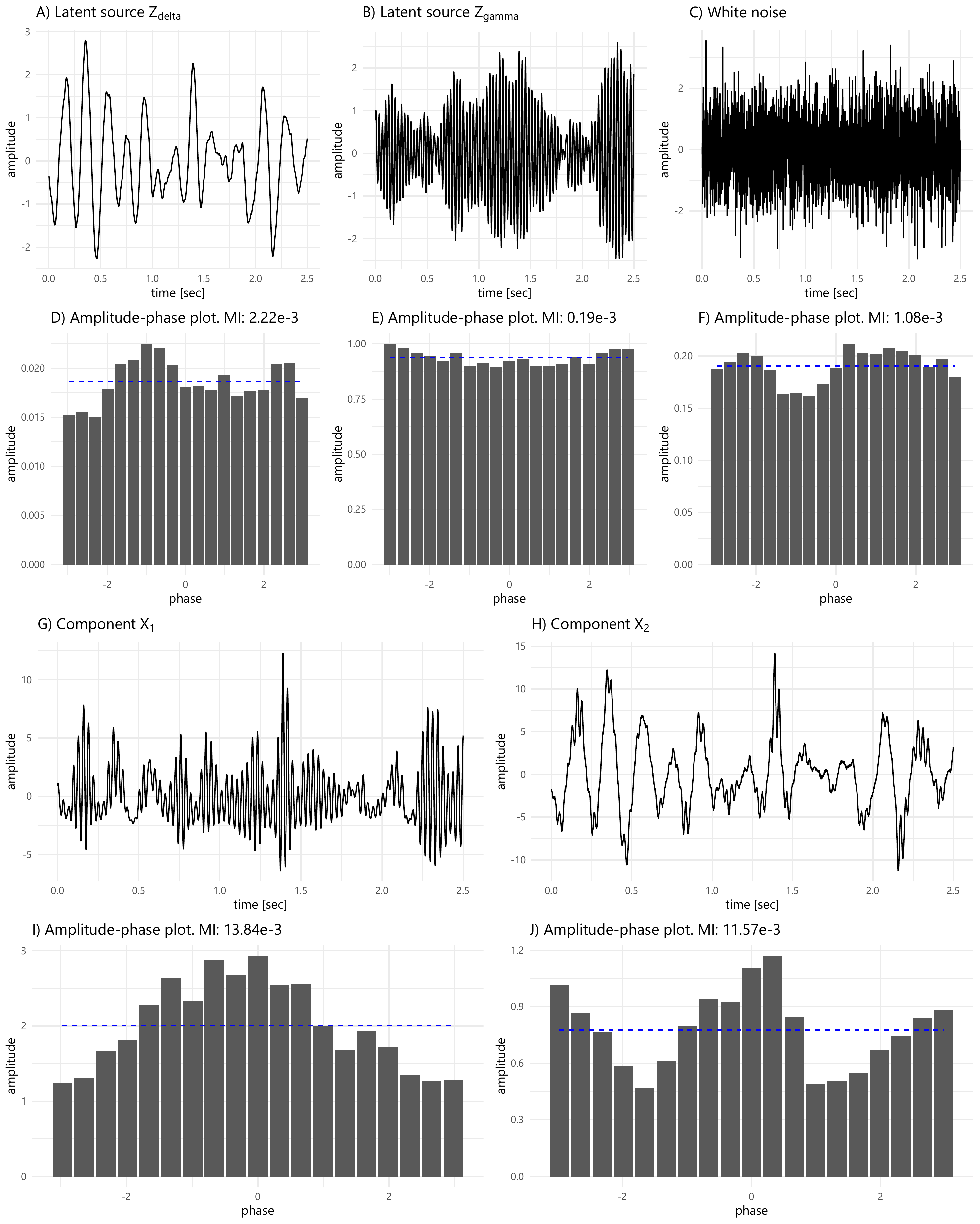}
\caption{%
Phase-amplitude coupling in Example \ref{ex:phase-amplitude-coupling}: phase-amplitude graphs $P(j)$ and modulation indexes (MIs) for the latent sources $Z_{\delta}(t)$,
$Z_{\gamma}(t)$ and the observations $X_{1}(t)$ and $X_{2}(t)$.
Note that MI values in $X_{1}(t)$ and $X_{2}(t)$ are up to five times greater than MI magnitudes of the latent sources or the noise $\varepsilon(t)$.
}
\label{fig:phase-amplitude-coupling}
\end{figure}

\begin{example}[Phase-amplitude coupling]\label{ex:phase-amplitude-coupling}
Define the theta-band and gamma-band latent sources to be $Z_{\theta}$ and 
$Z_{\gamma}$, respectively. Let assume that some modulation effects are observed
as a result of non-linear mixtures of the latent sources:
\begin{align*}
\left(\begin{matrix}%
X_{1}(t)\\
X_{2}(t)
\end{matrix}\right)
& =%
\left(\begin{matrix}%
A_{1,\delta}(t) & A_{1,\gamma}(t)\\
A_{2,\delta}(t) & A_{2,\gamma}(t)
\end{matrix}\right)
\left(\begin{matrix}%
Z_{\theta}(t)\\
Z_{\gamma}(t)
\end{matrix}\right)%
+%
\varepsilon\bigr(t\bigl)
\end{align*}
where $\varepsilon\bigr(t\bigl)\sim\mathcal{N}\bigr(0,\Sigma_\varepsilon\bigl)$ with a covariance matrix
$\Sigma_\varepsilon=0.1{\bf I}$ such that ${\bf I}$ is an identity matrix.

Let us assume that two phase-amplitude coupling effects are observed:
\begin{itemize}
\item $X_{1}(t)$ denotes an amplitude-modulation effect where the amplitude of $Z_{\theta}(t)$ 
instantaneously leads to the amplitude's changes in the $\gamma$-oscillations. Therefore, the mixture functions 
are defined as
$A_{1,\delta}\left(t\right)=Z_{\gamma}\left(t\right)+1$ and
$A_{1,\gamma}\left(t\right)=2$.
\item $X_{2}(t)$ shows a modulation effect where the $Z_{\gamma}(t)$ component has a low impact on
the $\delta$-oscillations: $A_{2,\delta}\left(t\right)=4$ and $A_{2,\gamma}\left(t\right)=Z_{\theta}$.
\end{itemize}

Figure \ref{fig:phase-amplitude-coupling} shows a simulation of this process along with the modulation indexes for $Z_{\delta}(t)$, $Z_{\gamma}(t)$, $\varepsilon(t)$ and the observed components $X_{1}(t)$ and $X_{2}(t)$. It is visually apparent that $P(j)$ is closer to $P_{U}(j)$ in scenarios without modulation effects (Figure \ref{fig:phase-amplitude-coupling}.A-F). In addition, we can remark that the modulated process $X_{2}(t)$ implies an alteration on the extreme values of the process, and it could also be modeled using statistical models for extreme values \citep{ConexConnectLearningPatterns-Guerrero-2021}.

\end{example}

\section{Partial directed coherence and spectral causality}

All previously mentioned spectral measures of dependence: 
coherence, partial coherence, dual-frequency coherence, 
and evolutionary dual-frequency coherence all ignore the 
lead-lag dependence between oscillatory components. 
This notion of lead-lag is very important in neuroscience, 
particularly in identifying effective connectivity between 
brain regions or channels. In fact, many pioneering models 
for brain connectivity were applied to functional magnetic 
resonance imaging data and therefore took into account 
the spatial structure in the brain data (\cite{BowmanBayesian}).
To address the computational issues for spatial covariance 
\cite{Castruccio} introduced a scalable multi-scale approach 
(local for voxels within a region of interest; global for 
regions of interest in the entire network). In another 
approach, \cite{KangSpatioSpec} the temporal covariance structure 
is diagonalized (and hence sparsified) by applying a Fourier transform 
on the voxel-specific time series while taking into account 
variation across subjects through a mixed-effects modeling 
framework. In a recent work, \cite{Luo-Granger} developed a method 
for mediation analysis in fMRI data,
and \citep{SpatiotemporalNonParamBayesian-Zhang-2016} proposed a variational Bayes algorithm
with reduced computational costs.

There have also been a number of statistical methods for  
modeling dynamic connectivity in fMRI: dynamic correlation in 
\cite{Cribben-2012}; switching vector autoregressive (VAR) 
model in \cite{Balqis2017}, regime-switching factor models in 
\cite{Ting2018A} and \cite{Shappell}; Bayesian model of brain 
networks in \citep{SpatiotemporalNonParamBayesian-Zhang-2016,BayesianModelsF-Zhang-2015,Mumford, NPBayesfMRINonParamBayesian-Kook-2019,BayesianGraphicalNetwork-Rembach-2015,BayesianModelsF-Zhang-2015}; models for high dimensional networks in
\cite{Ting2018B};
detection of dynamic community structure in \citep{Bassett}, 
\cite{MultiScaleFactor-Ting-2020} and \cite{Ting2021}. In \cite{ZheYu-JASA}, a Bayesian 
VAR model was used to assess stroke-induced changes in the 
functional connectivity structure.

In this section, we will cover approaches to identify lead-lag 
structures in brain signals using the vector autoregressive (VAR) 
models. Modeling these lead-lag dynamics is crucial to understanding 
the nature of brain systems, the impact of shocking events (such as stroke) 
on the dynamic configuration of such systems, and the 
downstream effect of such disruption on cognition and behavior.  
However, the emphasis of this section will be on investigating 
frequency-band specific lead-lag dynamics. Thus, while the 
classical VAR model characterizes the effect of the past observation 
of the signal $X_q$ on a future observation on the signal $X_q$, 
our emphasis here will be on modeling and assessing the impact 
of the previous oscillatory activity $X_{q, \omega_1}$ on the 
future oscillations $X_{p, \omega_2}$. 

\subsection{Vector autoregressive (VAR) models}

A $P$-variate time series ${\bf X}$ is said to be a vector autoregressive process 
of order $L$ (denoted VAR$(L)$) if it is weakly stationary and can be expressed as 
$\Phi(B) {\bf X}(t) = {\bf W}(t)$ where $\bigl\{ {\bf W}(t) \bigr\}$ is white noise with 
$\EX{{\bf W}(t)} = 0$ for all time $t$ and $\COV{{\bf W}(t)} = \Sigma_{\bf W}$
and 
\begin{equation*}
\Phi(B)  =  I - \Phi_1 B  - \Phi_2 B^2  - \ldots - \Phi_L B^{L} 
\end{equation*}
and $\Phi_{\ell}$, $\ell=1, \ldots, L$ are the $P \times P$ VAR coefficient matrices.
A comprehensive exposition of VAR models, 
including conditions for causality and methods for estimation of the coefficient 
matrices $\{\Phi_{\ell}\}$, are provided in \citep{NewIntroductionMultiple-Lutkepohl-2005}.
Note that from classical 
time series literature, the notion of "causality" is different from that of Granger 
causality. A time series ${\bf X}$ is causal if ${\bf X}(t)$ depends only on the 
current or past white noise $\{ {\bf W}(s), \ s = t, t-1, t-2, \ldots\}$. 

Consider the two components $X_p$ and $X_q$. From the point of view of forecasting, 
we say that ``$X_q$ Granger-causes $X_p$" (we write $X_q \longrightarrow X_p$) 
if the squared error for the forecast of  $X_p$ that uses the past values of $X_q$ is lower 
than that for the forecast that does not use the past values of $X_q$. Under the context of 
VAR models, note that 
\begin{equation}
X_p(t) = \sum_{\ell = 1}^{L} \Phi_{pq, \ell} X_q(t-\ell) + \sum_{r \ne q} \sum_{\ell=1}^{L} 
\Phi_{pr, \ell} X_r(t-\ell)
\end{equation}
where $\Phi_{pq, \ell}$ is the $(p,q)$ element of the matrix $\Phi_{\ell}$. For $X_p(t)$, 
it is the coefficient associated with the past value $X_q(t-\ell)$. Thus, $X_q\longrightarrow X_p$) 
if there exists some lag $\ell*$ where $\Phi_{pq, \ell*} \ne 0$. There is a large body of 
work on causality, starting with the seminal paper by \cite{Granger}. This was further 
studied in \cite{Geweke1982}, \cite{Geweke1984}, \cite{Hosoya1991}.
Additional applications for subject- and group-level analysis was also analyzed in \citep{BayesianVectorAR-Chiang-2017}.
Recently, under non-stationarity, the nature of Granger-causality could evolve over time and this was 
investigated in \cite{YanLiu}. Despite the fact that the concept of causality are derived from the 
spectral representation, the focus on the interpretations for causality has not been on 
the actual oscillatory activities. The goal is this section is to refocus the spotlight on this 
very important role of the oscillations in determining causality and, in general, directionality 
between a pair of signals.

\subsection{Partial directed coherence}

As noted, all previously discussed measures of coherence lack the important 
information on directionality. The concept of partial directed coherence (PDC), 
introduced in \cite{partial_directed_coherence_1} and \cite{UnifiedAsymptoticTheory-Baccala-2013}, 
gives this additional information. Define the VAR(L) transfer function to be 
\begin{equation*}
\Phi(\omega) = I - \sum_{\ell=1}^{P} \Phi_{\ell} \exp(-i 2 \pi \omega \ell)
\end{equation*}
and denote $\Phi_{pq}(\omega)$ to be the $(p,q)$ element of the transfer 
function matrix $\Phi(\omega)$. Then the PDC, from component $X_q$ to 
$X_p$, at frequency $\omega$ is 
\begin{equation*}
\pi_{pq}(\omega) = \frac{\vert \Phi_{pq}(\omega) \vert^2}{\sum_{rq}\vert \Phi_{rq}(\omega) \vert^2}.
\end{equation*}
Note that $\pi_{pq}(\omega)$ lies in $[0,1]$ and measures the amount of information flow, 
at frequency $\omega$, from component $X_q$ to $X_p$, \underline{relative} to the 
total amount of information flow from $X_q$ to all components. When $\pi_{pq}(\omega)$  
is close to $1$ then most of the $\omega$-information flow from $X_q$ goes directly to 
component $X_p$. Under this framework, we estimate PDC fitting a VAR model to the 
data ${\bf X}(t), t=1, \ldots, T\}$ where the optimal order $L$ can be selected using 
some objective criterion like the Akaike information criterion (AIC) or the Bayesian-Schwartz 
information criterion (BIC). 

\begin{figure}
\centering%
\includegraphics[width=0.9\textwidth,height=\textheight]{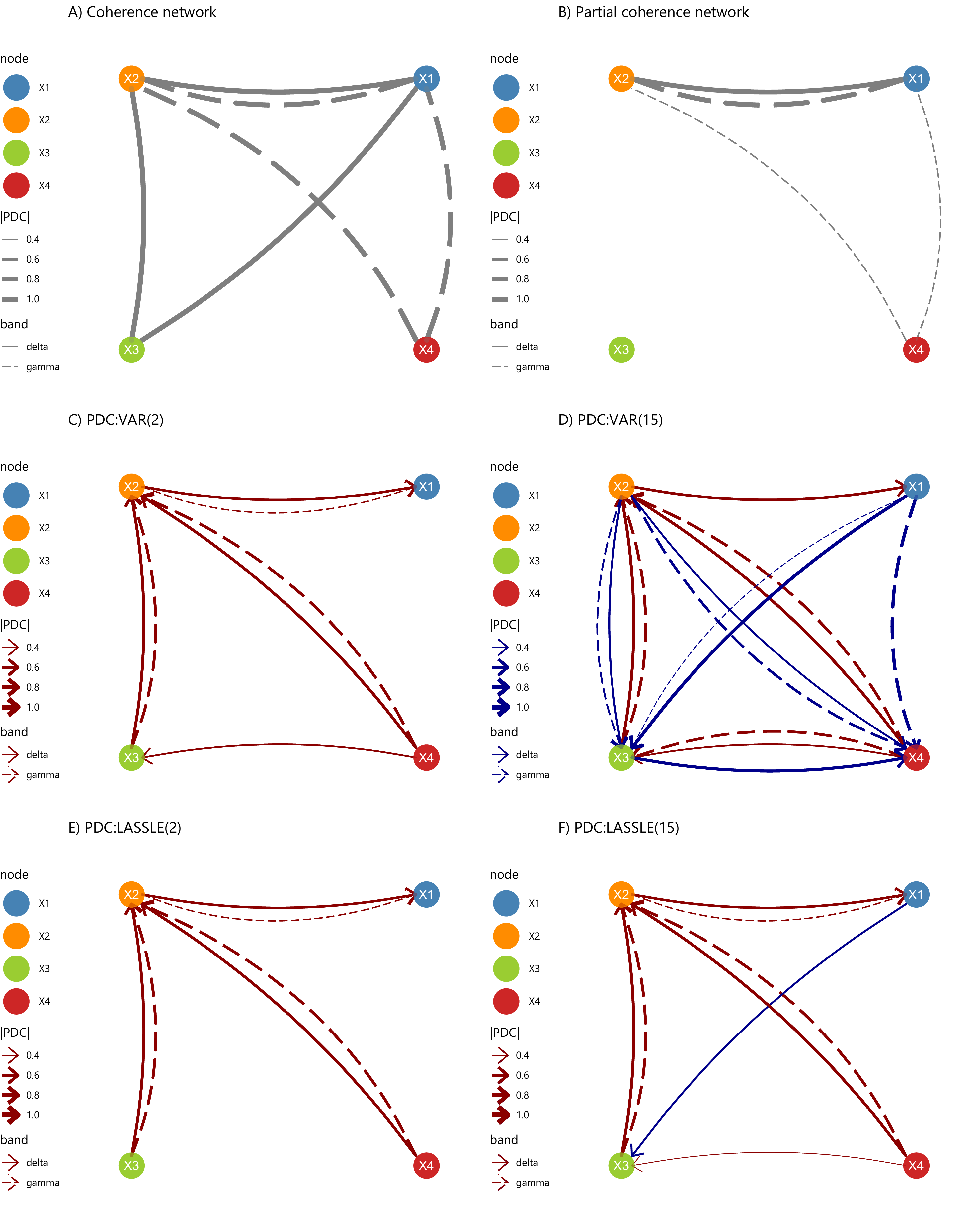}
\caption{%
Connectivity networks for the system in Example \ref{ex:comparison-coh-pcoh-pdcoh}.
Dependence metrics were obtained from six methods: coherence (COH), partial coherence (PCOH), 
partial directed coherence (PDC) estimated with an OLS-estimated VAR(2) and VAR(15) models,
and PDC estimated with a regularized second- and fifteenth-order LASSLE models.
\label{fig:comparison-coh-pcoh-pdcoh}
}
\end{figure}

\begin{example}[Connectivity network comparison]\label{ex:comparison-coh-pcoh-pdcoh}
Assume a system with four channels that can be described by the following sparse VAR$(2)$ model:
\begin{align*}
\left(\begin{matrix}X_{1}(t)\\
X_{2}(t)\\
X_{3}(t)\\
X_{4}(t)
\end{matrix}\right) & =\left(\begin{matrix}\phi_{\beta,1} & \frac{1}{2} & 0 & 0\\
0 & 0 & 1 & 0\\
0 & 0 & \phi_{\delta,1} & 0\\
0 & 0 & 0 & \phi_{\gamma,1}
\end{matrix}\right)\left(\begin{matrix}X_{1}(t-1)\\
X_{2}(t-1)\\
X_{3}(t-1)\\
X_{4}(t-1)
\end{matrix}\right)+\left(\begin{matrix}\phi_{\beta,2} & 0 & 0 & 0\\
0 & 0 & 0 & 1\\
0 & 0 & \phi_{\delta,2} & 0\\
0 & 0 & 0 & \phi_{\gamma,2}
\end{matrix}\right)\left(\begin{matrix}X_{1}(t-2)\\
X_{2}(t-2)\\
X_{3}(t-2)\\
X_{4}(t-2)
\end{matrix}\right)+\varepsilon\left(t\right)
\end{align*}
where $\left(\phi_{\delta,1},\phi_{\delta,2}\right)$, $\left(\phi_{\gamma,1},\phi_{\gamma,2}\right)$ and $\left(\phi_{\beta,1},\phi_{\beta,2}\right)$ were calculated as described in Equation \ref{eq:AR2-oscillator} with $M=1.049787$ and $\psi_{\delta}=\frac{2}{128}$, $\psi_{\gamma}=\frac{40}{128}$ and $\psi_{\beta}=\frac{20}{128}$, respectively. The noise $\varepsilon\left(t\right)\sim\mathcal{N}\left(0,\Sigma\right)$ where $\Sigma$ is a diagonal matrix. Consequently, $X_{3}(t)$ and $X_{4}(t)$ are independent delta and gamma components.

In this multivariate system, direct and indirect lagged dependence links are denoted: $X_{3}$ and $X_{4}$ \emph{directly} lead $X_{2}$: $X_{4}(t-2)\Longrightarrow X_{2}(t)$, $X_{3}(t-1)\Longrightarrow X_{2}(t)$, while both \emph{indirectly} affect $X_{1}$ through $X_{2}$: $X_{2}(t-1)\Longrightarrow X_{1}(t)$. Previously, coherence (COH) and partial coherence (PCOH) were applied to identify \emph{instantaneous} direct dependencies. Figure \ref{fig:comparison-coh-pcoh-pdcoh} and Table \ref{tbl:comparison-coh-pcoh-pdcoh} show the connectivity networks that can be estimated using PDC in addition to both dependence metrics along with their magnitudes.

We should emphasize that VAR mismodeling can considerably affect the dependence metrics that rely on them (as can be observed in Figure \ref{fig:comparison-coh-pcoh-pdcoh}.F). However, these effects can be mitigated with regularization techniques that are discussed in Section \ref{sec:regularized-VAR}. For a comprehensive empirical analysis of the mismodeling phenomena in connectivity, we refer to \citep{RegularizedStructuralEquation-Stephanie-2017}.
\end{example}

\begin{table}\centering
\scalebox{0.8}{{
\begin{tabular}{>{\centering}p{1.75cm}>{\centering}p{1.75cm}>{\centering}p{1.75cm}>{\centering}p{1.75cm}>{\centering}p{1.9cm}>{\centering}p{1.9cm}>{\centering}p{1.9cm}>{\centering}p{1.9cm}}
\toprule 
\textbf{Band} & \textbf{Link} & \textbf{COH} & \textbf{PCOH} & \textbf{PDC: VAR(2)} & \textbf{PDC: VAR(15)} & \textbf{PDC: LASSLE(2)} & \textbf{PDC: LASSLE(15)}\tabularnewline
\midrule
$\delta$ & $X_{2}\Rightarrow X_{1}$  & \textbf{0.9996} & \textbf{0.9749} & \textbf{0.5096} & \textbf{0.5334} & \textbf{0.5097} & \textbf{0.5097}\tabularnewline
0-4 & $X_{1}\Rightarrow X_{2}$  &  &  & 0.0000 & 0.0070 & 0.0000 & 0.0000\tabularnewline
Hertz & $X_{3}\Rightarrow X_{1}$  & \textbf{0.9959} & \textbf{0.2074} & 0.0001 & 0.0438 & 0.0000 & 0.0000\tabularnewline
 & $X_{1}\Rightarrow X_{3}$  &  &  & \textbf{0.1754} & \textbf{0.6465} & \textbf{0.1878} & \textbf{0.4634}\tabularnewline
 & $X_{4}\Rightarrow X_{1}$  & 0.0257 & 0.0347 & 0.0001 & 0.0440 & 0.0000 & 0.0000\tabularnewline
 & $X_{1}\Rightarrow X_{4}$  &  &  & 0.0061 & 0.3040 & 0.0115 & 0.0031\tabularnewline
 & $X_{3}\Rightarrow X_{2}$  & \textbf{0.9961} & \textbf{0.3320} & \textbf{0.5774} & \textbf{0.5786} & \textbf{0.5774} & \textbf{0.5773}\tabularnewline
 & $X_{2}\Rightarrow X_{3}$  &  &  & 0.2799 & 0.4561 & 0.0900 & 0.1495\tabularnewline
 & $X_{4}\Rightarrow X_{2}$  & 0.0259 & 0.0336 & \textbf{0.5773} & \textbf{0.5789} & \textbf{0.5773} & \textbf{0.5773}\tabularnewline
 & $X_{2}\Rightarrow X_{4}$  &  &  & 0.0061 & 0.4376 & 0.0041 & 0.0096\tabularnewline
 & $X_{4}\Rightarrow X_{3}$  & 0.0239 & \textbf{0.2715} & \textbf{0.4152} & \textbf{0.4152} & 0.0448 & \textbf{0.3549}\tabularnewline
 & $X_{3}\Rightarrow X_{4}$  &  &  & 0.0052 & 0.6050 & 0.0000 & 0.0000\tabularnewline
 & $X_{2}\Rightarrow X_{1}$  & 0.9857 & \textbf{0.9863} & \textbf{0.3854} & \textbf{0.2901} & \textbf{0.3854} & \textbf{0.3855}\tabularnewline
 & $X_{1}\Rightarrow X_{2}$  &  &  & 0.0000 & 0.0547 & 0.0000 & 0.0000\tabularnewline
\midrule
$\gamma$ & $X_{3}\Rightarrow X_{1}$   & 0.0280 & 0.0404 & 0.0001 & 0.1165 & 0.0000 & 0.0000\tabularnewline
30-50 & $X_{1}\Rightarrow X_{3}$  &  &  & 0.0009 & \textbf{0.3611} & 0.0009 & 0.0036\tabularnewline
Hertz & $X_{4}\Rightarrow X_{1}$  & \textbf{0.9860} & \textbf{0.3938} & 0.0000 & 0.1170 & 0.0000 & 0.0000\tabularnewline
 & $X_{1}\Rightarrow X_{4}$  &  &  & 0.1233 & \textbf{0.6449} & 0.0599 & 0.0882\tabularnewline
 & $X_{3}\Rightarrow X_{2}$  & 0.0232 & 0.0394 & \textbf{0.5773} & \textbf{0.5800} & \textbf{0.5774} & \textbf{0.5773}\tabularnewline
 & $X_{2}\Rightarrow X_{3}$  &  &  & 0.0012 & 0.4270 & 0.0006 & 0.0023\tabularnewline
 & $X_{4}\Rightarrow X_{2}$  & \textbf{0.9996} & \textbf{0.3967} & \textbf{0.5774} & \textbf{0.5808} & \textbf{0.5773} & \textbf{0.5773}\tabularnewline
 & $X_{2}\Rightarrow X_{4}$  &  &  & 0.0212 & \textbf{0.5365} & 0.0216 & 0.0323\tabularnewline
 & $X_{4}\Rightarrow X_{3}$  & 0.0236 & 0.0179 & 0.0017 & \textbf{0.5550} & 0.0003 & 0.0025\tabularnewline
 & $X_{3}\Rightarrow X_{4}$ &  &  & \textbf{0.2119} & \textbf{0.3476} & 0.0000 & 0.0000\tabularnewline
\bottomrule
\end{tabular}
}}
\caption{%
Comparison of the connectivity metrics between channels and frequency in Example \ref{ex:comparison-coh-pcoh-pdcoh}:
coherence (COH), partial coherence (PCOH), and partial directed coherence (PDC).
PDC was estimated through four different vector autoregressive models: VAR(2) and VAR(15)
and regularized models: LASSLE(2) and LASSLE(15).
\label{tbl:comparison-coh-pcoh-pdcoh}
}
\end{table}

\subsection{Time-varying PDC}

A natural question to ask would be how to characterize and estimate 
PDC when it is changing over time. As noted, during the course of 
a trial, an experiment or even within an epoch, the brain functional 
network is dynamic (\cite{Ting2021}). Thus, one may characterize 
a time-varying PDC through a time-varying VAR model 
\begin{equation*}
\Phi_t(B) = I - I - \Phi_{t,1} B  - \Phi_{t,2} B^2  - \ldots - \Phi_{t,L} B^{L}
\end{equation*}
where $\Phi_{t,\ell}$ is the VAR coefficient matrix for lag $\ell$ at time 
$t$. The dimensionality of the parameters for at any time $t$ for a
time-varying VAR$(L)$ model is $P^2 L$. Thus, in order to have 
a sufficient number of observations at any time $t$, one can estimate 
the time-varying coefficient matrices $\{\Phi_{t,\ell}, \ell=1, \ldots, L\}$
by fitting a local conditional least squares estimate to a local data 
at time $\{t-\left (\frac{N}{2}-1 \right)$, $\ldots, \frac{N}{2}\}$. In addition 
to borrowing information from observations within a window (to increase 
the number of observations), it is also advisable to apply some  
regularization which will be discussed in Section~\ref{sec:regularized-VAR}. 
After obtaining estimates $\{\widehat{\Phi}_{t, \ell} \}$, we compute the 
estimate of the time-varying transfer function $\widehat{\Phi}(t,\omega)$, 
which then leads to a time-varying PDC estimate 
\begin{equation*}
\pi_{pq}(t,\omega) = \frac{\vert \Phi_{pq}(t,\omega) \vert^2}{\sum_{rq}\vert \Phi_{rq}(t,\omega) \vert^2}.
\end{equation*}

PDC has provided the important frequency-specific information of 
directionality from one component to another - which is already 
beyond what coherence, partial coherence, and dual-frequency 
coherence can offer.  One can compare the PDC from $X_q$ to $X_p$ 
of the alpha-band vs. the gamma-band since this relative strength is a
directed flow of communication that can vary across frequency oscillations. 

\subsection{Spectral causality}\label{sec:spectral-causality}

There are limitations with PDC as a measure of dependence. 
First, it does not indicate the phase or the physical time lag between 
oscillations. While it is useful to know the direction $X_q$ 
$\Longrightarrow$ $X_p$ at frequency $\omega$, it would be crucial 
to identify the time lag $\tau_{pq, \omega}$ via some relationship 
such as 
\begin{equation*}
X_{p, \omega}(t) = A_{pq, \omega} X_{q, \omega}(t - \tau_{pq, \omega}) + 
\epsilon_{p, \omega}(t)
\end{equation*}
for some coefficient $A_{pq, \omega}$ and 
time lag $\tau_{pq, \omega}$ that could vary depending on the 
channels and also on the frequency (or frequency bands). 
Second, PDC only captures directionality only for the {\it same} 
frequency (or frequency bands). It only models how the past 
of $\omega$-oscillation in channel $X_q$ could impact the 
future $\omega$-oscillation in channel $X_p$. It would be more 
desirable to capture between-frequencies directionality (as in 
the dual-frequency setting), for example, 
\begin{equation*}
X_{p, \omega_1}(t) = A_{pq, \omega_1} X_{a, \omega_1}(t-\tau_{pq, \omega_1}) + 
A_{pq, \omega_2} X_{a, \omega_2}(t-\tau_{pq, \omega_2}) + \epsilon_{p, \omega_1}(t)
\end{equation*}
The third limitation is that it captures only the linear associations 
between the oscillations. We
overcome the first and second limitations through the spectral 
vector autoregressive (Spectral-VAR) model. This current form 
of the model is linear and non-linear variants that are based on 
biophysical models will be reported in the future. An initial estimation
approach is introduced in \citep{SCAUSpectral-Pinto-2021}.

Consider the situation where we want to investigate the potential 
causality from channel $X_q$ to the gamma-oscillation of 
channel $X_p$. As used in previous examples, we will denote 
the delta, theta, alpha, beta, and gamma oscillations  
of $X_q$ to be, respectively, 
\begin{equation*}
X_{q, \delta}, X_{q, \theta}, X_{q, \alpha}, X_{q, \beta} \ {\mbox{and}} \ X_{q, \gamma}.
\end{equation*}
The oscillations for channel $X_p$ are denoted in a similar manner. The 
key distinction here is that these oscillations are obtained from a 
{\bf one-sided} filter
\begin{equation*}
X_{q, \delta}(t) = \sum_{j=0}^{\infty} C_{\delta, j} X_q(t-j) 
\end{equation*}
in order to properly capture these lead-lag relationships.

\begin{figure}
\centering%
\includegraphics[width=1\textwidth,height=\textheight]{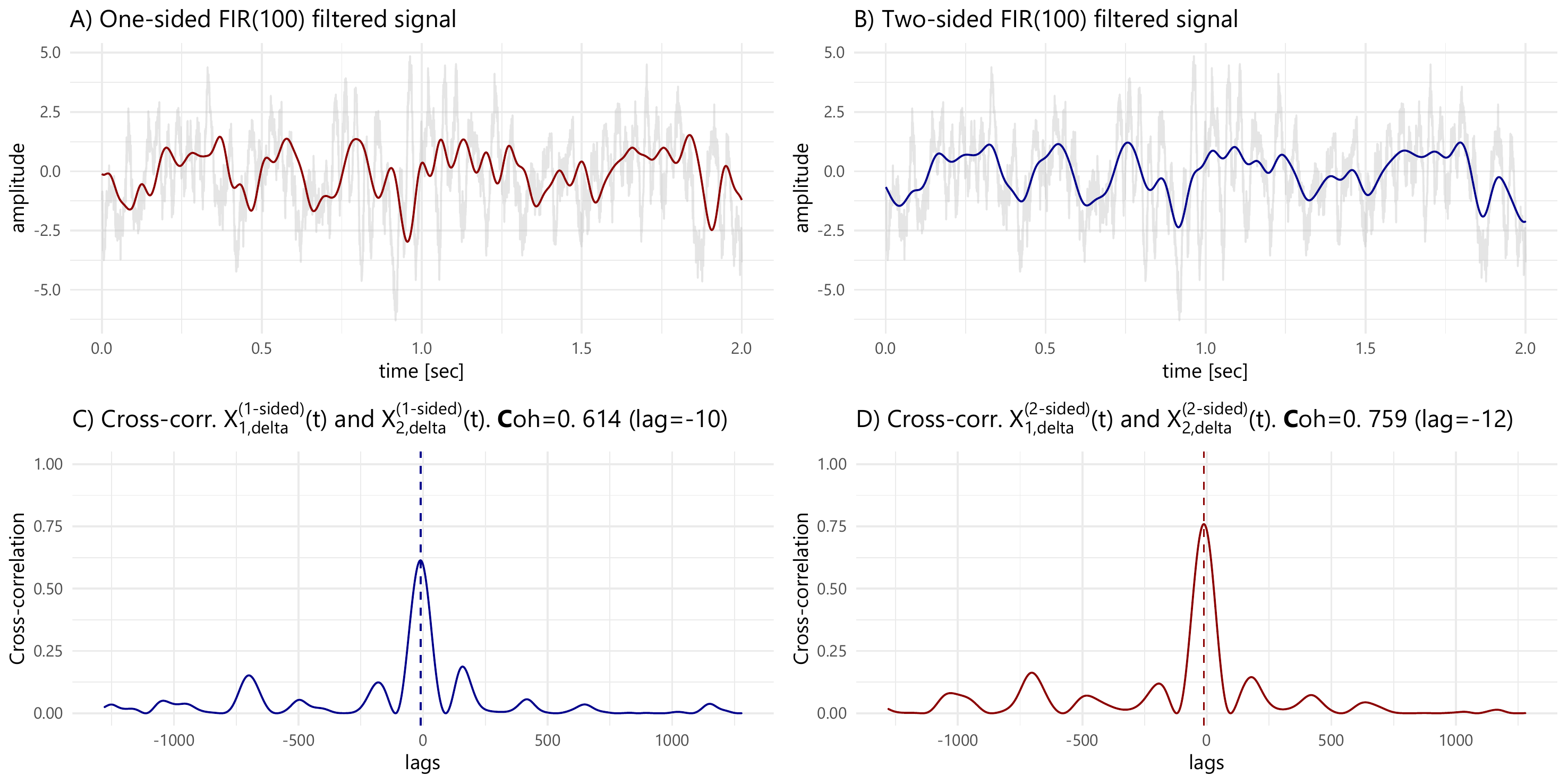}
\caption{%
Effects of an one-sided and two-sided filter in the lead-lag relationship (Example \ref{ex:filter-causal-noncausal}).
\label{fig:filter-causal-noncausal}
}
\end{figure}

\begin{example}[Two-sided filter lead-lag distortion]\label{ex:filter-causal-noncausal}
Let $Z_{\delta}(t)$, $Z_{\beta}(t)$ and $Z_{\gamma}(t)$ be three latent oscillatory signals with main frequencies at 2, 15 and 30 Hertz, respectively. Now assume that two signals are observed:
\begin{align*}
\left(\begin{matrix}X_{1}(t)\\
X_{2}(t)
\end{matrix}\right) & =\left(\begin{matrix}Z_{\delta}(t)\\
Z_{\delta}(t-10) & +Z_{\beta}(t-10) & +Z_{\gamma}(t-10)
\end{matrix}\right)+\frac{1}{2}\left(\begin{matrix}\varepsilon_{1}(t)\\
\varepsilon_{2}(t)
\end{matrix}\right)
\end{align*}
where $\varepsilon_{1}(t)$ and $\varepsilon_{2}(t)$ are uncorrelated white noise with unit variance.

Assume an one- and two-sided FIR(100) filter and denote their output filtered signals as $X_{k,delta}^{(1-sided)}$ and $X_{k,delta}^{(2-sided)}$ for a given channel $X_{k}$.
Even though that coherence between the delta-filtered $X_{1}(t)$ and $X_{2}(t)$ maintains a reasonable similar magnitude, it can be observed that the lead-lag relationship is not kept (Figure \ref{fig:filter-causal-noncausal}.C-D)

\end{example}

The spectral-VAR model of order $L$ for predicting the gamma-oscillation of 
$X_p$ is defined to be 
\begin{align*}\label{Eq:SpecVAR}
X_{p, \gamma}(t)
    & = \sum_{\ell=1}^{L}
        \left\{
          A_{(p, \gamma), (p, \delta), \ell} X_{p,\delta}(t-\ell)
          + A_{(p, \gamma), (q, \delta)} X_{q,\delta}(t-\ell)
        \right\} \\
{} & + \ldots \\
{} & + \sum_{\ell=1}^{L}
       \left\{
          A_{(p, \gamma), (p, \gamma), \ell} X_{p,\gamma}(t-\ell)
          + A_{(p, \gamma), (q, \gamma)} X_{q,\gamma}(t-\ell)
          + \epsilon_{p, \gamma}(t)
      \right\}. 
\end{align*}
Under this set-up, we say that there is a Granger-causality relation from 
the past alpha-oscillatory activity in $X_q$ to the future gamma-oscillation in
$X_p$ if there exists a time lag $\ell*$ such that $A_{(p, \gamma), (q, \alpha), \ell*} 
\ne 0$.


As a final remark, the usual VAR model does not address the need for 
frequency-specific lead-lag (or Granger-causality) relationships. Suppose 
that, from the usual VAR model, we conclude that $X_q \Longrightarrow$
$X_p$. This information is ``coarse" in the sense that it lacks the information 
on the specific frequency band (or bands) in $X_q$ that Granger-cause 
the specific band(s) in the $X_p$ as well as the precise channel-specific and
frequency bands-specific time-lag between these oscillations.

%
%
%
%
%
%
%

\section{High-Dimensional Signals}

Most brain signals are high-dimensional in space and time. For example, fMRI data are typically recorded over $10^5$ brain voxels and across hundreds of time points sampled at a speed of one-complete image 
every 2 seconds (the sampling rate is $0.50$ Hertz). Therefore, fMRI data are highly dense in space
(though they have poor temporal resolution). In contrast, EEG data are more sparse in space as the 
number of channels can vary from about 20 to 256, depending on the adopted recording system. 
However, the range of the number of temporal observations can be in the millions (with sampling 
frequency typically ranging from $100 -1000$ Hertz). In this section, we will present two approaches 
to dealing with high-dimensional signals: (a.) by creating low-dimensional signal summaries such 
as principal component analysis; and (b.) by including a penalization component in estimation. 

\subsection{Spectral principal components analysis}\label{Sec:SpecPCA}

It is common for components of a $P$-variate signal $X$ to display 
some multi-collinearity, especially between biomedical signals recorded as spatially close locations. It is therefore natural to obtain a low-dimensional summary that captures the main characteristics of these signals. One way is through the classical principal component analysis (PCA) -- or linear auto-encoder/decoder in machine learning jargon -- which is described as follows.

The auto-encoder algorithm described in \citep{LearningInternalRepresentations-Rumelhart-1988} is a general approach to learning compressed (low-dimensional) representations of the input data, which in this particular scenario are high-dimensional brain signals. The algorithm consists of two parts, namely, the (a.) encoder and (b.) decoder.
The encoder function $F: {\bf X} \rightarrow {\bf Y} $ is a mapping from the original high-dimensional space ${\bf X}$ to lower dimension space ${\bf Y}$. Due to its dimensionality reduction purpose, $F$ is also called compression step. The decoder or reconstruction function, defined as $D: {\bf F}\rightarrow {\bf X}$ is a mapping from the encoded (low-dimensional) space to the original high-dimensional space. 
Consider a time series $\{{\bf X}(t),
t=1, \ldots, T\}$ generated by some process $p({\bf X})$. The optimal mapping encoder $F$ and decoder $D$ minimize the expected reconstruction error defined as
\begin{equation}\label{eq:expected_loss_function}
L\left(F, D\right) = \ex_{p({\bf X})}
          \left\Vert
            {\bf X} - D(F({\bf X}))
          \right\Vert^2_F
\end{equation}
where  $\left\Vert \cdot \right\Vert_F$ is the Frobenius norm, which is defined as $\left\Vert| E \right\Vert|_F = \sqrt{\text{Trace}(EE')}$.

Here, we will consider only the special cases where both the encoder and decoder are linear transformations of the original signal, which can be either instantaneous mixing or filtering. For these types of functions, the solution is closely related to principal component analysis (PCA) defined under the aforementioned Frobenius norm based on the squared error of reconstruction.

Let us consider the first family of encoder-decoders: instantaneous mixture processes of the observed signal $\{ {\bf X}(t), \ t=1, \ldots, T\}$. Under this model, the compressed signal is obtained as ${\bf Y}(t) = {\bf A}' {\bf X}(t)$. Denote the dimension of ${\bf Y}(t)$ to be $Q$; the dimension of 
${\bf X}(t)$ to be $P$ and $Q < P$. Similarly, the signal is reconstructed by the decoder $\widehat{\bf X}(t) = {\bf B} {\bf Y}(t)$.
For identifiability purposes, ${A}' {A} = I_{Q}$, and $\cov {\bf Y}(t)$ is diagonal so that the components of the compressed signal X are uncorrelated. The optimal low-dimensional representation maximizes the best reconstruction accuracy (or minimizes the squared error loss) via the following steps:
\begin{itemize}
  \item[Step 1.]
  Obtain the eigenvalues-eigenvectors of the covariance matrix of ${\bf X}$: $\Sigma^{\bf X}(0)$. Denote e-value and e-vector pair  as  $\{ (\lambda_{p}, {\bf v}_{p} )\}_{p=1}^{P}$ where $\lambda_{1} > \ldots, > \lambda_{P}$ and $\| {\bf v}_{p} \| = 1$ for all $p$. When $\Sigma^{\bf X}(0)$ is not known, we obtain an estimator from the observed signal $\{ {\bf X}(t), \ t=1, \ldots, T\}$.
  \item[Step 2.]
  The solution for the optimal encoder-decoder is derived as 
  \begin{equation*}
  \widehat{{A}}'  = \widehat{{B}} = [{\bf v}_{1}, \ldots, {\bf v}_{Q}]
  \ \ \ {\mbox{and}} \ \ \  
  \widehat{\bf Y}(t) = \widehat{{A}}{\bf X}(t) \ .
  \end{equation*}
\end{itemize}
Under the squared reconstruction error as the loss function, the solution is identical to applying PCA on the input signals using the covariance matrix
at lag zero. Indeed, the solution accounts for most of the variation of the time series (or gives the minimum squared reconstruction error), among all instantaneous linear projections with the same dimension. 

When the goal is to obtain summaries from time series data, it is important 
for the encoded (compressed) and decoded (expanded) components to
capture the entire temporal dynamics (lead-lag structure) of the signal. The previous approach is a contemporaneous mixture and thus could miss 
important dynamics in the data. The second category of linear encoder-decoders relies on the idea of applying linear filters on $\{ {\bf X}(t) \}$ instead of applying a merely instantaneous mixture. In contrast to the contemporaneous mixture, 
this encoder is more flexible and its lower-dimensional representation is written as
\begin{equation}\label{eq:compression}
{\bf Y}(t) = \sum_{\ell=\infty}^{\infty} A'(\ell) {\bf X}(t-\ell)
\end{equation}
where $A'(\ell) \in \mathbb{C}^{Q \times P}$ with $Q < P$. The components 
of the summarized signal, ${Y}_p(t)$ and ${Y}_q(t)$, have zero coherency. That is, the 
components $Y_p(t)$ and $Y_q(s)$ are uncorrelated at all time points $t$ and $s$, 
and hence also for all lags. The reconstruction (decoder) function has the following form
\begin{equation}\label{eq:reconstruction}
\widehat{\bf X}(t) = \sum_{\ell=-\infty}^{\infty} B(\ell) {\bf Y}(t-\ell)
\end{equation}
where $B(\ell) \in \mathbb{C}^{P \times Q}$ is the transformation 
coefficient matrix.

The optimal values of $A(\ell)$ and $B(\ell)$ are chosen to minimize the reconstruction error defined in Equation \ref{eq:expected_loss_function}. The solution is obtained via principal components analysis of the spectral matrix ${\bf f}(\omega)$ of the process ${\bf X}$ -- rather than the lag-0 covariance 
matrix. Denote the eigenvalues of the spectral matrix at frequency $\omega$ to be $\{\lambda_1(\omega) > \lambda_2(\omega), \dots, \lambda_P(\omega)\}$, and the corresponding eigenvectors to be $\{{\bf v}_1(\omega), {\bf v}_2(\omega), \dots, {\bf v}_n(\omega)\}$.
Then, the solution is 
\begin{eqnarray}
{A}(\ell) &=& \int_{-1/2}^{1/2} {A}(\omega) \exp(i2\pi \ell \omega) d \omega \\
{B}(\ell) &=& \int_{-1/2}^{1/2} {B}(\omega) \exp(i2\pi \ell \omega) d \omega
\end{eqnarray}
where ${B}(\omega) = {A}^*(\omega)$ and ${A}(\omega) = [{\bf v}_1(\omega), \dots, {\bf v}_Q(\omega)]$.

This dimension reduction procedure was originally described in \citep{FreqApproachTechniques-Brillinger-1964}, and in this paper, we shall refer to this as the ``Spectral-PCA" method (SPCA). This method was extended to 
various nonstationary settings, including \cite{SLEXMultivariate-Ombao-2005}, where the stochastic representation of a 
multichannel signal was selected from a library of orthogonal localized Fourier waveforms (SLEX).
In \cite{TimedependentFreqDomain-Ombao-2006}, the time-varying spectral PCA was developed under the context of the 
Priestley-Dahlhaus model, which was further refined in \cite{EvolutionaryFactor-Motta-2012} for the experimental setting where there are replicated multivariate nonstationary signals. In practical data analysis, the interest is on the magnitude of the 
components of the eigenvector (or eigenvectors) with the largest eigenvalues because they represent the 
loading or weights given by the components of the observed signal. We motivate this in the example below. 

\begin{figure}
\centering%
\includegraphics[width=1\textwidth,height=\textheight]{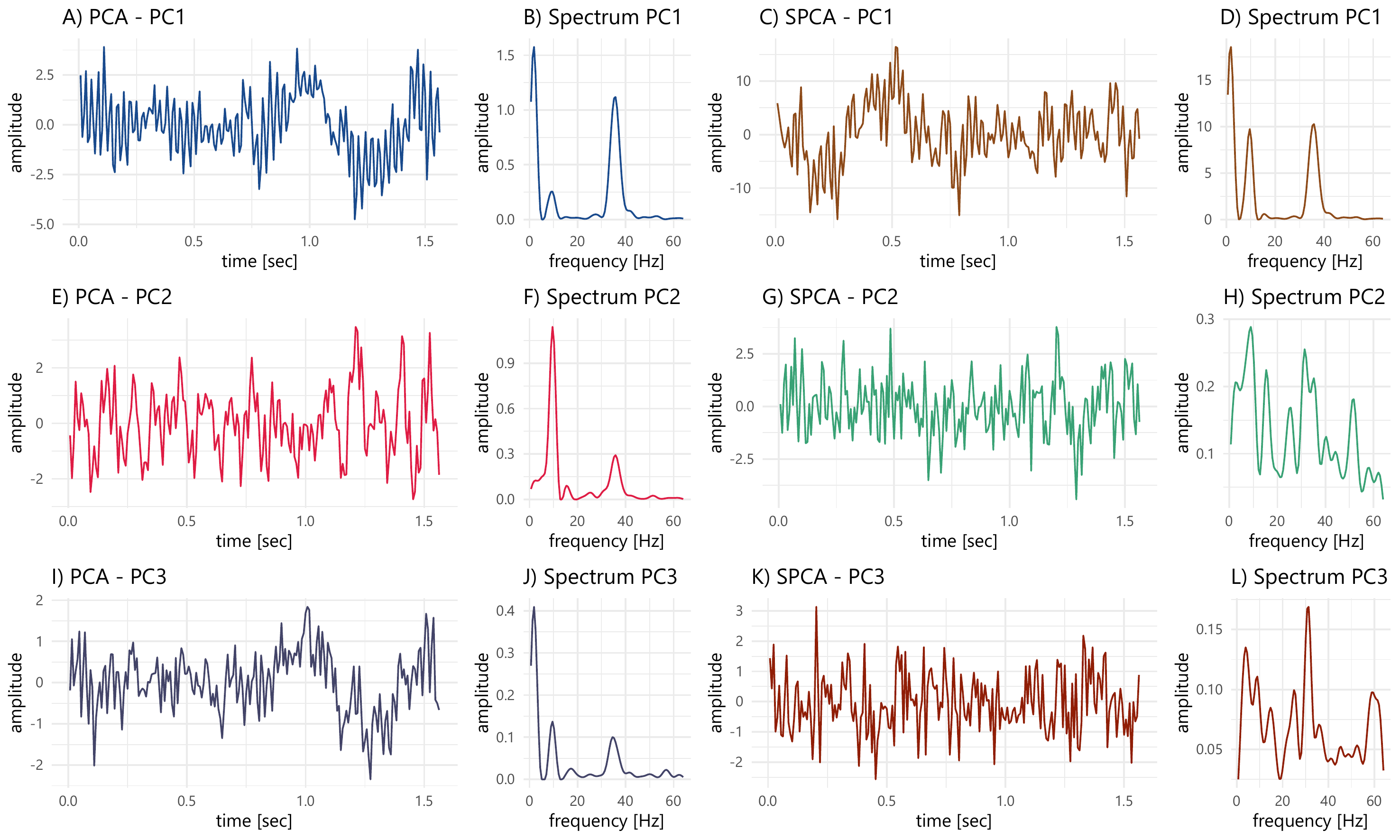}
\caption{%
Summary signals ${\bf Y}(t)$ using PCA and SPCA (Example \ref{ex:pca}). The components in the time and frequency domain are shown for both encoding algorithms. %
}
\label{fig:SPCA-PCA}
\end{figure}

\begin{figure}
\centering%
\includegraphics[width=.75\textwidth,height=\textheight]{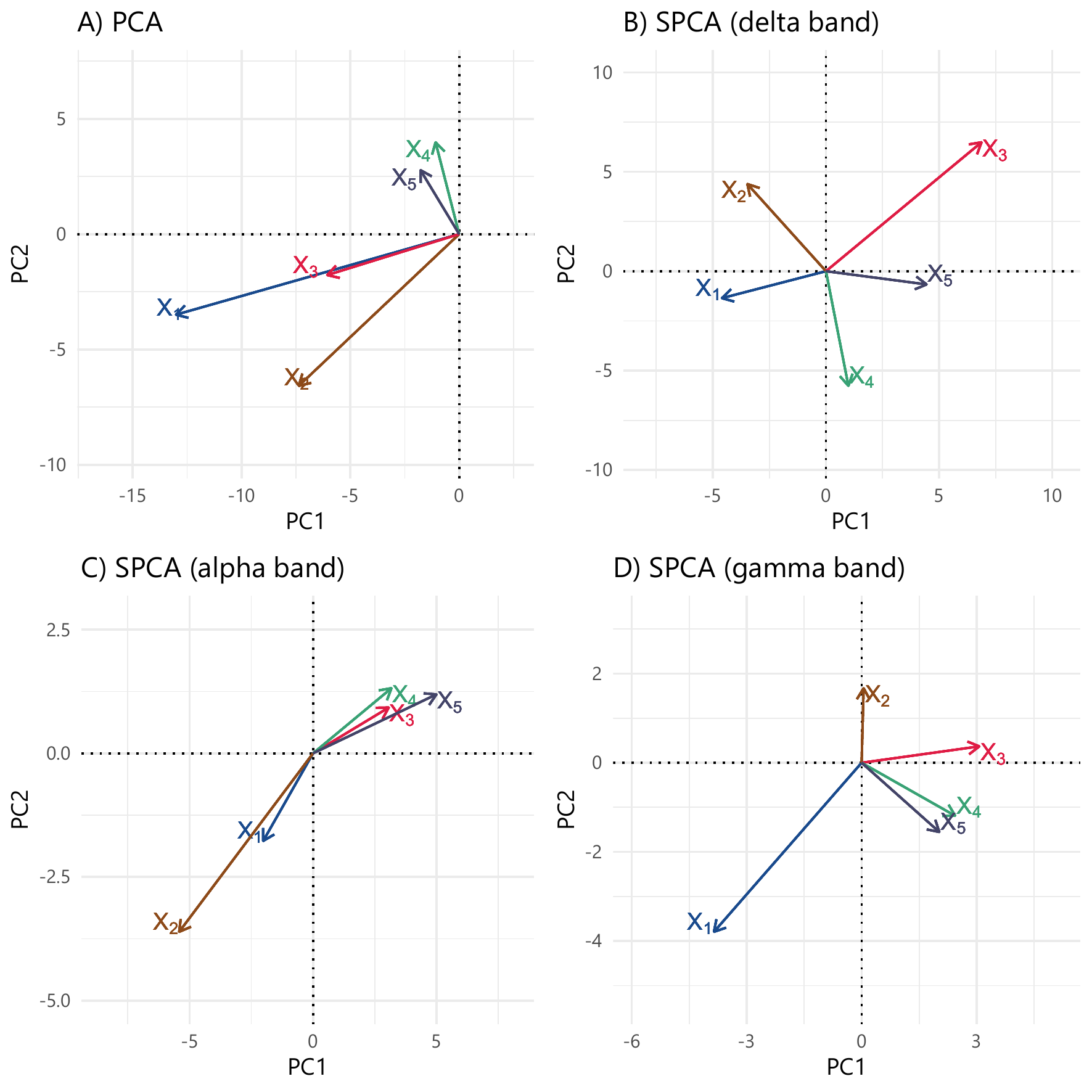}
\caption{%
Two-dimensional loadings of Example \ref{ex:pca}:
PCA loadings ($\widehat{A}$) and mean SPCA loadings ($A(\omega)$) in the frequency bands: delta (0-4 Hertz), alpha (8-12 Hertz), and gamma (30-45 Hertz).
}
\label{fig:loadings-SPCA-PCA}
\end{figure}

\begin{example}[Example on spectral PCA]\label{ex:pca}
Suppose that we 
have latent sources that are oscillations at the delta, alpha, and gamma bands, denoted 
by $Z_{\delta}$, $Z_{\alpha}$ and $Z_{\gamma}$, respectively. The observed signal
is a mixture ${\bf X}(t)$ of the latent sources 
\begin{eqnarray}
\left( 
\begin{matrix}
X_1(t) \\
X_2(t) \\
X_3(t) \\
X_4(t) \\
X_5(t)
\end{matrix}
\right) & = & 
\left(
\begin{matrix}
1 & 0 & 1 \\
0 & 0 & 1 \\
1 & 0 & 1 \\
0 & 1 & 0 \\
1 & 1 & 0 
\end{matrix}
\right)
\left(
\begin{matrix}
Z_{\delta}(t) \\
Z_{\alpha}(t) \\
Z_{\gamma}(t)
\end{matrix}
\right) \ + \ 
\left( 
\begin{matrix}
\epsilon_1(t) \\
\epsilon_2(t) \\
\epsilon_3(t) \\
\epsilon_4(t) \\
\epsilon_5(t)
\end{matrix}
\right)
\end{eqnarray}

\end{example}

The first two compressed components of the instantaneous mixture (PCA) and the spectral PCA (SPCA) are shown in Figure \ref{fig:SPCA-PCA}. Note that all three first PCA principal components capture a mixture of the latent sources, and therefore, contain the delta, alpha, and gamma oscillations in different proportions. This oscillation mixture is made evident from its loadings (Figure~\ref{fig:loadings-SPCA-PCA}.A) where $X_1,\ldots,X_6$. Nevertheless, the first component emphasizes the delta and gamma sources, whereas the 
second component highlights the alpha and gamma bands. 

On the other hand, the first SPCA component captures all the main spectral information of the latent oscillations (Figure \ref{fig:SPCA-PCA}.D). Due to the properties of SPCA, mean loadings can be evaluated as a function of the frequency, and the contribution of the components can be evaluated for each frequency band (through the loadings in Figure \ref{fig:loadings-SPCA-PCA}.B-D). For instance, consider the encoding processes on the alpha band: the main descriptive components are $X_4$ and $X_5$, but SPCA compensates the effects of the other signals in such a way that the sum of $X_1$ and $X_2$ (delta and gamma) will mitigate the effect of $X_3$ (also delta and gamma).


\subsection{Brain Connectivity Analysis through Low-Dimensional Embedding}

As noted, neuronal populations behave in a coordinated manner both during resting-state 
and while executing tasks such as learning and memory retention. One of the 
major challenges to modeling connectivity in brain signals %
is the high dimensionality. In the case of fMRI data with $V=10^5$ voxels, one would 
need to compute connectivity from in the order of $10^{10}$ pairs of voxels (or 
$V(V-1)/2$ pairs). To alleviate this problem, one approach in fMRI is to parcellate the 
entire brain volume into distinct regions of interest 
(ROIs) and hence connectivity is computed between ROIs rather than between voxels. 
This approach effectively reduces the dimensionality since connectivity is computed 
between broad regions rather than at the voxel level. This is also justified by the fact that 
neighboring voxels tend to behave similarly and thus it would be redundant to calculate 
connectivity between all pairs of voxels. Motivated by the ROI-based approach in fMRI, 
one procedure to study connectivity in EEG signals is to first create groups of channels 
using some anatomical information.  Depending on the parcellation of the brain cortex, a 
cortical region could correspond to 15-25 EEG channels \citep{wu2014resting}. Within 
each group, we compute the signal summaries using spectral principal components 
analysis. In the second step, we model connectivity between groups of channels by 
computing dependence between the summaries. More precisely, suppose that the 
$P$-dimensional time series ${\bf X}$ is segmented into $R$ groups denoted by 
${\bf X}_r$, $r=1, \ldots, R$. In each group ${\bf X}_r$, summaries are computed,
which we denote by ${\bf Y}_r$. Thus, connectivity between groups $r$ and $r'$ will be 
derived from the summaries ${\bf Y}_r$ and ${\bf Y}_{r'}$

There are many possible methods for computing summaries.
When biomedical signals can be modeled as functional data,
principal components analysis (PCA) extensions have been formulated, as it is shown in \citep{PrincipalComponents-Kokoszka-2019}.
Under sampled time series, a na\"ive solution is to 
compute the average across all channels within a group.
In fact, connectivity analyses of functional magnetic resonance imaging (fMRI) are 
usually conducted by taking the time series averaged across voxels in pre-defined ROIs (\citep{QuantifyingTemporalOscs-Fiecas-2013, EstimatingPopulationLocal-Gott-2015}).
However, simple averaging is problematic, especially when some of the signals are out of 
phase often observed in EEGs due to averaging can lead to signal cancellation.
Sato et al. \citep{AnalyzingConnectivityRegions-Sato-2010} already pointed the pitfalls and suggests a data-driven approach 
via conventional PCA, which essentially provides an instantaneous (or contemporaneous) mixing of time series.
Other approaches for modeling brain connectivity from high-dimensional brain imaging data include
Dynamic Connectivity Regression (DCR) \citep{DynamicConnectivityReg-Cribben-2012}, Dynamic Conditional Correlation (DCC) \citep{EvaluatingDynamicBivariate-Lindquist-2014}, group independent component analysis (ICA) \citep{ChronnectomeTimevaryingConnectivity-Calhoun-2014, MultisubjectIndependentComponent-Calhoun-2012}, and
sparse vector autoregressive (VAR) modeling \citep{SparseVectorAR-Davis-2016}.
Here, we propose 
extract summaries ${\bf Y}_r$ from each group of channels via the spectral PCA method in 
Equations~\ref{eq:compression} and \ref{eq:reconstruction} above.

\subsection{Regularized vector autoregressive models} \label{sec:regularized-VAR}

Recall that connectivity measures such as coherence, partial coherence and 
partial directed coherence are based on the frequency domain but they can be 
motivated under the context of parametric models. In fact, PDC was developed 
within the framework of a vector autoregressive (VAR) model. Here, we discuss 
the challenges of fitting a VAR model when the number of channels $P$ and the 
VAR order $D$ are large. In this setting there will be $P^2D$ number of VAR 
parameters that have to be estimated.  The goal here is to introduce some of the 
regularization procedures. 

The classic method for estimating the VAR parameters is via the least squares 
estimator (or conditional likelihood for Gaussian signals), which is generally 
unbiased.
However, the least-squares estimators (LSE) are problematic because of the 
high computational demand and that it does not possess specificity for 
coefficients whose true values are zero. In many applications, brain networks are 
high-dimensional structures that are assumed to be sparse but interconnected. 
To address the problem of high dimensional parameter space, a common estimation approach 
is by penalized regression (\cite{tibshirani1996regression}, \cite{fu1998penalized}, \cite{zhao2009composite}, \cite{hesterberg2008least}), and a specific method is the LASSO (least absolute shrinkage and selection operator). Compared to the LSE approach, the LASSO 
has a lower computational cost (\cite{mairal2012complexity}) and has higher specificity of zero-coefficients. However, the main limitation of the LASSO (and, in general, most regularization methods) 
is the bias of the non-zero coefficients' estimators. Thus, it could lead to misleading results when investigating the true strength of brain connectivity. By leveraging the strengths of each of the
 LSE and the LASSO, Hu et al. \citep{LASSL} proposes a (hybrid) two-step estimation procedure: 
the LASSLE method. This approach suggest a two-phase estimation process.
In the first stage, the LASSO is applied to identify coefficients whose 
estimates are set to 0. In the second stage, the coefficients that survived the thresholding from
the first stage are re-estimated via LSE. LASSLE was shown to have inherent low-bias for non-zero estimates, high specificity for zero-estimates, and significantly lower mean squared error (MSE) in the simulation study.

An example of the impact of using regularized models in connectivity estimation was visually shown in Figure \ref{fig:comparison-coh-pcoh-pdcoh}.E-F (from Example \ref{ex:comparison-coh-pcoh-pdcoh}).

\section{Modeling dependence in non-stationary brain signals}

Contrary to common intuition, the brain operates actively even during resting periods, and therefore this is also reflected in the brain signal dynamics \citep{CmptaticDynamic-Menon-2019}.
Brain signals from
various modalities, such as magnetoencephalograms, electroencephalograms, local field 
potentials, near-infrared optical signals, and functional magnetic resonance imaging, all 
show statistical properties that evolve over time during rest and various task-related 
settings (memory, somatosensory, audio-visual). Such time-evolving characteristics are also observed across species: 
laboratory rats, macaque monkeys, and humans.  These changes are seen in the variance, 
auto-correlation, cross-correlation, coherence, partial coherence, partial directed coherence, or
graph-network properties \citep{TimedependenceGraphTheory-Chiang-2016}.
It is worth noting that in some experimental settings, changes in the 
cross-channel dependence may be more pronounced than changes within a channel (e.g., 
auto-spectrum and variance). This phenomenon was observed in \citep{fiecas2016}, where 
changes in cross-coherence were observed in a macaque monkey local field potentials and 
correlated with the learning task. 

In this section, an overview of the different approaches to analyzing non-stationary 
signals will be discussed. The first class of approaches gives stochastic representations 
using the Fourier waveforms or some multi-scale orthonormal basis such as
wavelets, wavelet packets, and the smoothed localized complex exponentials (SLEX). 
In the second class of methods, the signals are segmented into quasi-stationary 
blocks, and the time-varying spectral properties such as the auto-spectra, coherence, and partial coherence are computed within each time block. This class of approaches 
produces a specific tiling of the time-frequency plane. The third class of approaches 
assumes that the dynamic brain activity fluctuates or switches between a finite number 
of "states". Each of these states, defined by a vector autoregressive model or a stochastic 
block model, gives a unique characterization of the brain functional network. This 
class of models depicts brain responses to a stimulus (or background activity during 
resting state) as switching between these states. 

\subsection{Stochastic representations}

%
%
%
%

\noindent {\bf Priestley-Dahlhaus model.} The major theme in this paper is the characterization of brain signals as
mixtures of randomly oscillating waveforms. So far, the emphasis has been 
on stochastic representations in terms of the Fourier waveforms. As already 
noted, the Cram\'er representation of a $P$-dimensional stationary time series 
${\bf X}_t$  
\[
{\bf X}(t) = \int_{-0.50}^{0.50} {\bf A}(\omega) \exp(i 2 \pi \omega t) d{\bf Z}(\omega)
\]
where ${\bf A}(\omega)$ is a $P \times P$ the transfer function matrix and 
$\{ d{\bf Z}(\omega) \}$ is a random increment process with $\ex d{\bf Z}(\omega) = 0$ 
and $\cov[ d{\bf Z}(\omega), d{\bf Z}(\omega')] = {\bf I} \delta(\omega - \omega)$ where 
$\delta$ is the Dirac-delta function. To illustrate the role of the transfer function, define 
$A_{pq}(\omega)$ to be the $(p,q)$ element of ${\bf A}(\omega)$; \ and \ $dZ_p(\omega)$ 
to be the $p$-th element of the random vector $d{\bf Z}(\omega)$. The spectral 
matrix is ${\bf f}(\omega) = {\bf A}(\omega) {\bf A}*(\omega)$. 

The time series at channels $p$ and $q$ can be written as 
\begin{align*}
X_p(t) & = \sum_{r=1}^{P} \int_{-0.50}^{0.50} A_{pr}(\omega) \exp(i 2 \pi \omega t) dZ_r(\omega) \\
X_q(t) & = \sum_{r=1}^{P} \int_{-0.50}^{0.50} A_{qr}(\omega) \exp(i 2 \pi \omega t) dZ_r(\omega).
\end{align*}
They are coherent at frequency $\omega$ if  there exists some $r$ where 
$A_{pr}(\omega) \ne 0$ and $A_{qr}(\omega) \ne 0$. In this case, ${\bf A}_p(\omega) 
{\bf A}_q^*(\omega) \ne 0$ \ (where ${\bf A}_r(\omega)$ is the $r$-th row of the 
transfer function matrix ${\bf A}(\omega)$.

Under stationarity, spectral cross-dependence between signals (e.g., coherence) is constant over time. 
For brain signals, dependence between channels varies across time. A time-dependent 
generalization of the Cram\'er representation is the Priestly-Dahlhaus model (\cite{Priestley:1965} and 
\cite{Dahlhaus:1997}). For a time series of length $T$,  the Dahlhaus-Priestley model uses a 
time-dependent transfer function where 
\begin{equation*}
{\bf X}(t) = \int_{-0.50}^{0.50} {\bf A}(\frac{t}{T},\omega) \exp(i 2 \pi \omega t) d{\bf Z}(\omega).
\end{equation*}
where the transfer function ${\bf A}(\frac{t}{T},\omega)$ is defined on rescaled time $\frac{t}{T}$ 
\ $\in (0,1)$ and frequency $\omega \in (-0.50, 0.50)$. 
Under this model, the mixture changes over time because the random coefficient vector 
$ {\bf A}(\frac{t}{T},\omega) \ d{\bf Z}(\omega)$ also changes with time. The time-varying 
spectrum is ${\bf f}(u, \omega) =  {\bf A}(u,\omega) {\bf A}*(u,\omega)$ and, consequently, the 
time-varying coherence between channels $p$ and $q$ is 
\begin{equation*}
\rho_{pq}(u, \omega)  = \frac{\vert f_{pq}(u, \omega) \vert^2}{f_{pp}(u, \omega) f_{qq}(u, \omega)}.
\end{equation*}
To estimate the time-varying spectrum at a particular rescaled time $u \in (0,1)$, local 
observations around this time point are used to form local periodograms, which are then smoothed 
over frequency. Alternatively, one can fit a localized semi-parametric estimator as shown in 
\cite{fiecas2011generalized}. Note that a change-point detection extension was introduced in \citep{ChangePointDetection-Gorecki-2018}.

\noindent {\bf Locally stationary wavelet process.} An alternative to the Priestley-Dahlhaus model is the locally stationary wavelet process (LSW)
proposed in \cite{Nason:2000}, which uses the wavelets as building blocks. Under the LSW 
model, a scale- and time-dependent wavelet spectrum is defined. The original LSW model 
is univariate and has been extended to the multivariate setting in \cite{Park2014} as follows. 
A time series ${\bf X}(t), t=1, 2, \ldots, T$ is a multivariate LSW process if it has the 
representation
\begin{equation*}
{\bf X}(t) = \sum_j \sum_k  {\bf V}_j(\frac{k}{T}) \psi_{j, t-k} {\bf Z}_{jk}
\end{equation*}
where $\{ \psi_{j, t-k} \}$ is a set of discrete non-decimated wavelets; 
${\bf V}_j(\frac{k}{T})$ is the transfer function matrix which is lower-triangular; and 
$ {\bf Z}_{jk}$ are uncorrelated zero mean with covariance matrix equals to the 
identity matrix. Note that $ {\bf Z}_{jk}$ in the multivariate LSW is the analog of 
$d{\bf Z}(\omega)$ in the Fourier-based stochastic representation. The classification 
procedure for the LSW model was developed in \cite{LSW-Class} for univariate time 
series and in \cite{Park:2018} for multivariate time series. Given training data (signals 
with known group membership), these methods extract the wavelet scale-shift features 
or projections that separate the different classes of signals. These features are then 
used to classify a test signal. These wavelet-based classification methods are 
demonstrated to be asymptotically consistent, i.e., the probability of misclassification 
decreases to zero as the length of the test signal increases. 

\begin{figure}[t]
\centering%
\includegraphics[width=0.5\textwidth]{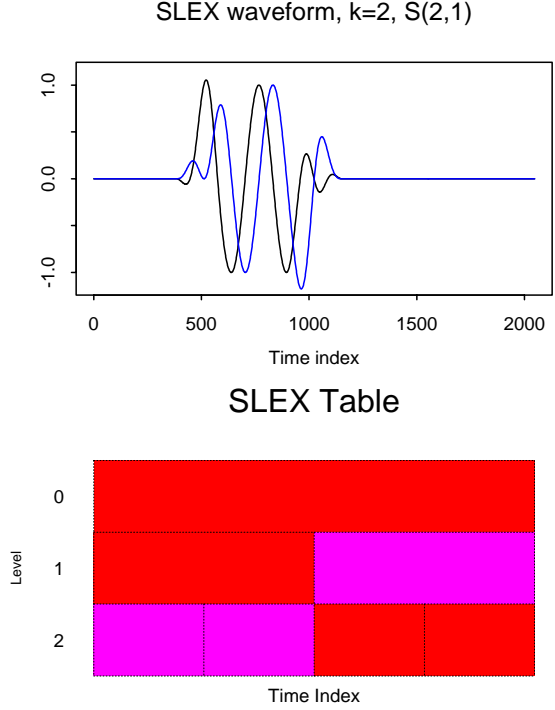}
\caption{Top: SLEX waveform (real and imaginary parts) with local support 
at dyadic block $B(2,1)$. Bottom: SLEX library with level $J=2$. The SLEX 
library consists of a finite number of bases and each basis gives a dyadic 
segmentation of the time axis. The set of magenta-colored blocks represent 
one particular basis with time blocks $B(2,0) B(2,1), B(1,1)$ which corresponds 
to the segmentation on rescaled time: $\bigl[0, \frac{1}{4}\bigr) \cup 
bigl[\frac{1}{4}, \frac{1}{2}\bigr) \cup \bigl[\frac{1}{2}, 1\bigr]$. 
\label{fig:SLEX}
}
\end{figure}

\begin{figure}[t]
\centering%
\includegraphics[width=0.85\textwidth]{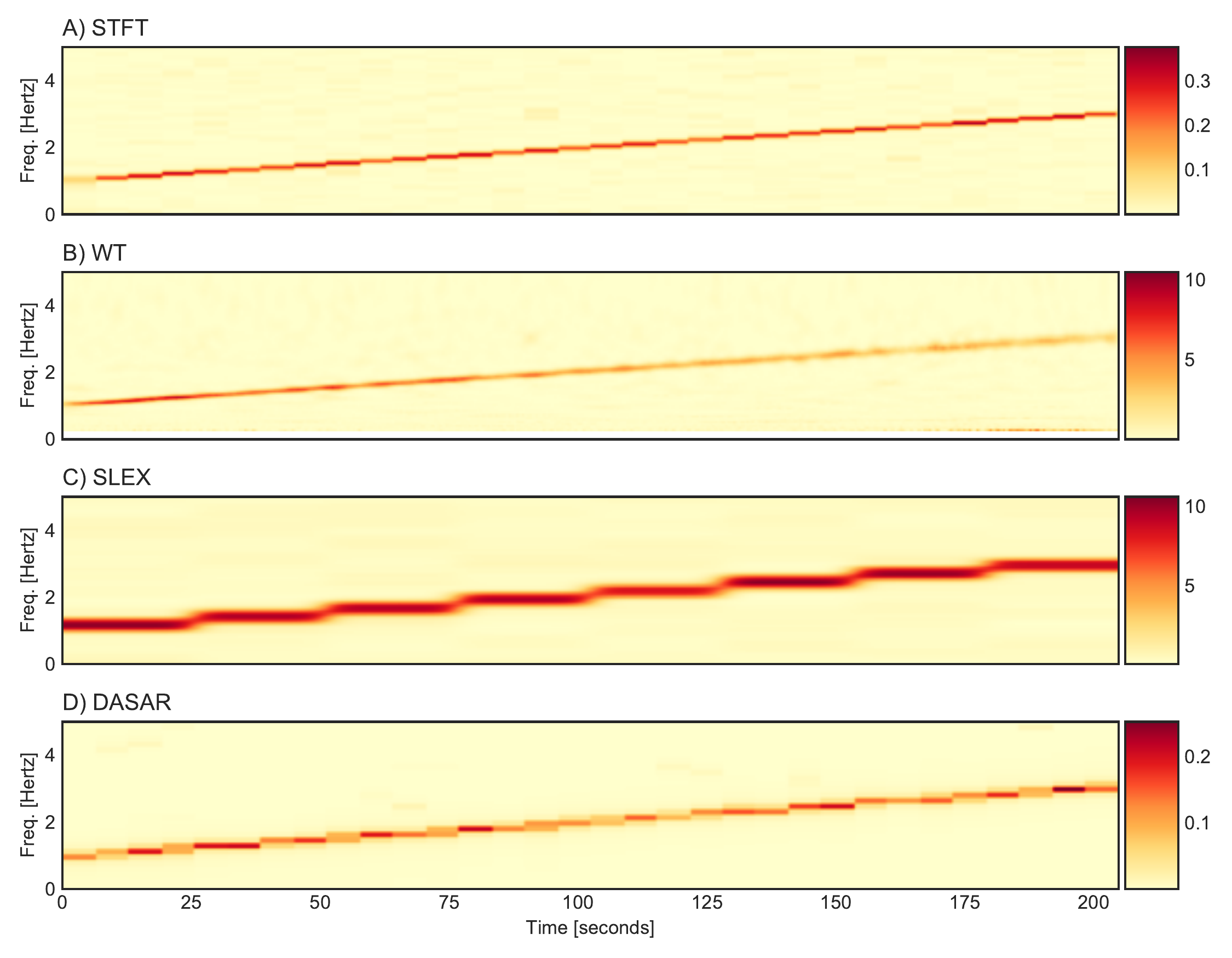}
\caption{
Spectral estimation of a linearly frequency evolving signal using four representation methods: short-time Fourier transform, wavelet transform, SLEX and DASAR.
\label{fig:slex-comparison-chirp}
}
\end{figure}

\begin{figure}
\centering%
\includegraphics[width=0.85\textwidth]{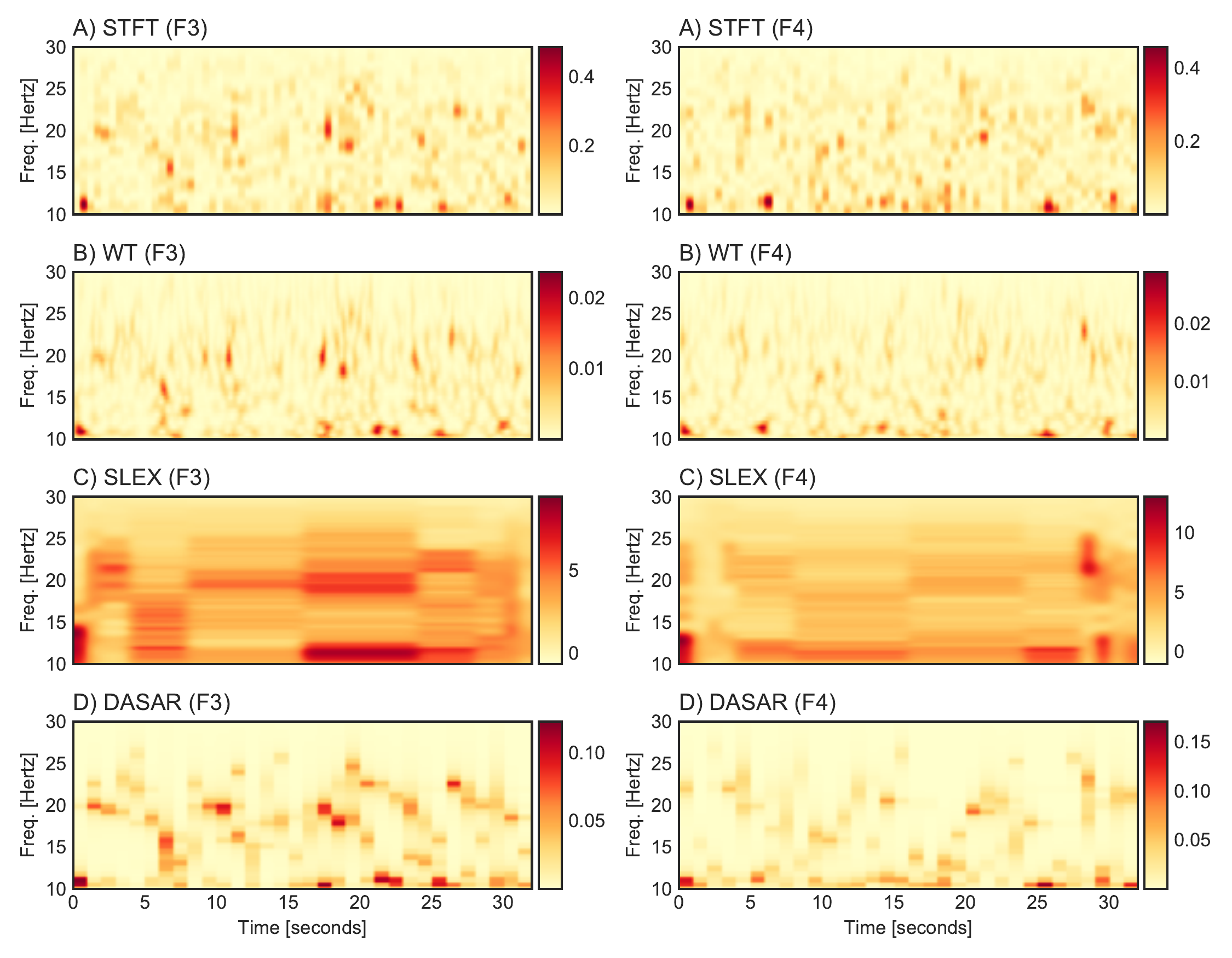}
\caption{
Time-frequence representation of 32-second EEG recordings from channels on the frontal left (F3) and right (F4) hemisphere using short-time Fourier transform, wavelet transform, SLEX and DASAR.
\label{fig:slex-comparison-eeg}
}
\end{figure}

\noindent {\bf The SLEX model.} There are other time-localized bases that are well suited for representing 
non-stationary time series. In particular, the SLEX (smooth localized complex 
exponentials) waveforms are ideal for a comprehensive analysis that is a 
time-dependent generalization of Fourier-based methods. The SLEX waveforms 
are time-localized versions of the Fourier waveforms and they are constructed 
by applying a projection operator (\cite{Ombao:2005}). The starting point 
is to build the SLEX library which is a collection of many bases. These bases 
consist of functions defined on a dyadic support on rescaled time. 
In Figure~\ref{fig:SLEX}, see the plot of a specific SLEX waveform with support 
on the second quarter of the rescaled time and with approximately 2 complete 
cycles within that block. In this example, a tree is grown to level $J=2$ 
(i.e., the finest blocks have support with width $\frac{1}{2^2}$). There are 5 bases 
in this library and each basis corresponds to a specific dyadic segmentation of 
$[0,1]$. One particular basis is represented by the magenta-colored blocks which 
are denoted by $B(2,0), B(2,1), B(1,1)$. This basis also corresponds to the 
specific segmentation $[0, \frac{1}{4}) \cup [\frac{1}{4}, \frac{1}{2}) \cup [\frac{1}{2}, 1]$.
Define $\mathcal{B}$ to be a set of indices $(j,b)$ that make up 
one particular basis. In this particular example, $\mathcal{B} = \{(2,0), (2,1), (1,1) \}$. 
Denote an SLEX waveform with support on block $B(j,b)$ oscillating at 
frequency $\omega$ is denoted as $\Psi_{j,b,\omega}(t)$. Then the SLEX model 
corresponding to this particular basis $\mathcal{B}$ is 
\begin{equation}
{\bf X}(t)
  = \sum_{(j,b) \in \mathcal{B}}
         \int_{-0.50}^{0.50}
         \Theta_{j,b}(\omega)%
         \psi_{j,b,\omega}(t)
         d{\bf Z}_{j,b}(\omega)
\end{equation}
where $\Theta_{j,b}(\omega)$ is the SLEX transfer function defined on time 
block $B(j,b)$ and $\{d{\bf Z}_{j,b}(\omega)\}$ is an increment random 
process that is orthonormal across time blocks $(j,b)$ and frequencies $\omega$. 
For a given SLEX basis, the time-dependent SLEX spectral matrix is
\[
{\bf f}_s(t/T, \omega) = \Theta_{j,b}(\omega) \Theta_{j,b}^*(\omega)
\]
for $t$ in the time block $B(j,b)$. Thus, the SLEX auto-spectra and the 
SLEX-coherence are defined in a similar way as the classical Fourier approach. 

The main advantage of the SLEX methods for analyzing non-stationary 
signals is the flexibility offered by the library of bases. Depending on the 
particular problem of interest, a "best" basis is selected from this collection 
of bases.
Each SLEX basis gives 
rise to a unique segmentation and an SLEX model. In \cite{Ombao:2001automatic}, 
a penalized Kullback-Leibler criterion was proposed for selecting the SLEX 
basis. This criterion jointly minimizes: (i.) the error of discrepancy between the 
empirical time-varying SLEX spectrum and the candidate true SLEX 
spectrum, and (ii.) the complexity as measured by the number of blocks 
for each candidate basis. For the problem of modeling high dimensional time series, \citep{Ombao:2005} develop a procedure for model selection and dimension reduction 
through the SLEX principal components analysis. In some applications, the 
goal is to discriminate between classes of signals and to classify the test signal. 
Under the SLEX framework, there is a rich set of time-frequency features 
derived from the many potential bases. The SLEX method for discrimination 
selects the basis that gives the maximal discrepancy between classes of signals 
(e.g., signals from healthy controls vs. disease groups). A classification procedure 
for univariate signals was proposed in \citep{DiscriminationClassifNonstationary-Huang-2004} and for multivariate signals 
in \cite{ClassifMultivariateNonstationary-Bohm-2010}. The SLEX method for classification is also demonstrated to be consistent, i.e., 
the probability of misclassification converges to zero as the length of the test 
signal increases.

\subsection{Change-points approach}

One broad class of methods for analyzing non-stationary signals identifies 
the change-points and thus segments the signals into quasi-stationary time 
blocks. There are many of these change-points methods and a few most 
relevant to brain signal analysis are discussed here. 

One line of segmentation methods is based on dyadic segmentation. 
In \cite{Adak:1998time},  a dyadic segmentation of a non-stationary univariate 
signal is developed. The signal is split dyadically up to some specified level $J$ 
so that at each level $j$ there are exactly $2^j$. Each block at level $j$ (mother block) 
is split into two blocks (children blocks). Starting from the deepest level of the 
tree, the spectra at the adjoining children blocks are estimated and then compared. 
When they are similar (based on some discrepancy metric), then these children blocks are 
combined to form one mother block.  Otherwise, they are kept as distinct blocks. 
A large discrepancy between the spectral estimates indicates that there is a 
change-point. The Adak method also imposes a penalty for complexity in order 
to prevent a stationary block from unnecessarily splitting into two. The SLEX methods in 
\citep{Ombao:2005} and \cite{Ombao:2001automatic} can also be viewed as 
dyadic segmentation methods even though the segmentation is merely a 
by-product of the best model (or best SLEX basis) selected from the penalized 
Kullback-Leibler criterion.
Another method based on a similar dyadic division is presented in \citep{DyadicAggregatedAR-Pinto-Orellana-2020} and \citep{DyadicAggregatedAR-Pinto-2021}: the dyadic aggregated autoregressive model (DASAR). In this representation, where each dyadic block at level $j$ is modeled using an aggregated autoregressive model (with degrees of freedom in the number of maximum components and the maximum order of each autoregressive component). Rules for its tree expansion are similar to the abovementioned for SLEX. By construction, DASAR blocks represents one resonating component through two parameters: central frequency and bandwidth (exponential of the negative root modulus $M$ in Equation \ref{eq:AR2-oscillator}). This compact representation shows notable relevance for classification purposes \citep{DyadicAggregatedAR-Pinto-Orellana-2020} and frequency tracking \citep{DyadicAggregatedAR-Pinto-2021}.

For completeness, we also enumerate the other interesting change-point methods that 
have been either applied to brain signals or are potentially useful in analyzing brain 
signals: binary segmentation for transformed autoregressive conditionally heteroscedastic 
models in \cite{BASTA};  group lasso in \cite{NHChan}; score-type test statistics for 
VAR-based models in \cite{Kirch}; test of non-stationarity based on the discrete Fourier 
transform in \cite{SubbaRao-Jentsch}. 

\begin{example}[Non-stationary signal with linearly frequency evolution]\label{ex:slex-comparison-chirp}
Let us assume a non-stationary time series $x(t)$ that has a continuous evolving frequency, i.e. a chirp signal, with a frequency starting at $\omega_0$ and incrementing $\Delta\omega$ Hertz each second:
\begin{equation*}
x(t) = 2\sin\bigr(
           2\pi\bigr(
               \omega_0 + \Delta\omega t
           \bigl)t
       \bigl) + \varepsilon(t)
\end{equation*}
where $\varepsilon(t)\sim\mathcal{N}(0,1)$ is unit-variance white noise.
Recall that these types of signals have been observed in seismic events 
\citep[p.10-11]{ExplorationsTimeFreq-Flandrin-2018}, and also during epileptic episodes as Schiff et al. observed \citep{BrainChirpsSpectrographic-Schiff-2000}.

Figure \ref{fig:slex-comparison-chirp} shows the spectral estimations obtained through short-time Fourier transform, wavelet transform, SLEX and DASAR. Note that the four of them were capable of detecting the time-evolving resonator.

\end{example}

\begin{example}[Time-frequency EEG representation]\label{ex:slex-comparison-eeg}
Recall the ADHD-EEG dataset described in Section \ref{sec:eeg-dataset}.
In Figure \ref{fig:slex-comparison-eeg}, we show the time-varying spectral response of the signals at channel F3 and F4 in the frequency interval 10--30 Hertz (beta band). Remark that the power at the 10-Hertz and 20-Hertz components are changing over time: 20-Hertz oscillations appear stronger at 8-12 seconds and 16-21 seconds with higher magnitudes in the left hemisphere.
\end{example}

\subsection{Switching Processes and Community Detection }

Another class of models characterizes changes in brain connectivity networks via a 
Markov-switching between states. Many approaches that use the sliding window 
or fit some time-varying coefficient models have difficulty in capturing abrupt changes. 
A Markov-switching dynamic factor model was proposed in \citep{Ting2018} which 
captures dynamic connectivity states multivariate brain signals driven by lower-dimensional 
latent factors. A regime-switching vector autoregressive (SVAR) factor process was used 
to measure the dynamic directed connectivity.  There is also an emerging line of models for brain connectivity that are based on stochastic block models. Indeed there is accumulating evidence 
that suggests that functional connectivity patterns are dynamically evolving over multiple 
time scales during rest and while performing a cognitive task. However, 
functional connectivity tends to be temporally clustered into a finite number of  connectivity states. 
These are distinct connectivity patterns that recur over the course of the experiment.  Recent 
work in \citep{Ting2021}  characterizes dynamic functional connectivity - specifically the 
state-driven changes in community organization of brain networks. A key contribution is that 
the approach takes into account variation across individuals. Many key innovations are being 
developed in this line of modeling, and they are anticipated to set the trend given the ability of these 
methods to handle high dimensionality datasets while providing results that are easily interpretable.

\subsection{Time-varying dependence between filtered signals}

Previously described methods for analyzing frequency-specific dependence between 
channels $X_p(t)$ and $X_q(t)$ can be extended to the non-stationary setting where 
this dependence can evolve over time. Recall that $X_{p, \omega}(t)$ and 
$X_{q, \omega}(t)$ are the $\omega$-oscillations derived from the two channels. 
Thus, the time-varying correlation between these oscillations is defined to be
\[
\rho_{pq, \omega}(t^*) = \corr \left ( X_{p, \omega}(t^*), X_{q, \omega}(t^*) \right).
\]
The natural approach for estimating $\rho_{pq, \omega}(t^*)$ is to form a sliding window and 
then compute the local correlation within each window. This approach is motivated 
by a time-dependent generalization of the mixture model in Equations~\ref{Eq:ModelMix1}, 
\ref{Eq:ModelMix2} and \ref{Eq:MultiAR2Mixture}, 
\begin{eqnarray}
\left ( 
\begin{matrix}
X_1(t) \\
X_2(t) \\
\ldots \\
X_P(t)
\end{matrix}
\right )  =  
\left ( 
\begin{matrix}
A_{11}(t) & A_{12}(t) & \ldots & A_{1K}(t) \\
A_{21}(t) & A_{22}(t) & \ldots & A_{2K}(t) \\
\ldots & {} & {} & {} \\
A_{P1}(t) & A_{P2}(t) & \ldots & A_{PK}(t)
\end{matrix}
\right ) \
\left( 
\begin{matrix}
Z_1(t) \\
Z_2(t) \\
\ldots \\
Z_K(t)
\end{matrix}
\right).
\end{eqnarray}\label{Eq:TV-MultiAR2Mixture}
The above model specifies the mixture matrix to have components $A_{pq}(t)$ that 
change with time. The time-varying mixture model in Equation~\ref{Eq:TV-MultiAR2Mixture} 
can be further generalized by allowing for temporal delay $A_{pq}(t)B^{d_{pq}}$, which will 
be captured by the backshift operator $B^{d_{pq}}$.

\section{Code availability and data reproducibility}
An R package (ECOSTASpecDepRef) was developed to provide a reference implementation for the algorithms discussed in the current paper, as well as the procedures for replicating the results shown in Examples 1--10. The source code is available at the repository \url{https://github.com/biostatistics-kaust/ecosta-spectral-dependence}.

\section{Conclusion}

This paper presents a general approach to modeling dependence in multivariate signals 
through the oscillatory activities extracted from each channel. The unifying theme is 
to explore the strength of dependence and possible lead-lag dynamics through filtering. 
The proposed framework is sufficiently comprehensive given its capability to model and 
study both linear and non-linear dependence in both contemporaneous and lagged 
configurations. Furthermore, it was demonstrated here that some of the most prominent 
frequency domain measures such as coherence, partial coherence, and dual-frequency 
coherence could be derived as special cases under this general framework. 

There are numerous open problems in analyzing time series from designed experiments,
with the spectral dependence approach, that were not addressed in this paper due to 
space constraints. 
Among them, we can mention the following possible experimental scenarios:
(i.) There could be well-defined labeled groups, with their own dependence 
characteristics, according to disease (controls vs. ADHD), 
stimulus types (reading vs. math), cognitive load (easy vs. difficult).
(ii.) There could be several participants within each group, and consequently, 
the models should account for the cross-subject variation of 
the brain's functional response within groups.
(iii.) There could be repeated expositions (or trials) to the same stimulus, 
and the model should be able to extract the common information across responses 
while simultaneously modeling trial-wise variations of brain response.
(iv.) There could be different temporal scales in the experiment, and the model 
should take into account transient responses (i.e., within a trial or epoch), 
long-term responses (across the trials in an experiment) while considering the 
developmental changes in the human brain on longitudinal scales.

Statistical models should also allow for assessing differences in the fixed effects 
(e.g., control vs. disease) while also accounting for between-subject variation. 
A candidate approach can be performed via mixed-effects models, such as the 
framework proposed in \cite{MEVAR} and \citep{BayesianVectorAR-Chiang-2017}, or fully Bayesian approaches,
as the methods proposed in \cite{ZheYu-JASA} and \citep{BayesianfMRI}. 
In addition to the need for formal statistical models, there is also a need to develop 
new approaches for studying possible non-linear dependence through high-dimensional models. 
These methods should leverage advances in dimensionality reduction, 
statistical/machine learning, and optimization in order to examine these complex 
dependence structures from high-dimensional models.

\bibliography{REF,Library}

\end{document}